\documentclass[aps]{revtex4}
\usepackage{eurosym}
\usepackage{amsfonts}
\usepackage{amsmath}
\usepackage{amssymb,epsf}
\usepackage{color}
\usepackage{graphicx}
\usepackage{epstopdf}
\usepackage{float}
\usepackage{caption}
\usepackage{subfig}

\begin{document}

\title{Thermal Stability, P-V Criticality and Heat Engine of \\
Charged Rotating Accelerating Black Holes}
\author{B. Eslam Panah$^{1,2,3}$\footnote{%
email address: eslampanah@umz.ac.ir}, and Kh. Jafarzade$^{1,2}$\footnote{
email address: khadije.jafarzade@gmail.com}}
\affiliation{$^{1}$ Department of Theoretical Physics, Faculty of Science, University of
Mazandaran, P. O. Box 47416-95447, Babolsar, Iran\\
$^{2}$ ICRANet-Mazandaran, University of Mazandaran, P. O. Box 47416-95447,
Babolsar, Iran\\
$^{3}$ ICRANet, Piazza della Repubblica 10, I-65122 Pescara, Italy}

\begin{abstract}
In this paper, we study thermodynamic features of the charged rotating accelerating black holes in anti-de Sitter spacetime. First, we consider these black holes as the thermodynamic systems and analyze thermal stability/instability through the use of heat capacity in the canonical ensemble. We also investigate the effects of angular momentum, electric charge and string tension on the thermodynamic quantities and stability of
the system. Considering the known relation between pressure and the cosmological constant, we extract the critical quantities and discuss how the mentioned parameters affect them. Then, we construct a heat engine by taking into account this black hole as the working substance, and obtain the heat engine efficiency by considering a rectangle heat cycle in the $P-V$ plane. We examine the effects of black hole parameters on the efficiency and analyze their effective roles. Finally, by comparing the engine efficiency with Carnot efficiency, we investigate conditions in order to have a consistent thermodynamic second law.
\end{abstract}

\maketitle

\section{Introduction}

Black holes provide practical environments for testing strong gravity. They are one of incredibly important theoretical tools for exploring General Relativity (GR). In this regard, some of black hole solutions such as Schwarzschild, Reissner-N\"{o}rdstrom, Kerr, and Kerr-Newman black holes
have been studied in literature. Among of these solutions, Kerr-Newman black hole is the most general solution. This black hole is characterized by mass ($M$), charge ($Q$) and angular momentum ($J$). However, there is another black hole with interesting properties which is called accelerating black hole. The accelerating black hole solution is extracted from $C$
-metric \cite{Cmetric1,Cmetric2,Cmetric3,Cmetric4,Cmetric5,Cmetric6}. This black hole has a conical deficit angle along one polar axis which provides the force driving acceleration. Conical singularity pulling the black hole can be replaced by a cosmic string \cite{Gregory}, or a magnetic flux tube \cite{Dowker}, and one can imagine that something similar to the $C$-metric
with its distorted horizon could describe a black hole which has been accelerated by an interaction with a local cosmological medium.

There are other reasons for the study of the accelerating black hole. The first reason is related to this fact that the accelerating black hole is described by $C$-metric \cite%
{Cmetric1,Cmetric2,Cmetric3,Cmetric4,Cmetric5,Cmetric6}, which has uncommon asymptotic behavior. In other words, its asymptotic behavior depends on various parameters such as the acceleration parameter, electrical charge, angular coordinate and the cosmological constant which lead to accelerating horizon and complicate the asymptotic structure. The second reason, the accelerating black hole presents at least one non-removable conical singularity on the azimuthal axis of symmetry, both in the rotating and static cases.

One of the exciting and challenging subjects in theoretical physics is black hole thermodynamics. The discovery of a profound connection between the laws of thermodynamics and the gravitational systems has been one of the remarkable achievements of theoretical physics \cite{MBardeen,SHawking}. The
possible identification of a black hole as the thermodynamic object was first realized by Bardeen, Carter, and Hawking \cite{MBardeen}. They clarified the laws of black hole mechanics and showed that these laws are identical to ordinary thermodynamics. In recent years, investigation of black hole thermodynamics in anti de Sitter (AdS) spacetime provided us with
deep insights in understanding the quantum nature of gravity which has been one of the most important theoretical subjects in physical communities \cite%
{AdS1,AdS2,AdS3,AdS4,AdS5,AdS6,AdS7,AdS8,AdS9,AdS10,AdS11,AdS12,AdS13,AdS14,AdS15}. According to AdS/CFT correspondence \cite{JMaldacena,SGubser,EWitten}, the thermodynamics of black hole in asymptotically AdS spacetime can be recognized by that of Conformal Field Theory (CFT) living on the boundary of AdS spacetime.

One of the interesting issues in the context of black hole thermodynamics is studying the phase transition. The issue of phase transition plays a significant role in understanding of the microscopic structure of black holes. The study on phase transition of black hole in an asymptotically AdS spacetime was initiated by Hawking and Page \cite{DNPage}, who demonstrated a
certain phase transition between thermal AdS and Schwarzschild AdS black hole. This phase transition can be interpreted as a
confinement/deconfinement phase transition in the dual strongly coupled gauge theory. Consideration of the cosmological constant as a thermodynamic pressure, opened up new avenues in studying the thermodynamic phase structure of the black holes \cite{GCognola,DKastor}. Such an idea modifies the first law of black hole thermodynamics and changes the role of black
hole mass from internal energy to enthalpy \cite{BPDolan1,BPDolan2}. With this interpretation, the mass of a black hole plays a significant role in the thermodynamic structure of black holes and also contains more information
regarding its phase structure \cite{MCvetic}. Considering the cosmological constant as the pressure of a system leads to a van der Waals-like phase transition for black holes (phase transition between small and large black holes). Thermodynamic critical behavior and van der Waals-like phase transition of black holes in the presence of different matter fields and various theories of gravity have been
investigated in Refs. \cite%
{PV1,PV2,PV3,PV3a,PV3b,PV4,PV5,PV6,PV7,PV8,PV9,PV9a,PV10,PV11,PV11a,PV11b,PV11c,PV12,PV12a,PV13,PV14,PV14a,PV15,PV5a,PV16,PV17,PV18,PV18a,PV19,PV20,PV21,PV22}. Also, some efforts have been made in the context of thermodynamics, $P-V$ criticality, phase transition, and geometrothermodynamics of accelerating black holes in Refs. \cite{Accel1,Accel2,Accel3,Accel4,Accel5,Accel6,Accel7}. Using the standard black hole thermodynamics, it was shown that the
Hawking temperature of accelerating black hole is more than Unruh temperature of the accelerated frame. Recently, Gregory and Scoins have introduced a new set of chemical variables for the accelerating black hole and suggested that conical defects emerging from a black hole can be considered as true hair \cite{GScoins}.
 
Considering black holes as thermodynamic systems in the extended phase space, it is natural to use them as heat engines. In fact, having thermodynamic pressure and volume at hand, it is possible to extract mechanical useful work from the heat energy via the $VdP$ term in the first law. The concept of the holographic heat engine was first proposed by Johnson in Ref. \cite{CVJohnson}. He used the charged AdS black hole as a heat engine working substance to construct a holographic heat engine and calculated the efficiency by defining a closed path in the $P-V$ plane.
Afterward, this issue has been gained a lot of attention in black hole thermodynamics and a number of attempts were conducted in different black hole backgrounds, such as the rotating black holes \cite{RAHennigar,Jafarzade1}, Horava-Lifshitz black holes \cite{Jafarzade2}, Born-Infeld black holes \cite{CVJohnson2}, charged BTZ black holes \cite{JXMo}, black holes in massive gravity \cite{Meng,Ghanaatian,Wang} and
gravity's rainbow \cite{Panah}. The holographic heat engine of the accelerating black hole is investigated in Refs. \cite{Zhang1,Zhang2}, and showed that the acceleration parameter increases the efficiency. Although our finding of charged rotating accelerating black hole will demonstrate
that the efficiency decreases by increasing the acceleration parameter.

The outline of our paper is as follows: in the next section, we give a brief review of the charged rotating accelerating black holes and investigate the admissible parameter space of the solutions. Then we explore the impact of black hole parameters on thermodynamic quantities and study the thermal stability of these black holes in the context of canonical ensemble. We
also will investigate the existence of van der Waals-like behavior of such black holes and discuss how the parameters of the black holes affect critical quantities. The heat engine efficiency is another interesting quantity that we will evaluate in section IV. We finish with concluding remarks in the last section.

\section{Charged rotating accelerating black hole}

The charged rotating accelerating black hole solutions are described by the following metric \cite{Andres} 
\begin{eqnarray}
ds^{2} &=&\frac{1}{\Omega ^{2}}\bigg\lbrace-\frac{f(r)}{\Sigma }\left[ \frac{%
dt}{\alpha }-a\sin ^{2}\theta \frac{d\varphi }{K}\right] ^{2}+\frac{\Sigma }{%
f(r)}dr^{2}  \notag \\
&&  \notag \\
&&\hspace{-10mm}+\frac{\Sigma r^{2}}{h(\theta )}d\theta ^{2}+\frac{h(\theta
)\sin ^{2}\theta }{\Sigma r^{2}}\left[ \frac{adt}{\alpha }-(r^{2}+a^{2})%
\frac{d\varphi }{K}\right] ^{2}\bigg\rbrace,  \label{Eq1}
\end{eqnarray}%
where $f(r)$, $h(\theta )$ and $\Sigma $ are in the following forms \cite%
{Andres} 
\begin{eqnarray}
f(r) &=&(1-A^{2}r^{2})\left[ 1-\frac{2m}{r}+\frac{a^{2}+e^{2}}{r^{2}}\right]
+\frac{r^{2}+a^{2}}{\ell ^{2}}, \\
&&  \notag \\
h(\theta ) &=&1+2mA\cos \theta +\left[ A^{2}(a^{2}+e^{2})-\frac{a^{2}}{\ell
^{2}}\right] \cos ^{2}\theta ,  \label{h} \\
&&  \notag \\
\Sigma  &=&1+\frac{a^{2}}{r^{2}}\cos ^{2}\theta ,  \label{hteta}
\end{eqnarray}%
and the corresponding gauge potential is expressed as \cite{Andres} 
\begin{equation}
E=d\mathcal{B},~~~~\&~~~\mathcal{B}=-\frac{e}{\Sigma r}\left[ \frac{dt}{%
\alpha }-a\sin ^{2}\theta \frac{d\varphi }{K}\right] +\Phi _{t}dt,
\label{Eq3}
\end{equation}%
where 
\begin{equation}
\Phi _{t}=\frac{er_{+}}{(a^{2}+r_{+}^{2})\alpha },  \label{Eq4}
\end{equation}%
in which $r_{+}$ is related to the event horizon of the black holes. Also, the parameters $A$ ($A>0$) and $\ell ^{2}=-3/\Lambda ^{2}$ (where $\Lambda $ is the cosmological constant) are the acceleration parameter and AdS radius, respectively.

The parameters $a$, $e$, and $m$, respectively, are related to the angular momentum, electric charge and total mass of the black hole in the following manner \cite{Andres} 
\begin{equation}
J=\frac{ma}{K^{2}},~~~~\&~~~Q=\frac{e}{K},~~~~\&~~~M=\frac{m(\Xi +\frac{a^{2}%
}{\ell ^{2}})(1-A^{2}\ell ^{2}\Xi )}{K\Xi \alpha (1+a^{2}A^{2})},
\label{Eq5}
\end{equation}%
where the factor of $\alpha $ is chosen for rescalling the time coordinate.

To have the correct thermodynamics, one should consider an appropriate
normalized time $\tau =\alpha t$ with 
\begin{equation*}
\alpha =\frac{\sqrt{(\Xi +\frac{a^{2}}{\ell ^{2}})(1-A^{2}\ell ^{2}\Xi )}}{%
1+a^{2}A^{2}},
\end{equation*}%
where 
\begin{equation*}
\Xi =1-\frac{a^{2}}{\ell ^{2}}+A^{2}(e^{2}+a^{2}).
\end{equation*}

Since we are interested in small rotating accelerating black holes, we can neglect all terms higher-order in $A^{2}$ and $a^{2}$, and consider $\alpha $ as 
\begin{equation*}
\alpha \simeq \sqrt{1+e^{2}A^{2}-A^{2}\ell ^{2}}.
\end{equation*}

The conical deficits on the north pole ($\theta _{+}=0$) and the south pole ($\theta _{-}=\pi $) are given by 
\begin{equation}
\delta _{\pm }=2\pi \left( 1-\frac{h(\theta _{\pm })}{K}\right) ,
\label{Eq6}
\end{equation}%
which corresponds to a cosmic string with tension \cite{Zhang2,Andres} 
\begin{equation}
\mu _{\pm }=\frac{\delta _{\pm }}{8\pi }=\frac{1}{4}\left[ 1-\frac{\Xi \pm
2mA}{K}\right] =\frac{1}{4}\left[ 1-\frac{K_{\pm }}{K}\right] .  \label{Eq7}
\end{equation}

To have positive tension defects, we require $0\leq \mu _{+}\leq \mu_{-}\leq \frac{1}{4}$. Here, by setting $K=K_{+}=\Xi +2mA$, one can remove the conical singularity on the north pole axis. In this case, only one string tension $\mu=\mu _{-}=\frac{mA}{K}$ (located at the south pole) pulls
on the black holes. 
%%%%%%%%%%%%%%%%%%%%%%%%%%%%%%%%%%%%%%%%%%%%%%%%%%%%%%%%%%%%%%%%%%%%%%%%%%%%%%%%%%%%%%%%%%

\begin{figure}[tbh]
\centering
\subfloat[$\beta=0.1$ and $\mu =0.15$]{
        \includegraphics[width=0.33\textwidth]{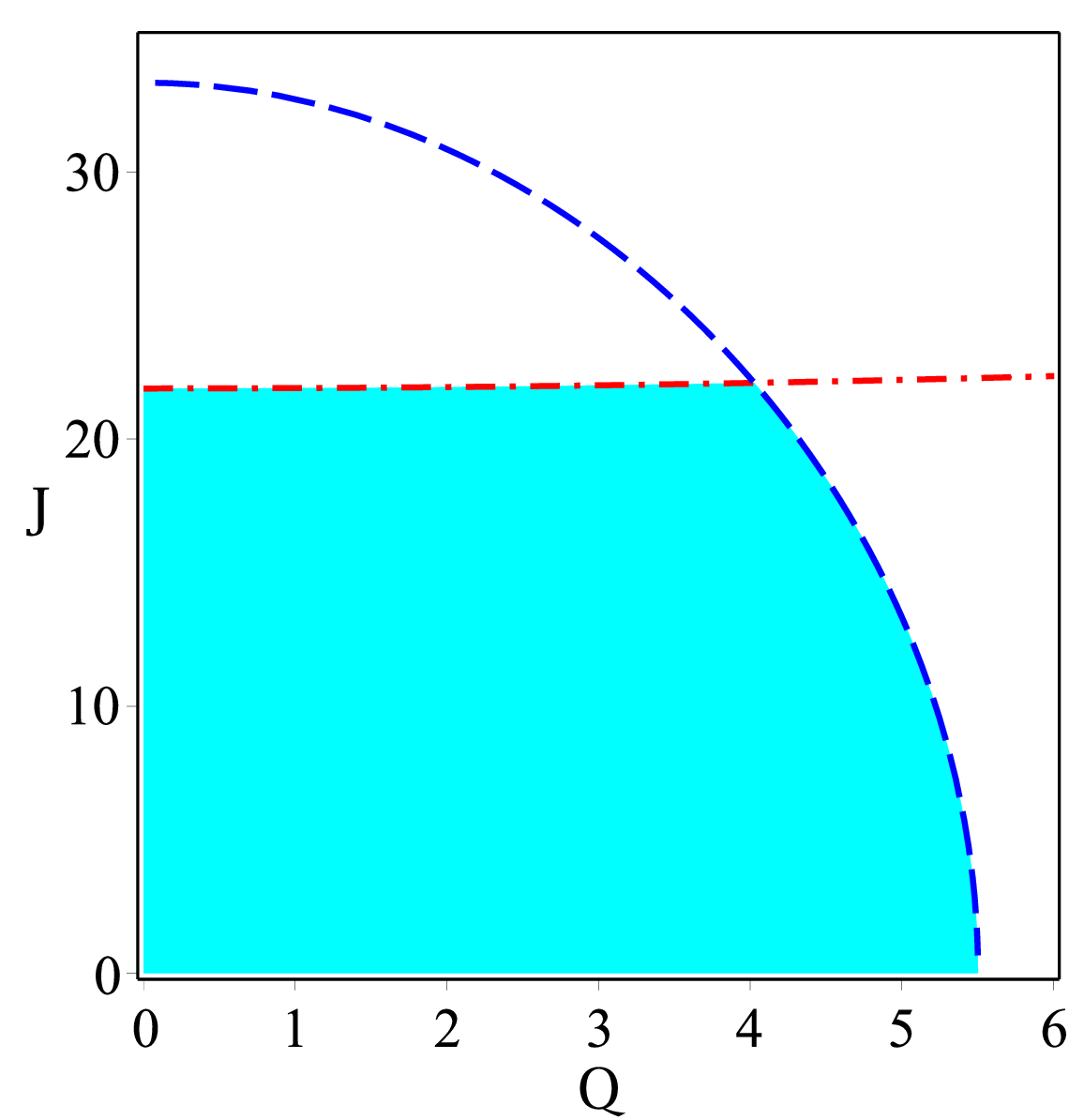}} 
\subfloat[$\beta=0.04$ and $\mu =0.15$]{
        \includegraphics[width=0.32\textwidth]{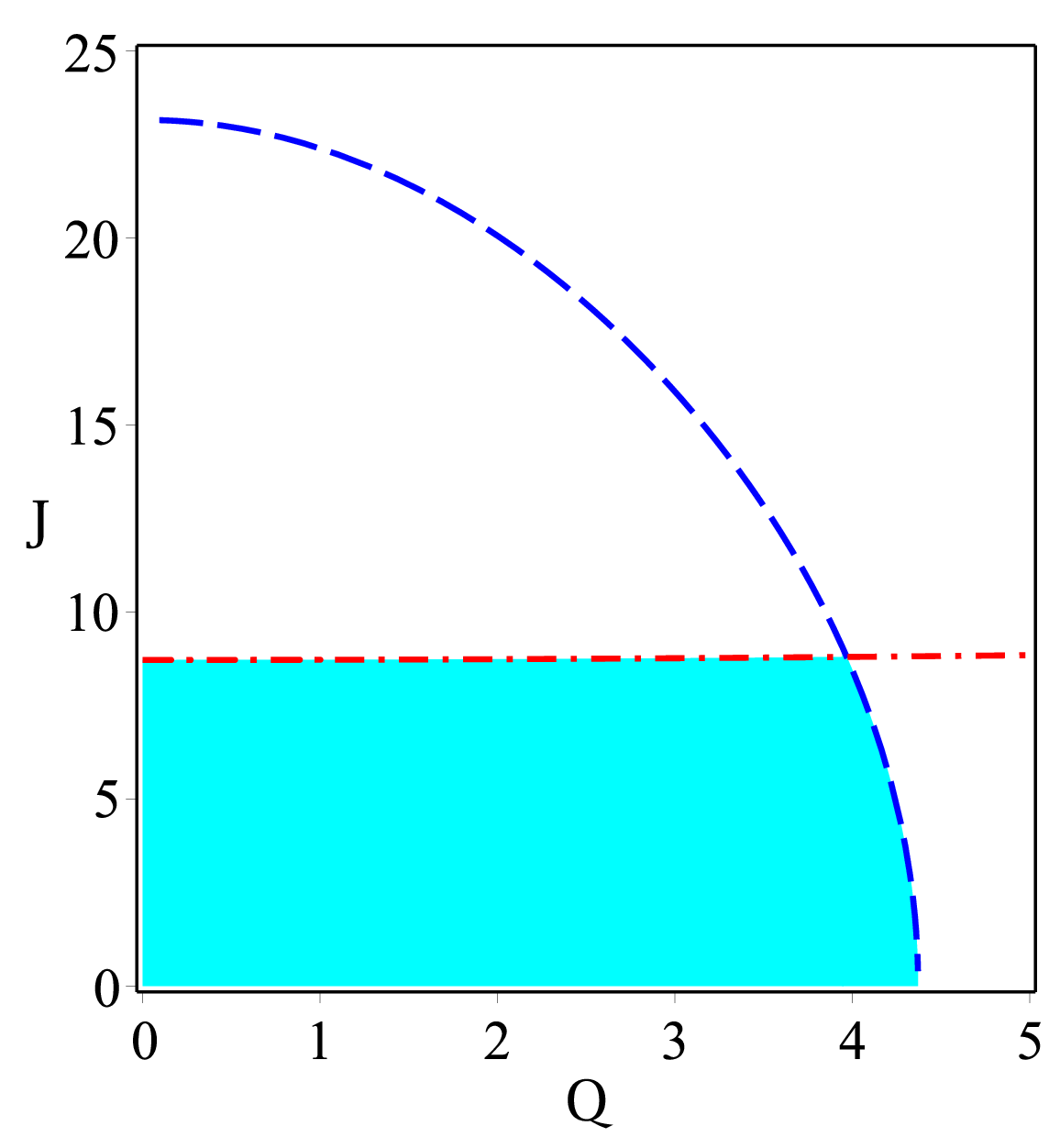}} 
\subfloat[$\beta=0.04$ and $\mu =0.1$]{
        \includegraphics[width=0.325\textwidth]{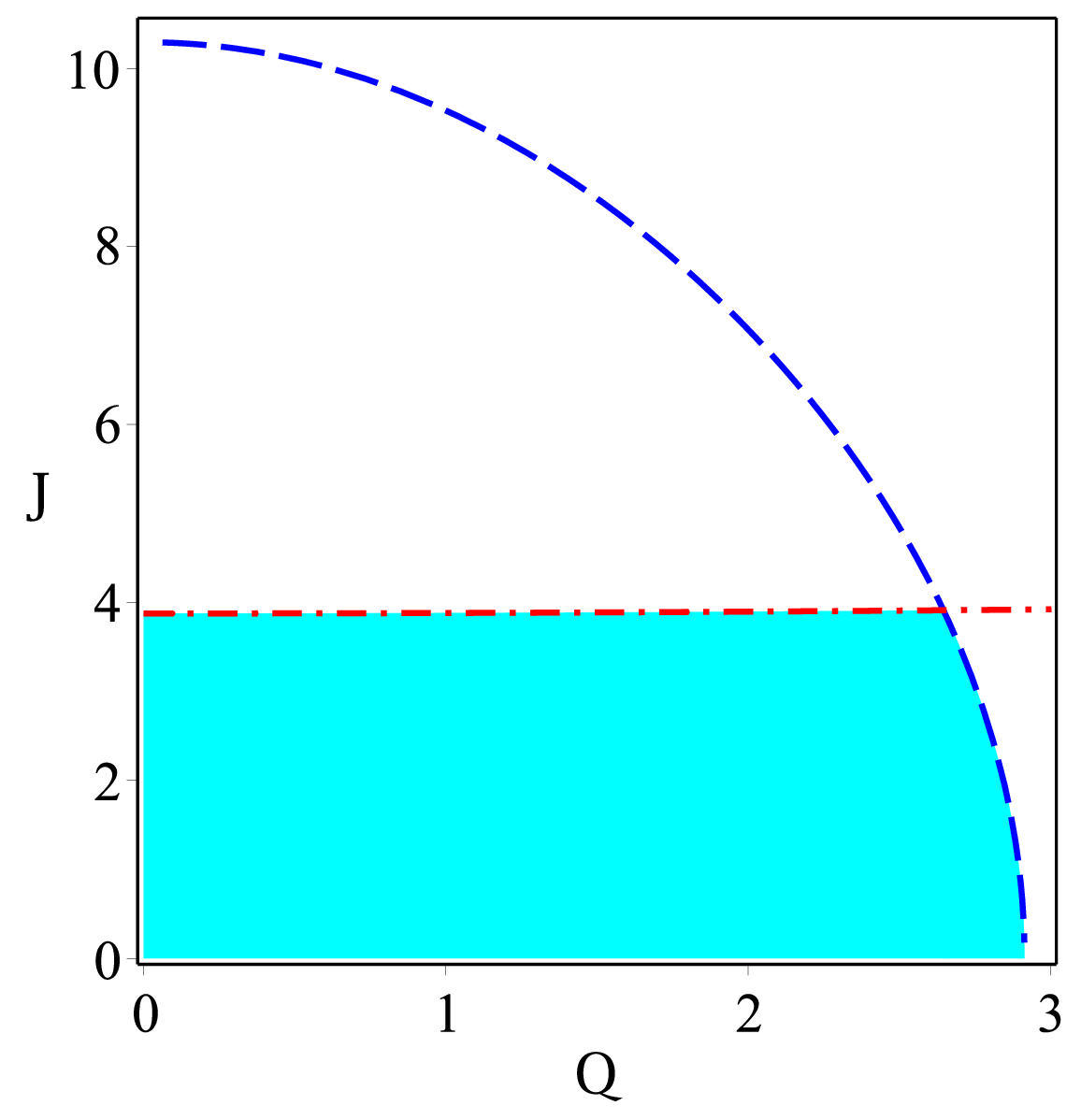}}\newline
\caption{ The admissible region (denoted by shaded areas) is displayed in
the $J$ - $Q$ plane for $B=mA=0.2$ and $A=0.02$. The red dash-dotted curve
outline the boundary of positive $h$. The blue dashed curve is the boundary
for the existence of black holes in the bulk, with extremal black holes
sitting on the curve. }
\label{Fig1}
\end{figure}
%%%%%%%%%%%%%%%%%%%%%%%%%%%%%%%%%%%%%%%%%%%%%%%%%%%%%%%%%%%%%%%%%%%%%%%%%%%%%%%%%%%%%%

It should be pointed out that the metric (\ref{Eq1}) will be well defined (correspond physically to a slowly accelerated black hole in the bulk) under the satisfaction of three conditions:

i) the function $h(\theta)$ must be positive in range of $[0,\pi ]$, which implies \cite{Andres} 
\begin{equation*}
mA<\left\{ 
\begin{array}{cc}
\frac{1}{2}\Xi ~~~for~~~\Xi \in (0,2] &  \\ 
\text{ \ \ \ \ \ \ \ \ \ \ \ \ \ } &  \\ 
\sqrt{\Xi -1}~~~for~~~\Xi >2 & 
\end{array}%
\right. ,
\end{equation*}
the boundary is displayed in Fig. \ref{Fig1}, by the red dash-dotted line.

ii) due to the requirement of slow acceleration, $f(-\frac{1}{A\cos \theta }) $ has to have no root.

iii) the function $f(r)$ must have at least one root in the range of $r\in(0,\frac{1}{A})$. To investigate it we appeal to the condition for the extremal black holes $f(r_{e})=0=f^{\prime }(r_{e})$, which results in the following relations 
\begin{eqnarray}
J &=&\sqrt{\frac{\mu ^{4}\left[ \left( 1-\beta ^{2}\right)
A^{5}r_{e}^{5}-2\left( 1-\beta ^{2}\right) A^{3}r_{e}^{3}+\beta ^{2}\left(
B-Ar_{e}\right) -B\beta ^{2}A^{2}r_{e}^{2}\left( 2-A^{2}r_{e}^{2}\right) %
\right] }{B^{2}A^{5}r_{e}}},  \notag \\
&&  \notag \\
Q &=&\sqrt{\frac{\mu ^{2}\left[ \left( 1-\beta ^{2}\right)
^{2}A^{5}r_{e}^{5}+\left( 1-\beta ^{2}\right) B\beta
^{2}A^{4}r_{e}^{4}+2\left( 1-\beta \right) \beta ^{2}A^{3}r_{e}^{3}+B\left(
1+2\beta ^{2}\right) \beta ^{2}A^{2}r_{e}^{2}-\beta ^{4}\left(
B-Ar_{e}\right) \right] }{B^{2}\beta ^{2}A^{3}r_{e}}},
\end{eqnarray}%
where $B=mA$ and $\beta =A\ell $, are considered as constant parameters throughout this paper. Also $r_{e}$ is related to the radius of extremal black holes.

The resultant curve is displayed in Fig. \ref{Fig1}, by the blue dashed line, denoting the extremal limit. Below this line, black holes (with two horizons) are present, whereas no black hole exists above it.

\section{THERMODYNAMICS}

In this section, we would like to investigate the effects of black hole parameters on the themodynamic quantities of the system. Then, we study the thermal stability of such black holes in the context of canonical ensemble. Furthermore, we study the van der Waals-like behavior of these black holes; and extract critical quantities. Then we discuss how these quantities change under the variation of different parameters.

\subsection{Thermodynamics quantities}

Thermodynamic quantities of the charged rotating black holes are given by 
\cite{Andres}%
\begin{eqnarray}
T &=&\frac{f_{+}^{\prime }r_{+}^{2}}{4\pi \alpha \left(
r_{+}^{2}+a^{2}\right) },  \label{T} \\
&&  \notag \\
S &=&\frac{\pi \left( r_{+}^{2}+a^{2}\right) }{K\left(
1-A^{2}r_{+}^{2}\right) },  \label{S} \\
&&  \notag \\
V &=&\frac{4\pi }{3K\alpha }\left[ \frac{r_{+}\left( r_{+}^{2}+a^{2}\right) 
}{\left( 1-A^{2}r_{+}^{2}\right) ^{2}}+\frac{m\left[ a^{2}\left( 1-A^{2}\ell
^{2}\Xi \right) +A^{2}\ell ^{4}\Xi \left( \Xi +\frac{a^{2}}{\ell ^{2}}%
\right) \right] }{\left( 1+A^{2}a^{2}\right) \Xi }\right] ,  \label{V123} \\
&&  \notag \\
\lambda _{\pm } &=&\frac{r_{+}}{\alpha \left( 1\pm Ar_{+}\right) }-\frac{m%
\left[ \Xi +\frac{a^{2}}{\ell ^{2}}+\frac{a^{2}}{\ell ^{2}}\left(
1-A^{2}\ell ^{2}\Xi \right) \right] }{\alpha \left( 1+A^{2}a^{2}\right) \Xi
^{2}}\mp \frac{A\ell ^{2}\left( \Xi +\frac{a^{2}}{\ell ^{2}}\right) }{\alpha
\left( 1+A^{2}a^{2}\right) }.  \label{EqTSV}
\end{eqnarray}%
where $f_{+}^{\prime }=\left. \frac{df\left( r\right) }{dr}\right\vert
_{r=r_{+}}$. Also, $\lambda _{\pm }$ are conjugate quantities of string
tensions $\mu _{\pm }$.

Temperature is the first quantity that we will investigate. Inserting Eq. (\ref{Eq5}) into temperature relation in Eq. (\ref{T}), one can rewrite the temperature in terms of $Q$ and $J$ as%
\begin{equation}
T=\frac{X_{1}^{2}}{4\pi \alpha r_{+}(X_{1}^{2}+4J^{2}B^{4}\ell ^{4})}\left(
1+\frac{3r_{+}^{2}}{\ell ^{2}}-\frac{\left( 1+4B\right) Q^{2}}{r_{+}^{2}}+%
\frac{4J^{2}B^{4}\ell ^{2}\left( r_{+}^{2}+2Q^{2}-\ell ^{2}\right) }{%
X_{1}^{2}}-A^{2}r_{+}^{2}X_{2}\right) ,  \label{Eq9}
\end{equation}%
where%
\begin{eqnarray*}
X_{1} &=&\mu ^{2}\left[ r_{+}^{4}+\ell ^{2}r_{+}^{2}+\left( 1+4B\right)
Q^{2}\ell ^{2}\right] , \\
&& \\
X_{2} &=&1-\frac{Q^{2}}{r_{+}^{2}}-\frac{2r_{+}^{2}}{\ell ^{2}}+\frac{2Q^{4}%
}{r_{+}^{4}}-\frac{4J^{2}B^{4}\ell ^{2}\left( 2r_{+}^{2}+\ell ^{2}\right) }{%
X_{1}^{2}}.
\end{eqnarray*}

To preserve the metric signature and remove the acceleration horizon, one
should consider $2B<1$ and $\beta <1$, respectively \cite{Anabalon}.

Analyzing the temperature for small and large values of the horizon radius,
provides us with interesting information regarding these black holes. It is
a matter of calculation to show that limiting behaviors of the temperature
are given by 
\begin{equation}
T~\Rightarrow \left\{ 
\begin{array}{ccc}
T\propto -\frac{\left( 1+4B+2A^{2}Q^{2}\right) \left( 1+4B\right)
^{2}Q^{6}\mu ^{4}\ell ^{4}}{4\pi \alpha r_{+}^{3}\left( 4J^{2}B^{4}\ell
^{4}+\left( 1+4B\right) ^{2}Q^{4}\mu ^{4}\ell ^{4}\right) }+O\left( \frac{1}{%
r_{+}}\right) , &  & \text{very small black holes} \\ 
&  &  \\ 
T\propto \frac{A^{2}r_{+}^{3}}{2\pi \alpha \ell ^{2}}+\frac{\left( 3-\beta
^{2}\right) r_{+}}{4\pi \alpha \ell ^{2}}+O\left( \frac{1}{r_{+}}\right) , & 
& \text{very large black holes}%
\end{array}%
\right. .
\end{equation}

\begin{figure}[!htb]
\centering
\subfloat[ $\mu =0.15$]{
        \includegraphics[width=0.32\textwidth]{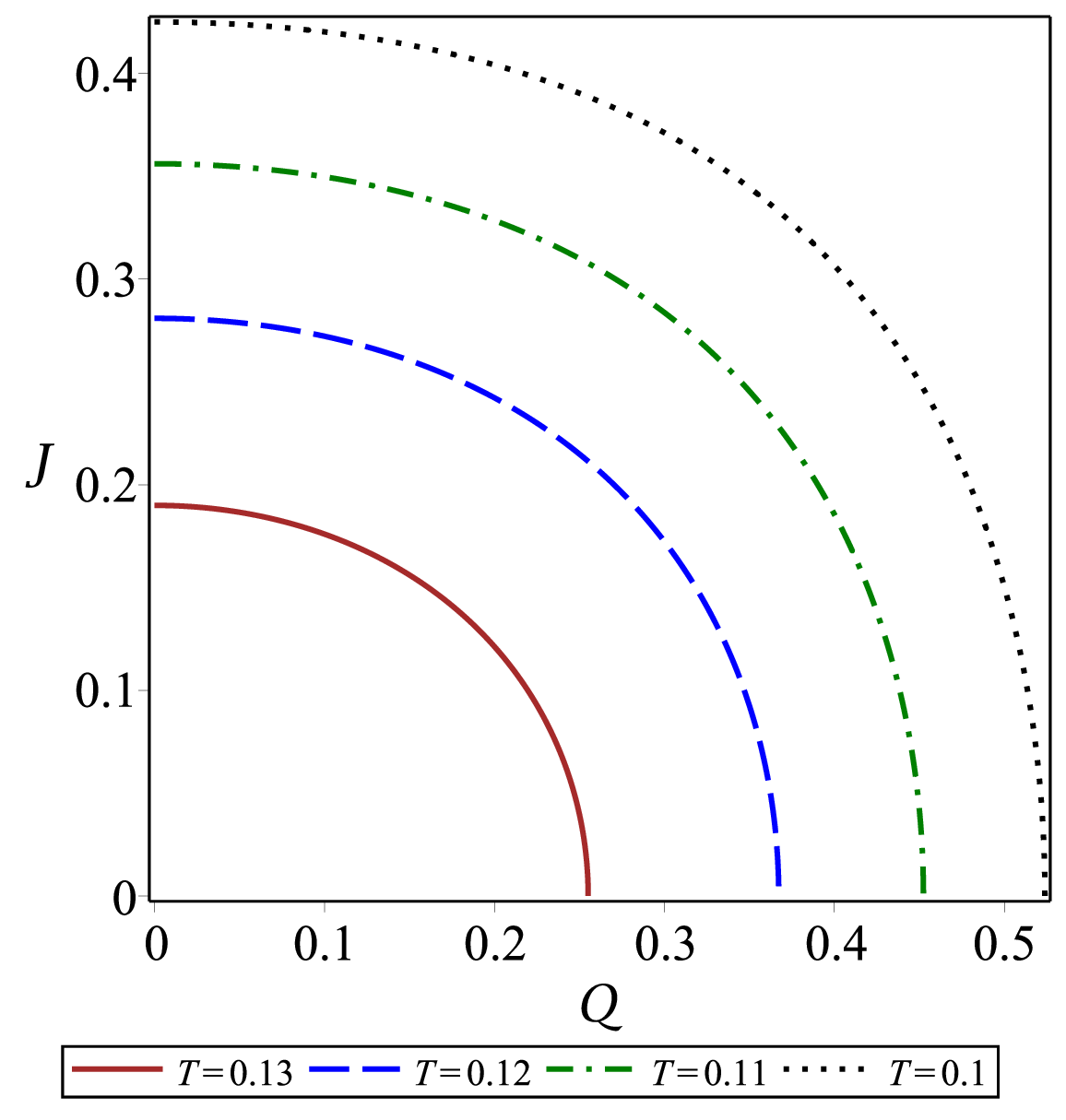}} 
\subfloat[$J=0.02$]{
        \includegraphics[width=0.33\textwidth]{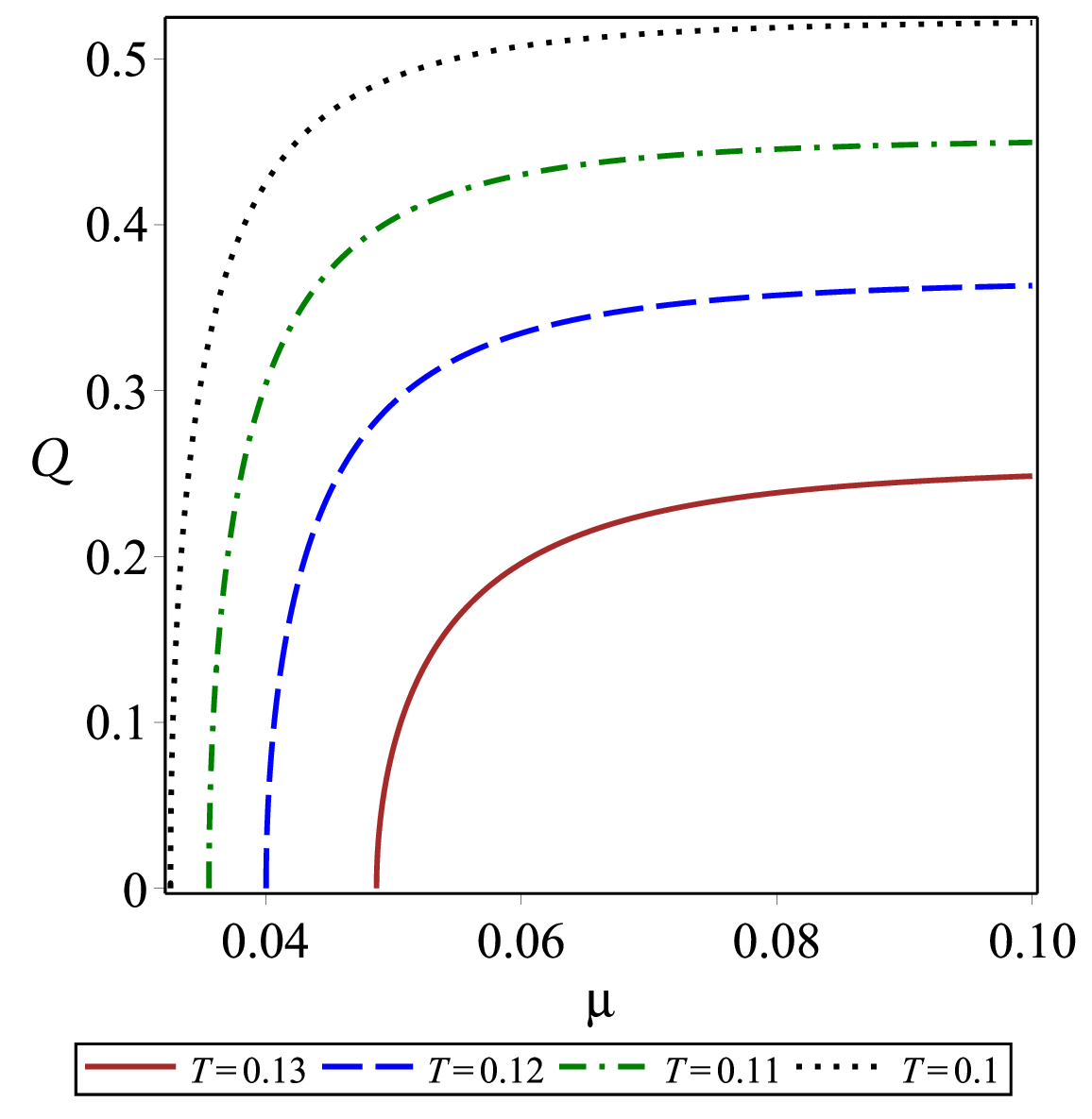}} \newline
\caption{\textbf{Left}: a plot of $J $ vs $Q $ for various values of $T $ at
fixed $\protect\mu $. \textbf{Right} : a plot of $Q $ vs $\protect\mu $ for
various values of $T $ at fixed $J $. We have set $r_{+}=1$, $B=0.2$, $%
\protect\beta =0.04$ and $A=0.02$.}
\label{Fig3}
\end{figure}

According to this fact that $B$ ($B=mA$, where $A>0$ and $m>0$) and $\alpha $ are always positive, so the temperature of very small black holes are always negatively valued. Since the negative temperature indicates the non-physical solutions, very small black holes are non-physical. On the other hand, for large values of the horizon radius, the temperature is positive which confirms the existence of physical solutions for large black holes. It is worth pointing out that although for small black holes, the temperature depends on all parameters, for large black holes, this quantity is dominated by the acceleration parameter and AdS radius. To study the behavior of temperature for medium black holes, we have plotted Fig. \ref{Fig3}. From Fig. \ref{Fig3}a, we observe that the temperature is a decreasing function of the angular momentum and electric charge. Whereas, the effect of string tension is to increase it (see Fig. \ref{Fig3}b).

%%%%%%%%%%%%%%%%%%%%%%%%%%%%%%%%%%%%%%%%%%%%%%%%%%%%%%%%%%%%%%%%%%%%%%%%%%%%%%%%%%%%%
%%%%%%%%%%%%%%%%%%%%%%%%%%%%%%%%%%%%%%%%%%%%%%%%%%%%%%%%%%%%%%%%%%%%%%%%%%%%%%%%%%%%%%

\begin{figure}[tbh]
\centering
\subfloat[ $Q=0.2$]{
        \includegraphics[width=0.32\textwidth]{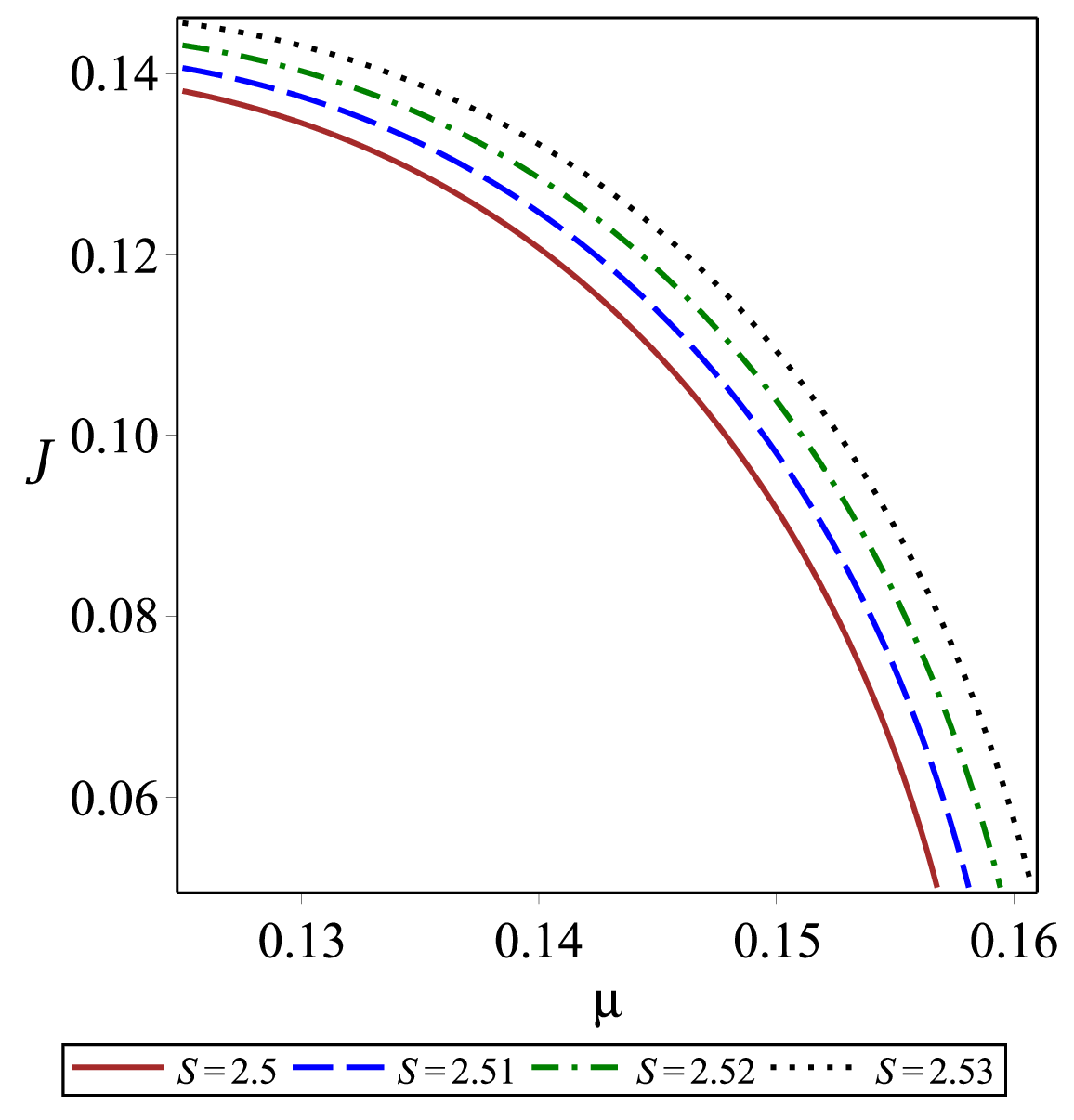}} 
\subfloat[$\mu=0.15$]{
        \includegraphics[width=0.33\textwidth]{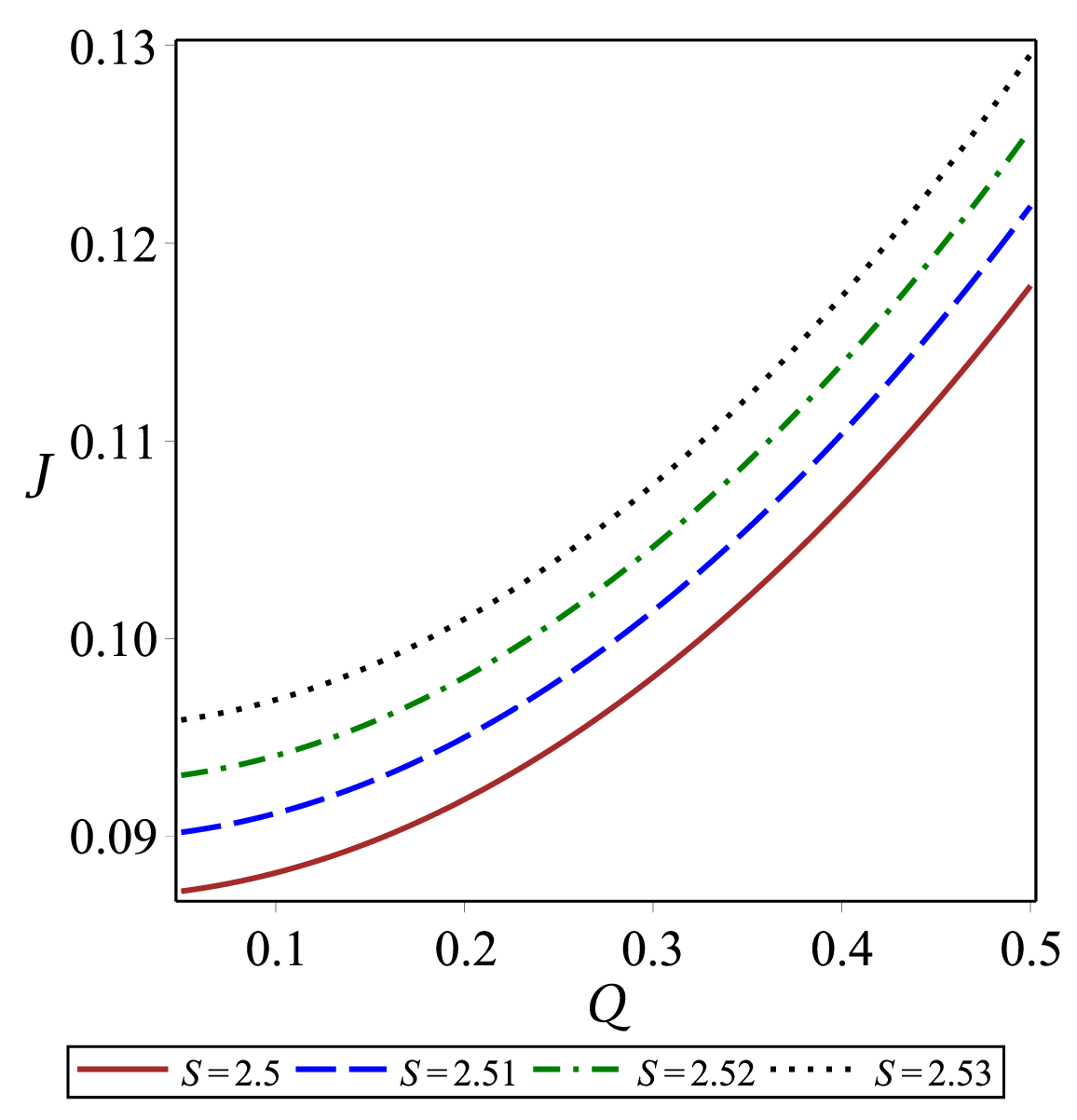}} \newline
\caption{\textbf{Left}: a plot of $J$ vs $\protect\mu $ for various values
of $S$ at fixed $Q$. \textbf{Right} : a plot of $J$ vs $Q$ for various
values of $S$ at fixed $\protect\mu $. We have set $r_{+}=1$, $B=0.2$, $%
\protect\beta =0.04$ and $A=0.02$.}
\label{Fig4}
\end{figure}

Another thermodynamic quantity of interest is entropy. The entropy relation in Eq. (\ref{S}) can be rewritten as the function of $Q$ and $J$ as 
\begin{equation}
S=\frac{\pi \mu r_{+}^{2}(X_{1}^{2}+4J^{2}B^{4}\ell ^{4})}{%
BX_{1}^{2}(1-A^{2}r_{+}^{2})},  \label{Eq11}
\end{equation}%
which gives us the following information:

I) the entropy is an increasing function of the angular momentum and string tension (see Fig. \ref{Fig4}a). While, from Fig. \ref{Fig4}b, the electric charge has a decreasing contribution to this quantity.

II) in the limit of very small value of the event horizon radius, the entropy is always positive 
\begin{equation}
S\propto \left( \frac{\pi \mu }{B}+\frac{4\pi J^{2}B^{3}}{\mu
^{3}(1+4B)^{2}Q^{4}}\right) r_{+}^{2}+O(r_{+}^{4}),
\end{equation}
as it was already mentioned, there is no physical solution in this case due to the negativity of temperature.

III) for very large value of the event horizon radius, the entropy is negative 
\begin{equation}
S\propto -\frac{\pi \mu }{BA^{2}}+O(\frac{1}{r_{+}^{2}}).
\end{equation}

IV) the entropy diverges at $Ar_{+}=1$. To avoid divergence and negativity, one should consider $Ar_{+}<1$.

%%%%%%%%%%%%%%%%%%%%%%%%%%%%%%%%%%%%%%%%%%%%%%%%%%%%%%%%%%%%%%%%%%%%%%%%%%%%%%%%%%%%%
%%%%%%%%%%%%%%%%%%%%%%%%%%%%%%%%%%%%%%%%%%%%%%%%%%%%%%%%%%%%%%%%%%%%%%%%%%%%%%%%%%%%%

The next thermodynamic quantity is the total mass of the black hole (Eq. \ref{Eq5}) which is rewritten as follows

\begin{equation}
M=\frac{\mu \alpha r_{+}}{2BX_{3}}\left( 1+\frac{\left( 1+4B\right) Q^{2}}{%
r_{+}^{2}}+\frac{2A^{2}Q^{4}}{r_{+}^{2}}+\frac{r_{+}^{2}}{\ell ^{2}\left(
1-A^{2}r_{+}^{2}\right) }+\frac{4\ell ^{2}J^{2}B^{4}\left[ r_{+}^{2}+\ell
^{2}-\beta ^{2}r_{+}^{2}-2Q^{2}\left( 1-A^{2}r_{+}^{2}\right) \right] }{%
X_{1}^{2}\left( 1-A^{2}r_{+}^{2}\right) }\right) ,  \label{M}
\end{equation}%
where 
\begin{equation*}
X_{3}=1+\left( 1+4B\right) A^{2}Q^{2}-\frac{4J^{2}\ell
^{2}B^{4}r_{+}^{2}\left( 1-\beta ^{2}\right) }{X_{1}^{2}}.
\end{equation*}

To understand the total mass in more details, we find its limiting behaviors as follows

\begin{equation*}
M~\Rightarrow \left\{ 
\begin{array}{ccc}
M\propto \frac{\mu \alpha \left[ \left( 1+4B\right) Q^{2}+2A^{2}Q^{4}\right] 
}{2B\left[ 1+\left( 1+4B\right) A^{2}Q^{2}\right] r_{+}}+O(r_{+}), &  & 
\text{very small black holes} \\ 
&  &  \\ 
M\propto -\frac{\mu \alpha r_{+}\left( 1-\beta ^{2}\right) }{2B\left[
1+\left( 1+4B\right) A^{2}Q^{2}\right] \beta ^{2}}+O\left( \frac{1}{r_{+}}%
\right) , &  & \text{very large black holes}%
\end{array}%
\right. .
\end{equation*}

\begin{figure}[tbh]
\centering
\subfloat[ $\mu =0.15$]{
        \includegraphics[width=0.32\textwidth]{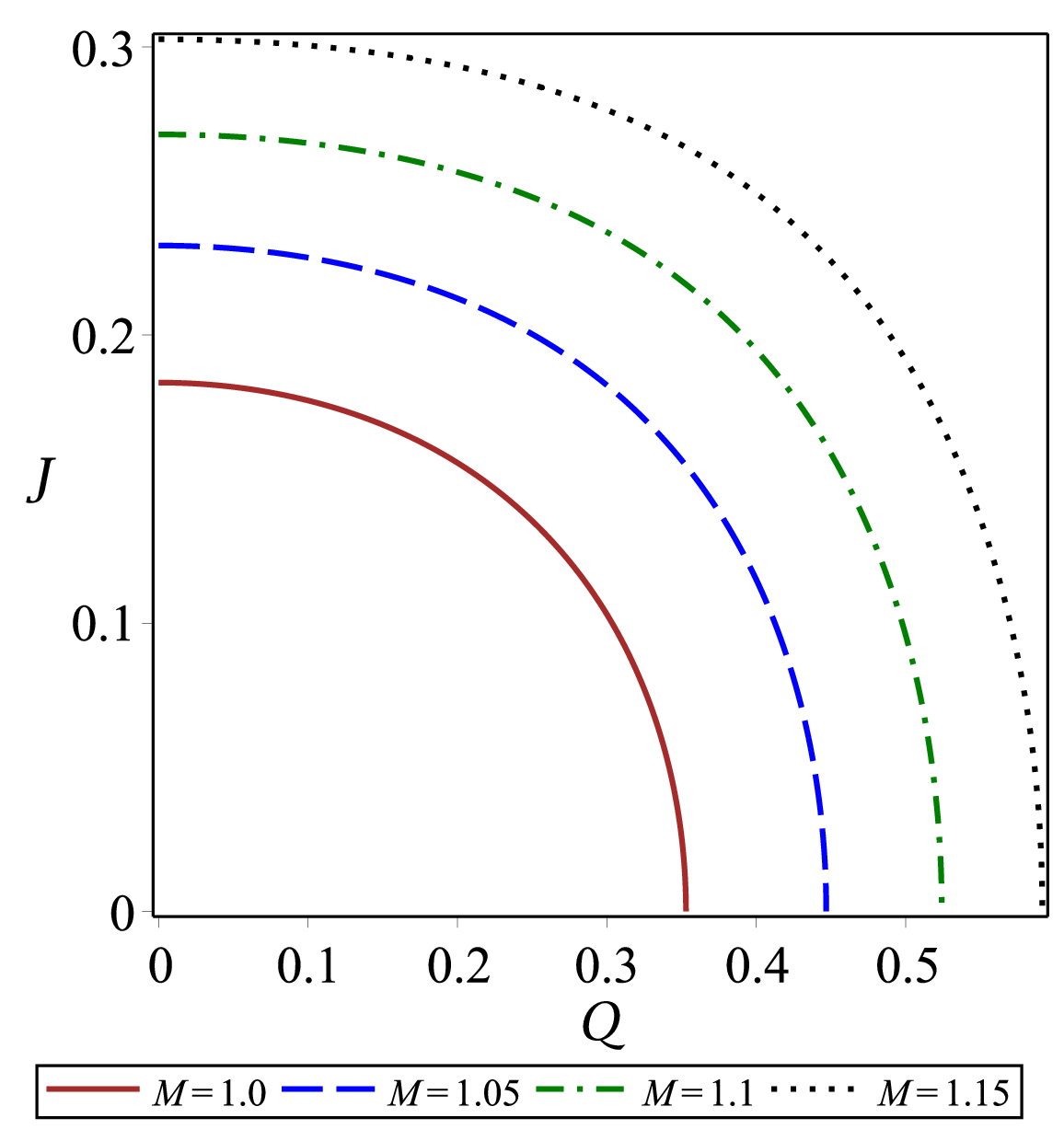}} 
\subfloat[$J=0.02$]{
        \includegraphics[width=0.32\textwidth]{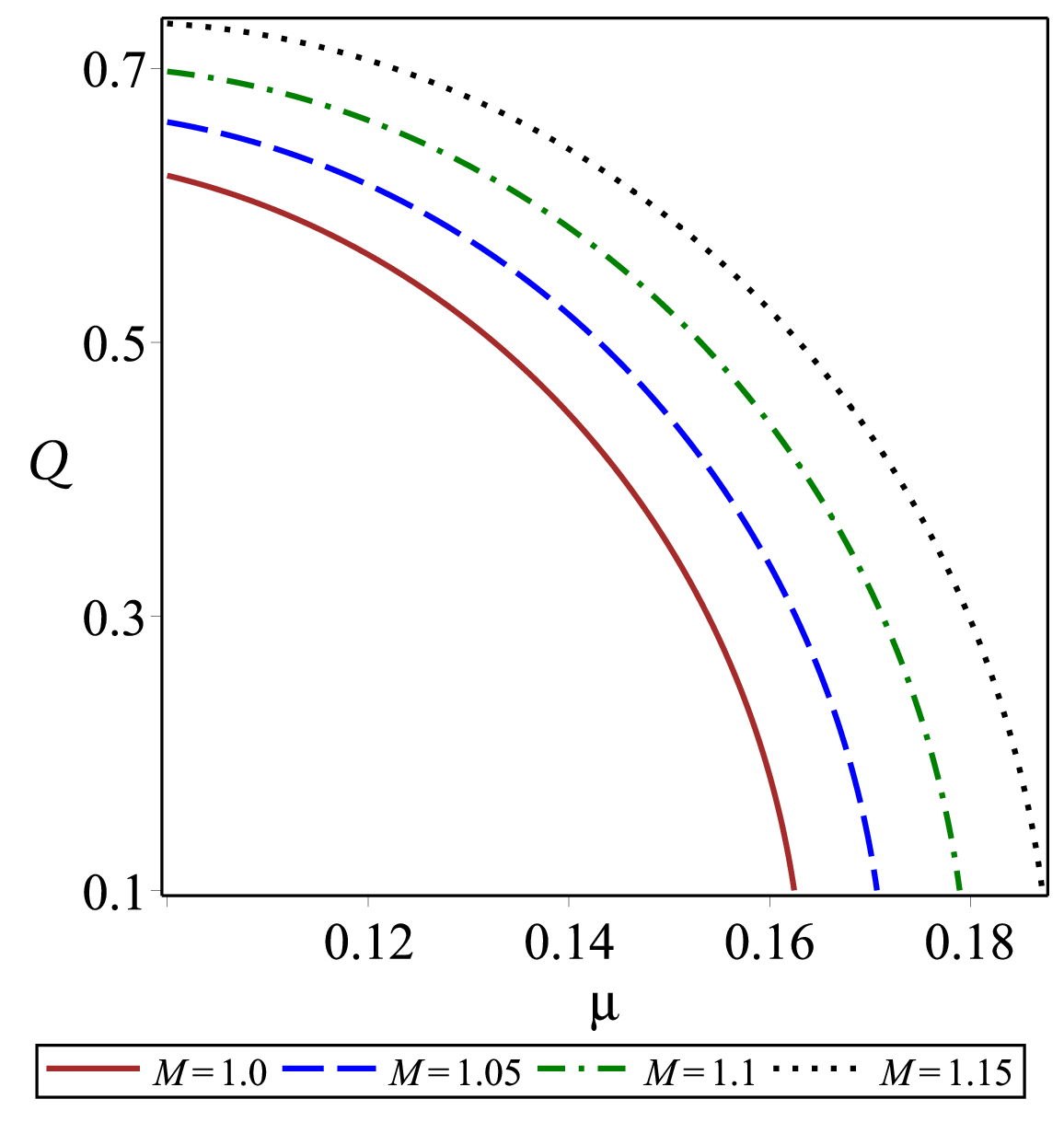}} \newline
\caption{\textbf{Left}: a plot of $J $ vs $Q $ for various values of $M $ at
fixed $\protect\mu $. \textbf{Right} : a plot of $Q $ vs $\protect\mu $ for
various values of $M $ at fixed $J $. We have set $r_{+}=1$, $B=0.2$, $%
\protect\beta =0.04$ and $A=0.02$.}
\label{Fig5}
\end{figure}

Evidently, for very small (large) black holes, the total mass is positive (negative) and independent of the angular momentum. As it was pointed out, no physical solutions exist for very small black holes. On the other hand, the total mass is negative for large black holes which is not physically acceptable. So, only medium black holes can be studied from the thermodynamic point of view. Regarding the medium black holes, as we see from Fig. \ref{Fig5}, the total mass is an increasing function of the angular momentum, electric charge and string tension. Since, the temperature (mass and entropy) of small (large) black holes is (are) negative, one can say that the medium ones can be physical objects.

The obtained quantities satisfy the first law of thermodynamic as 
\begin{equation}
dM=TdS+\Phi dQ+\Omega dJ-\lambda _{+}d\mu _{+}-\lambda _{-}d\mu _{-},
\label{1stlaw}
\end{equation}%
where $\Phi =\Phi _{t}$ is electric potential and $\Omega $ is angular velocity 
\begin{equation}
\Omega =\frac{aK}{\alpha (r_{+}^{2}+a^{2})}+\frac{aK(1-A^{2}\ell ^{2}\Xi )}{%
\ell ^{2}\Xi \alpha (1+a^{2}A^{2})}.  \label{Eq20}
\end{equation}%
%
%
%
%
%%%%%%%%%%%%%%%%%%%%%%%%%%%%%%%%%%%%%%%%%%%%%%%%%%%%%
%%%%%%%%%%%%%%%%%%%%%%%%%%%%%%%%%%%%%%%%%%%%%%%%%%%%%%%%%%%%%%%%%%%%%%%%%%%

\subsection{Thermal stability}

Now, we focus on the thermal stability/instability of solutions through the heat capacity. The signature of heat capacity determines the thermal stability/instability of the system. The positivity of heat capacity indicates black holes have a stable thermal phase, while the opposite corresponds to thermally unstable ones. The heat capacity is given by 
\begin{equation}
C_{Q,\mu,J}=T\left( \frac{\partial S}{\partial T}\right) _{Q,\mu,J}=\frac{%
\left( \frac{\partial M}{\partial S}\right) _{Q,\mu,J}}{\left( \frac{%
\partial ^{2}M}{\partial S^{2}}\right) _{Q,\mu,J}}.  \label{Eq14}
\end{equation}

Employing Eqs. (\ref{Eq9}), (\ref{Eq11}) and (\ref{Eq14}), one can find 
\begin{equation*}
C_{Q,\mu ,J}=\frac{2\pi \mu r_{+}X_{4}(4J^{2}B^{4}\ell
^{4}X_{5}+A^{2}r_{+}^{2}X_{1}^{3}+X_{1}^{3})}{BX_{1}^{3}(1-A^{2}r_{+}^{2})%
\left( X_{6}-\frac{2A^{2}r_{+}X_{7}}{X_{1}^{3}}-\frac{%
X_{4}(X_{1}^{3}+4J^{2}B^{4}\ell ^{4}(X_{1}-8\mu ^{2}r_{+}^{4}+4\ell ^{2}\mu
^{2}r_{+}^{2}))}{X_{1}r_{+}(X_{1}^{2}+4J^{2}B^{4}\ell ^{4})}\right) },
\end{equation*}%
where 
\begin{eqnarray*}
X_{4} &=&1+\frac{3r_{+}^{2}}{\ell ^{2}}-\frac{(1+4B)Q^{2}}{r_{+}^{2}}+\frac{%
4J^{2}B^{4}\ell ^{2}(r_{+}^{2}+2Q^{2}-\ell ^{2})}{X_{1}^{2}}%
-A^{2}r_{+}^{2}X_{2}, \\
&& \\
X_{5} &=&X_{1}+A^{2}r_{+}^{2}X_{1}-4\mu ^{2}r_{+}^{4}-2\mu ^{2}\ell
^{2}r_{+}^{2}, \\
&& \\
X_{6} &=&\frac{6r_{+}}{\ell ^{2}}+\frac{2(1+4B)Q^{2}}{r_{+}^{3}}+\frac{%
8J^{2}B^{4}\ell ^{2}r_{+}}{X_{1}^{2}}-\frac{16J^{2}B^{4}\ell ^{2}\mu
^{2}r_{+}(2Q^{2}-\ell ^{2}+r_{+}^{2})(2r_{+}^{2}+\ell ^{2})}{X_{1}^{3}}, \\
&& \\
X_{7} &=&X_{1}^{3}-\frac{4r_{+}^{2}X_{1}^{3}}{\ell ^{2}}-\frac{%
2Q^{4}X_{1}^{3}}{r_{+}^{4}}-4J^{2}B^{4}\ell ^{2}(4X_{1}r_{+}^{2}+X_{1}\ell
^{2}-8\mu ^{2}r_{+}^{6}-8\mu ^{2}\ell ^{2}r_{+}^{4}-2\mu ^{2}\ell
^{4}r_{+}^{2}).
\end{eqnarray*}

\begin{figure}[!htb]
\centering
\subfloat[$ Q=0.2 $ and $\mu =0.15$]{
        \includegraphics[width=0.32\textwidth]{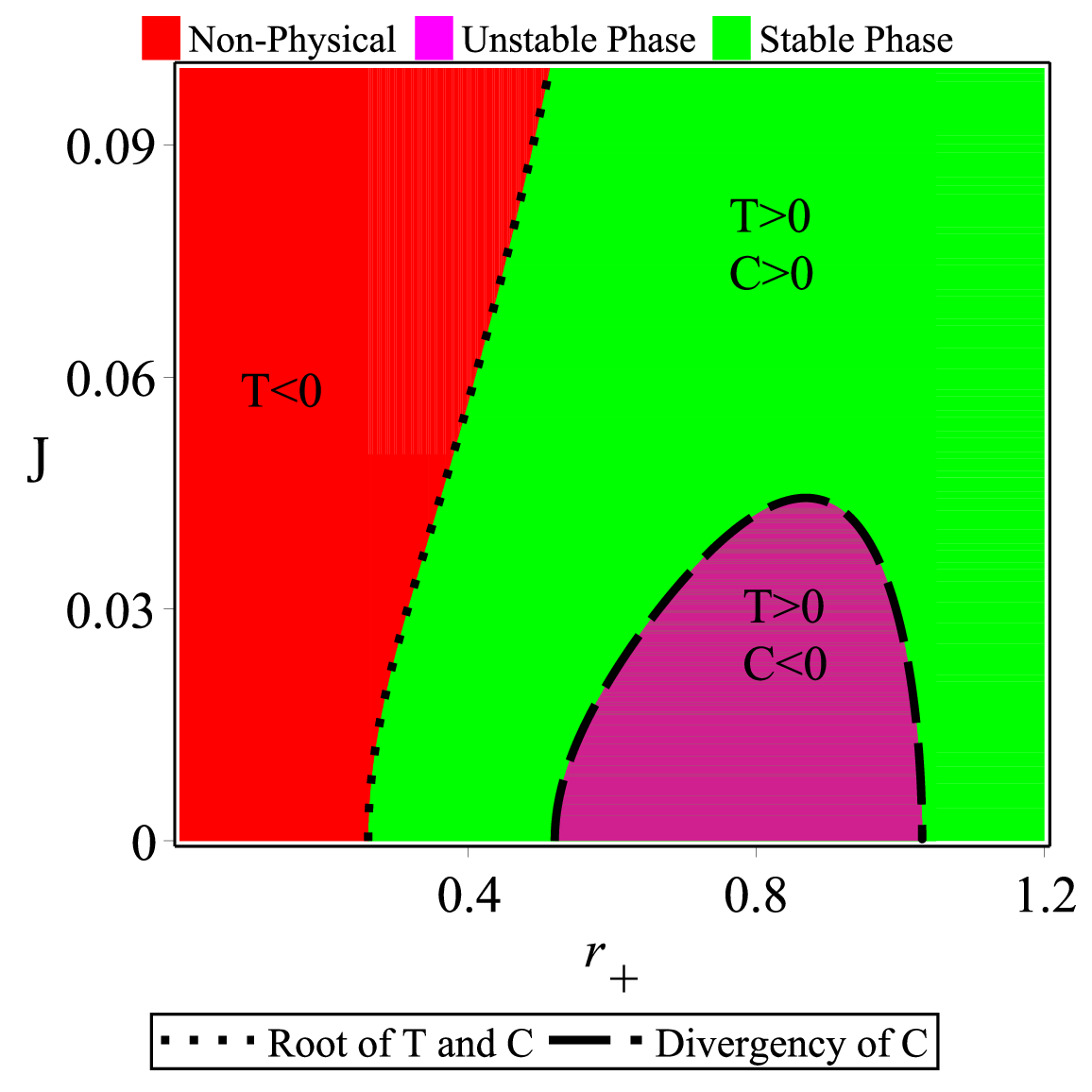}} 
\subfloat[$J=0.02$ and $\mu =0.15$]{
        \includegraphics[width=0.32\textwidth]{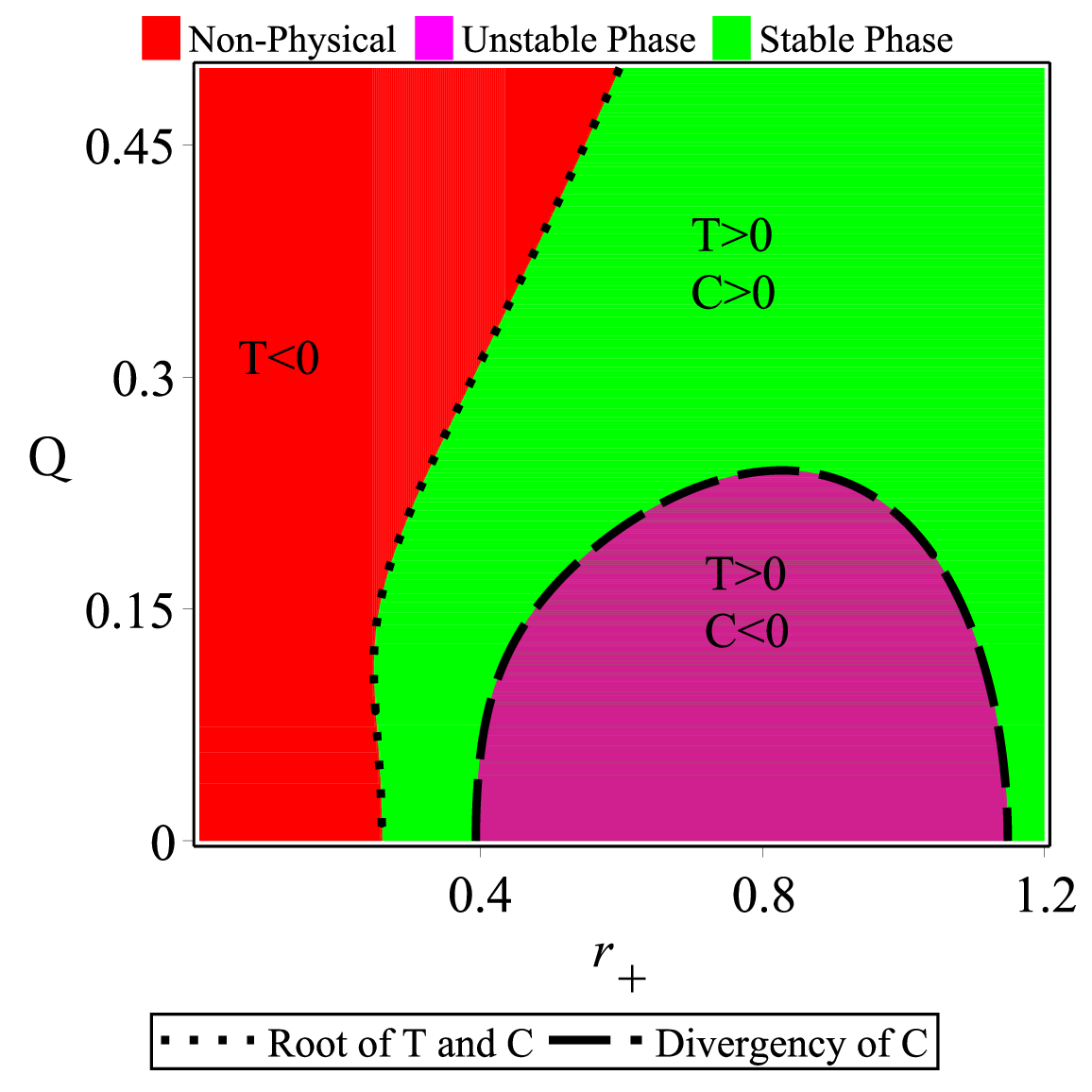}} 
\subfloat[ $J=0.02$ and $ Q=0.2 $]{
        \includegraphics[width=0.32\textwidth]{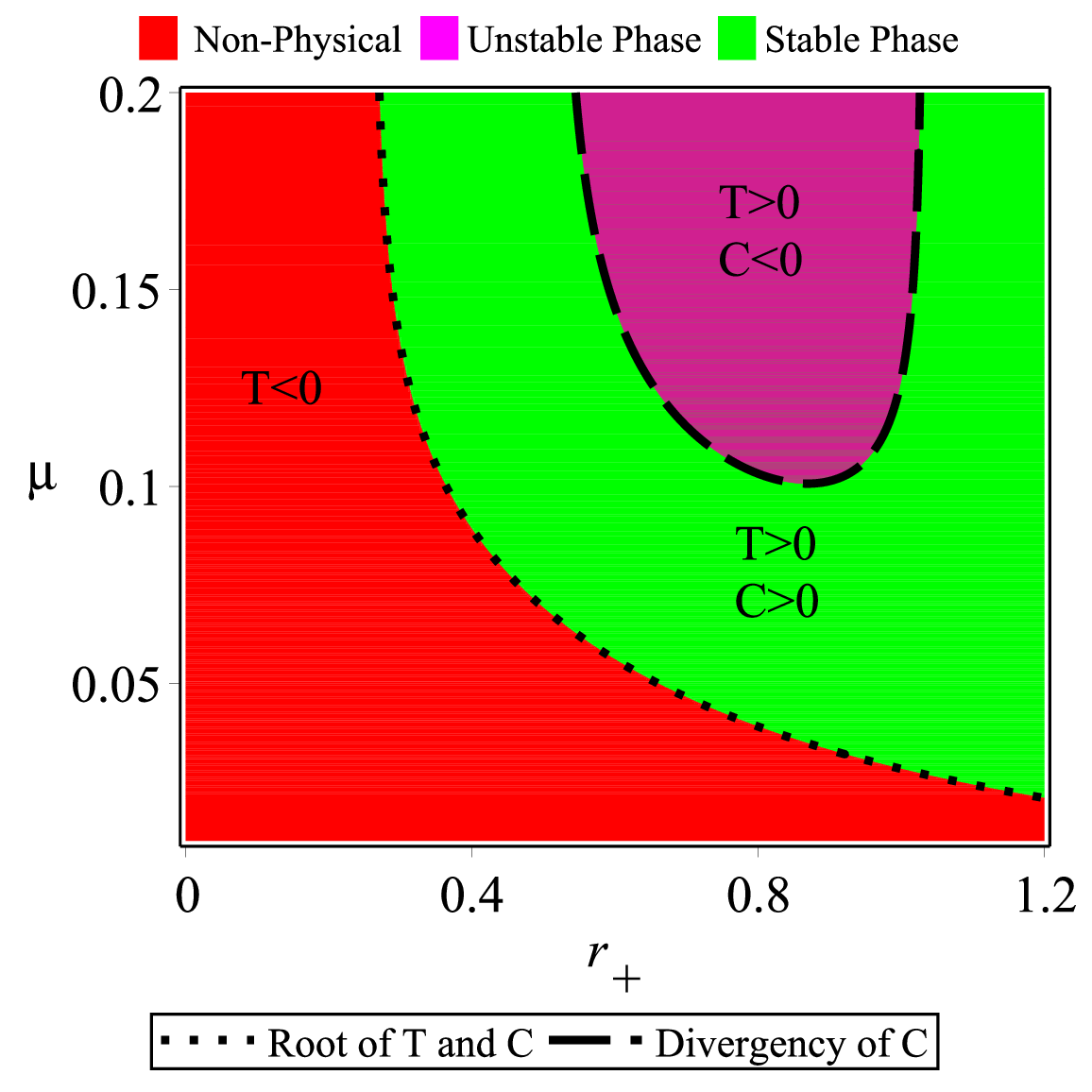}} \newline
\caption{Thermally stable and/or unstable regions of the black holes for $%
B=0.2$, $\protect\beta =0.04$ and $A=0.02$.}
\label{Fig6}
\end{figure}

To have a more precise picture of the thermal stability/instability, we have plotted Fig. \ref{Fig6}. Evidently, for large (small) values of the angular
momentum and electric charge (string tension), there are two phases of small and large black holes which are separated by the root of temperature/heat capacity. The small black holes have the negative temperature and are non-physical, while the large ones are thermally stable. For small (large) angular momentum and electric charge (string tension), there will be four
distinct phases of very small, small, medium and large black holes. Very small black holes are located before the root of temperature and are non-physical. The region between the root of $T$ and divergency of $C$ is related to small black holes which are thermally stable due to the positivity of the heat capacity. Between two divergencies (see the violet color region in Fig. \ref{Fig6}), the heat capacity has a negative value and medium black holes which are placed in this region are in the unstable phase. For region the after the larger divergency, the heat capacity is positive and large black holes are in the stable phase. Taking a closer look at Fig. \ref{Fig6}, one can find that by decreasing of the angular momentum and
electric charge the region related to the unstable phase increases. This reveals the fact that by decreasing $Q$ and $J$, the system achieves a stable state barely. Regarding the effect of string tension, as we see from Fig. \ref{Fig6}(c) its effect is opposite of that of the angular momentum and electric charge. This shows that a stable small/large accelerating black hole goes to an unstable phase by increasing this parameter. In other words, a small/large accelerating black hole exits in its stable state if it is pulled by a more powerful string tension.

As it was already mentioned, black holes with negative temperatures are non-physical. So, one can say that the root of temperature/heat capacity is a limitation point between physical and non-physical black hole solutions. The dotted curve in Fig. \ref{Fig6}, indicates a lower bound for the existence of the black holes in the bulk, with extremal black holes sitting on the curve. The right side of this line, black holes (with two horizons) are present, whereas no black hole exists on the left side. Studying the extremal limit of black holes can be significant as it is possible to set a threshold value for black hole parameters for which there are no physical black holes for smaller values. We will discuss the threshold values of parameters in more details in appendix A.

%%%%%%%%%%%%%%%%%%%%%%%%%%%%%%%%%%%%%%%%%%%%%%%%%%%%%%%%%%%%%%%%%%%%%%%%%%%%%%%%%%%%%%%%%%%%%%%%%%%%%%
%%%%%%%%%%%%%%%%%%%%%%%%%%%%%%%%%%%%%%%%%%%%%%%%%%%%%%%%%%%%%%%%%%%%%%%%%%%%%%%%%%%%%

\subsection{van der Waals-like Behavior}

Now we would like to investigate the possibility of the existence of van der
Waals-like phase transition for the charged accelerating black holes. We consider the cosmological
constant as a thermodynamic pressure, obtain the relation between horizon
radius and specific volume of the corresponding fluid and determine the
equation of state. We also extract critical thermodynamic quantities and
analyze the effects of black hole parameters on critical values.

The pressure associated with the cosmological constant is given by 
\begin{equation}
P=-\frac{\Lambda }{8\pi }=\frac{3}{8\pi \ell ^{2}},  \label{Eq16}
\end{equation}%
and its conjugate quantity called thermodynamic volume is expressed as 
\begin{equation}
V=\left( \frac{\partial M}{\partial P}\right) _{S,Q,J,\mu }.  \label{Eq17}
\end{equation}

It should be noted that by considering the cosmological constant as a
thermodynamic pressure, the identification of mass changes from internal
energy to enthalpy. With this new insight, free energy of the system is
given by 
\begin{equation}
F=H-TS=M-TS.  \label{Eq18}
\end{equation}

Also the first law of thermodynamics will be modified as follows 
\begin{equation}
dM=TdS+VdP+\Phi dQ+\Omega dJ-\lambda _{+}d\mu _{+}-\lambda _{-}d\mu _{-}.
\label{Eq19}
\end{equation}

To study van der Waals-like behavior of black holes, calculating the
equation of state is necessary. Employing Eqs. (\ref{Eq9}) and (\ref{Eq16}),
the pressure is obtained as 
\begin{eqnarray}
P &=&\frac{T}{2r_{+}}-\frac{\beta ^{2}T}{4r_{+}}+\frac{A^{2}Q^{2}T}{4r_{+}}-%
\frac{1}{8\pi r_{+}^{2}}-\frac{BA^{2}Q^{2}}{3\pi r_{+}^{2}}-\frac{5A^{2}Q^{2}%
}{24\pi r_{+}^{2}}+\frac{Q^{2}}{8\pi r_{+}^{4}}+\frac{BQ^{2}}{2\pi r_{+}^{4}}%
+\frac{A^{2}Q^{4}}{4\pi r_{+}^{4}}+\frac{5A^{2}}{24\pi }-\frac{TA^{2}r_{+}}{3%
}  \notag \\
&&  \notag \\
&&+\frac{3J^{2}B^{4}\left[ 8\pi Tr_{+}^{5}+4r_{+}^{4}-8\pi
TQ^{2}r_{+}^{3}+(1-4B)Q^{2}r_{+}^{2}-(2+8B)Q^{4}\right] }{8\pi \mu
^{4}r_{+}^{6}\left[ 2\pi Tr_{+}^{3}+r_{+}^{2}+(2+8B)Q^{2}\right] ^{2}}.
\label{Eq22}
\end{eqnarray}

To better understand properties of the pressure, we investigate its limiting
behavior as follows 
\begin{eqnarray*}
\lim_{r_{+}\longrightarrow 0}P &\propto &-\frac{3J^{2}B^{4}}{16\pi (1+4B)\mu
^{4}r_{+}^{6}}+\frac{A^{2}Q^{4}}{4\pi r_{+}^{4}}+\frac{(1+4B)Q^{2}}{8\pi
r_{+}^{4}}+\frac{3J^{2}B^{4}(3-4B)}{32\pi Q^{2}(1+4B)^{2}\mu ^{4}r_{+}^{4}}%
+O(\frac{1}{r_{+}^{3}}), \\
&& \\
\lim_{r_{+}\longrightarrow \infty }P &\propto &-\frac{TA^{2}r_{+}}{3}+\frac{%
5A^{2}}{24\pi }+\frac{T}{2r_{+}}+\frac{TQ^{2}A^{2}}{4r_{+}}-\frac{\beta ^{2}T%
}{4r_{+}}+O(\frac{1}{r_{+}^{2}}),
\end{eqnarray*}%
which shows that depending on the values of the different parameters, large
and small black holes could have positive or negative pressure. It is
worthwhile to mention that the pressure should be positively valued from a
classical thermodynamics perspective.

By rearranging the pressure in terms of specific volume, one can calculate
critical quantities. It should be noted that for rotating black holes, $%
r_{+} $ is not a linear function of the specific volume $\upsilon $. To do
so, we first rewrite thermodynamic volume in Eq. (\ref{V123}) in terms of $Q$
and $J$ as 
\begin{equation}
V=\frac{4\pi \mu }{3\alpha B}\left( r_{+}^{3}+2A^{2}r_{+}^{5}+\frac{%
2J^{2}B^{4}r_{+}}{\mu ^{4}x}+\frac{4J^{2}B^{4}r_{+}^{3}}{\mu ^{4}x^{2}}+%
\frac{9BA}{64\pi ^{2}P^{2}}\right) ,  \label{Eq23}
\end{equation}%
where $x=r_{+}^{2}+(1+4B)Q^{2}+\frac{8}{3}\pi Pr_{+}^{4}$. The specific
volume proportional to thermodynamic volume is determined in the following
form 
\begin{equation}
\upsilon =2\left( \frac{3V}{4\pi }\right) ^{\frac{1}{3}}\simeq 2r_{+}\left( 
\frac{\mu }{B}\right) ^{\frac{1}{3}}\left( 1+\frac{\left( \beta
^{2}-A^{2}Q^{2}\right) }{6}+\frac{2A^{2}r_{+}^{2}}{3}+\frac{2J^{2}B^{4}}{%
3\mu ^{4}r_{+}^{2}x}+\frac{4J^{2}B^{4}}{3\mu ^{4}x^{2}}+\frac{3BA}{64\pi
^{2}P^{2}r_{+}^{3}}\right) .  \label{Eq24}
\end{equation}

It is worth pointing out that since $A\ll 1 $ and $J\ll 1 $, we have
neglected all terms of higher-order of $J^{2}$ and $A^{2}$ in determining Eq
(\ref{Eq24}). Substituting the specific volume in Eq. (\ref{Eq22}), one can
arrange the pressure as follows 
\begin{eqnarray}
P &=&\frac{T\mu ^{\frac{1}{3}}}{\upsilon B^{\frac{1}{3}}}-\frac{T\beta
^{2}\mu ^{\frac{1}{3}}}{3\upsilon B^{\frac{1}{3}}}+\frac{TA^{2}Q^{2}\mu ^{%
\frac{1}{3}}}{3\upsilon B^{\frac{1}{3}}}+\frac{48TJ^{2}B^{\frac{11}{3}}}{%
\upsilon \mu ^{\frac{5}{3}}\chi _{1}^{2}}-\frac{\mu ^{\frac{2}{3}}}{2\pi
\upsilon ^{2}B^{\frac{2}{3}}}+\frac{2A^{2}Q^{2}\mu ^{\frac{2}{3}}}{3\pi
\upsilon ^{2}B^{\frac{2}{3}}}-\frac{\beta ^{2}\mu ^{\frac{2}{3}}}{6\pi
\upsilon ^{2}B^{\frac{2}{3}}}  \notag \\
&&  \notag \\
&&-\frac{4A^{2}Q^{2}B^{\frac{1}{3}}\mu ^{\frac{2}{3}}}{3\pi \upsilon ^{2}}-%
\frac{48J^{2}B^{\frac{10}{3}}}{\pi \upsilon ^{2}\mu ^{\frac{4}{3}}\chi
_{1}^{2}}+\frac{16TJ^{2}B^{3}}{\mu ^{2}\upsilon ^{3}\chi _{1}}+\frac{%
4Q^{2}\beta ^{2}\mu ^{\frac{4}{3}}}{3\pi \upsilon ^{4}B^{\frac{4}{3}}}+\frac{%
8A^{2}Q^{4}\mu ^{\frac{4}{3}}}{3\pi \upsilon ^{4}B^{\frac{4}{3}}}+\frac{%
384Q^{2}J^{2}B^{\frac{8}{3}}}{\pi \upsilon ^{4}\mu ^{\frac{2}{3}}\chi
_{1}^{2}}  \notag \\
&&  \notag \\
&&-\frac{16J^{2}B^{\frac{8}{3}}}{\pi \upsilon ^{4}\mu ^{\frac{5}{3}}\chi _{1}%
}+\frac{24B^{2}J^{2}}{\pi \mu ^{2}\upsilon ^{6}}+\frac{12A\upsilon Q^{2}B^{%
\frac{4}{3}}\mu ^{\frac{5}{3}}}{\pi \chi _{2}^{2}}+\frac{128B^{2}Q^{2}J^{2}}{%
\pi \mu \chi _{1}\upsilon ^{6}}-\frac{3A\mu B^{2}\upsilon ^{3}}{2\pi \chi
_{2}^{2}}+\frac{2Q^{2}\mu ^{\frac{4}{3}}}{\pi \upsilon ^{4}B^{\frac{4}{3}}} 
\notag \\
&&  \notag \\
&&+\frac{3AT\upsilon ^{4}\mu ^{\frac{2}{3}}B^{\frac{7}{3}}}{2\chi _{2}^{2}}+%
\frac{A^{2}}{24\pi }+\frac{3B^{2}J^{2}(3\mu ^{\frac{2}{3}}B^{\frac{4}{3}%
}\upsilon ^{4}-96\mu ^{2}Q^{4}-4Q^{2}B^{\frac{2}{3}}\mu ^{\frac{4}{3}}\chi
_{4}-B\pi T\upsilon ^{3}\chi _{3})}{2\pi \mu ^{2}\upsilon ^{6}\chi _{1}^{2}},
\label{Eq26}
\end{eqnarray}%
where 
\begin{eqnarray*}
\chi _{1} &=&B\pi T\upsilon ^{3}+\mu ^{\frac{1}{3}}B^{\frac{2}{3}}\upsilon
^{2}+8(1+4B)\mu Q^{2}, \\
&& \\
\chi _{2} &=&2B\pi T\upsilon ^{3}-\mu ^{\frac{1}{3}}B^{\frac{2}{3}}\upsilon
^{2}+4(1+4B)\mu Q^{2}, \\
&& \\
\chi _{3} &=&B\pi T\upsilon ^{3}-2\mu ^{\frac{1}{3}}B^{\frac{2}{3}}\upsilon
^{2}+32(1+2B)\mu Q^{2}, \\
&& \\
\chi _{4} &=&3\upsilon ^{2}+20B\upsilon ^{2}+160\mu ^{\frac{2}{3}}B^{\frac{1%
}{3}}Q^{2}+256\mu ^{\frac{2}{3}}B^{\frac{4}{3}}Q^{2}.
\end{eqnarray*}

\begin{figure}[tbh]
\centering
\subfloat[$J=0.02$, $ Q=0.2 $ and $\mu =0.15$]{
        \includegraphics[width=0.33\textwidth]{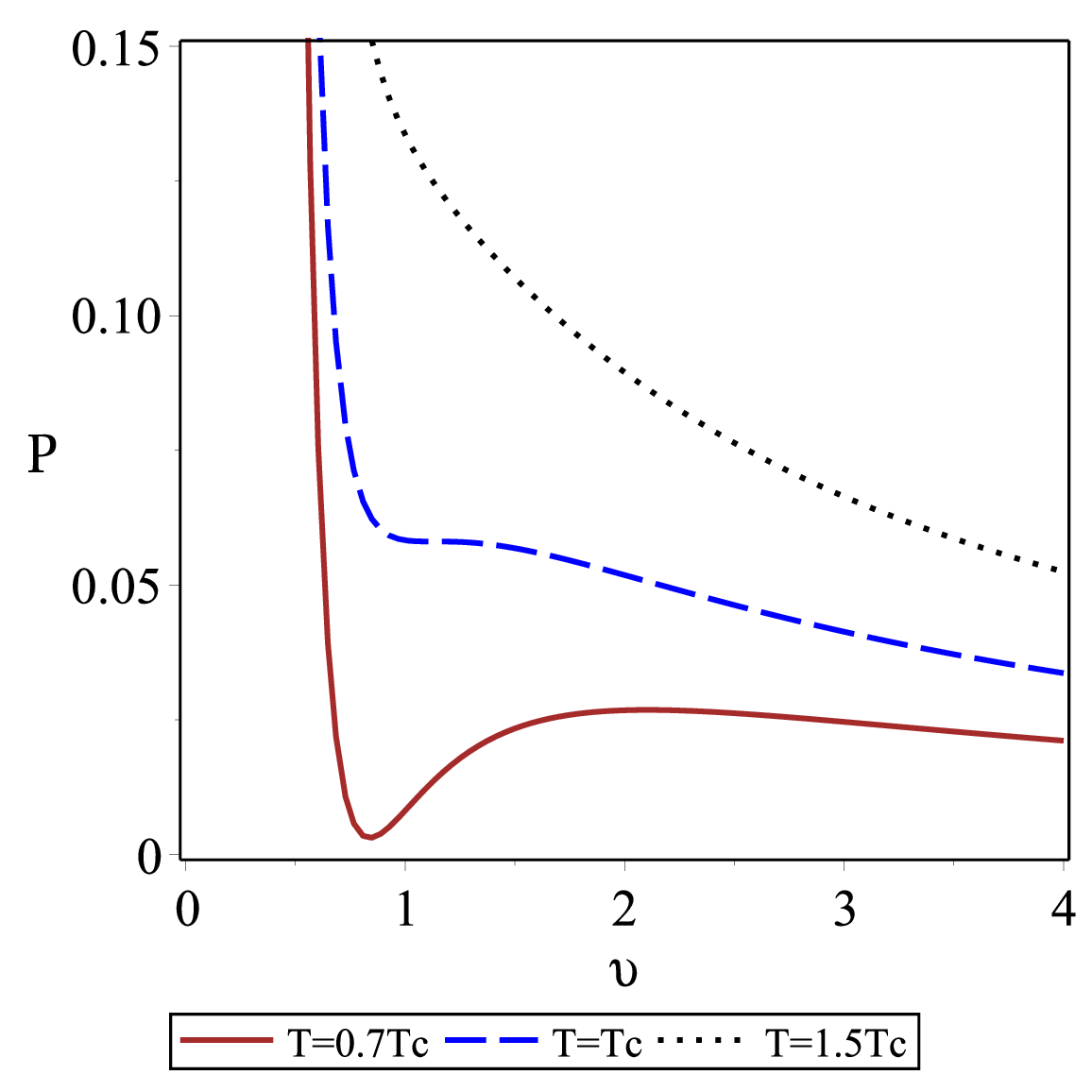}} 
\subfloat[$J=0.02$, $ Q=0.2 $ and $\mu =0.15$]{
        \includegraphics[width=0.33\textwidth]{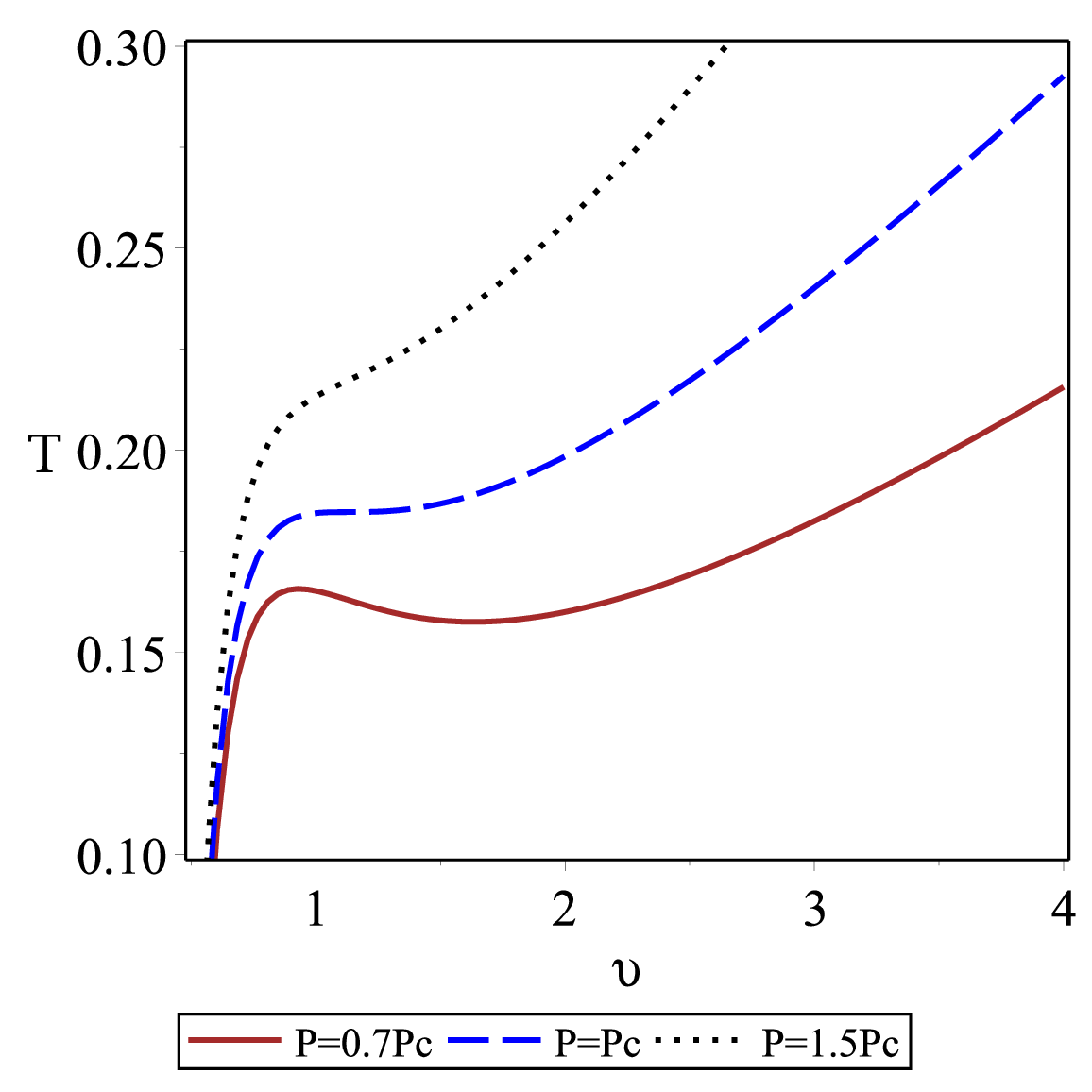}} 
\subfloat[$J=0.02$, $ Q=0.2 $ and $\mu =0.15$]{
        \includegraphics[width=0.31\textwidth]{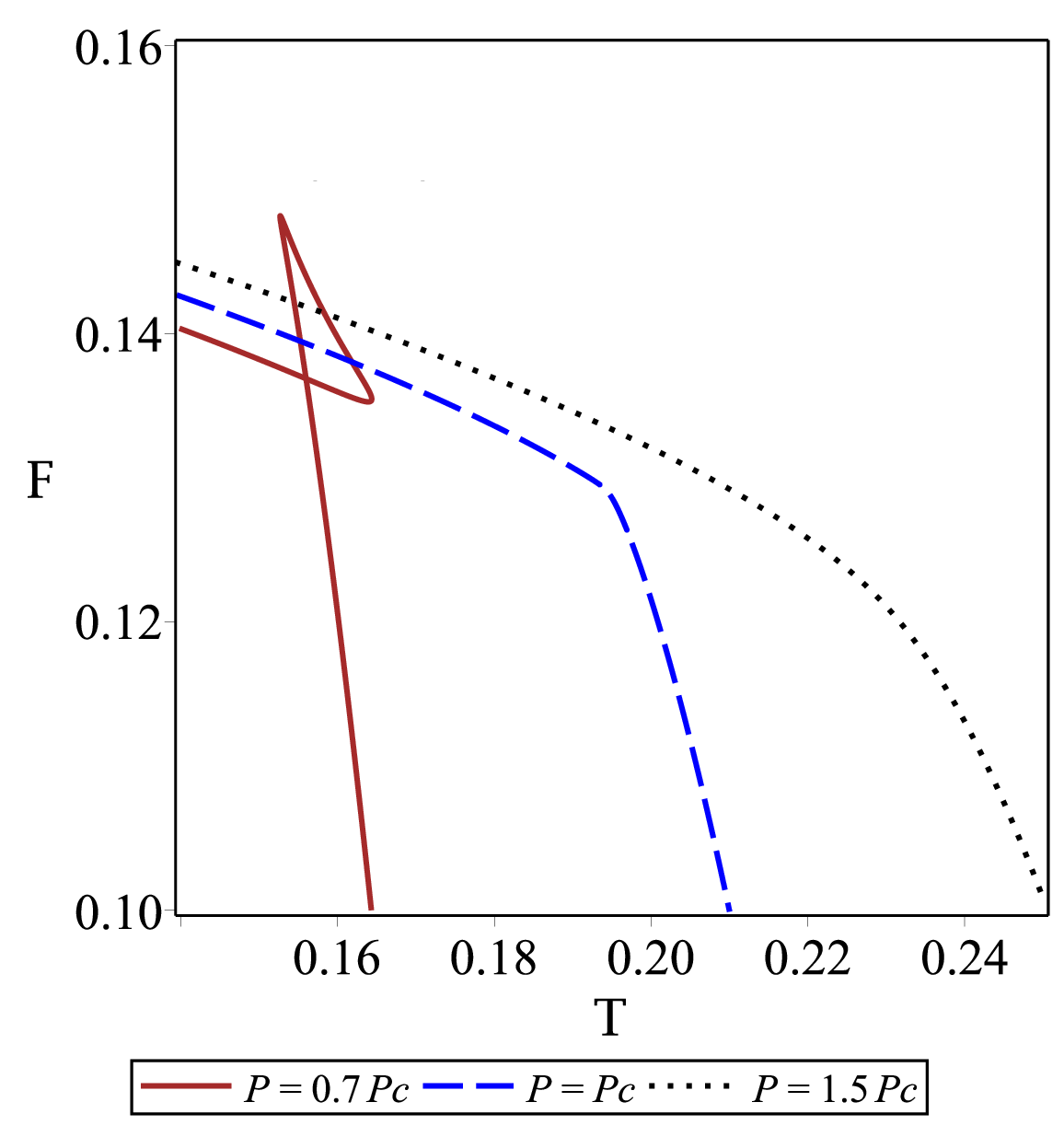}} \newline
\caption{Van der Waals-like phase diagrams for $B=0.2$, $\protect\beta =0.04$
and $A=0.02$. \textbf{Left}: $P$ versus $\protect\upsilon $ for different
temperatures. \textbf{Middle}: $T$ versus $\protect\upsilon $ for different
pressures. \textbf{Right}: $F$ versus $T$.}
\label{Fig7}
\end{figure}

As we know, the van der Waals fluid goes under a first-order liquid-gas
phase transition for temperatures smaller than the critical temperature ($%
T<T_{c}$). Whereas, at the critical temperature, its phase transition is a
second-order one \cite{PV2,MoLiu}. Fig. \ref{Fig7}, confirms the van der Waals-like behavior for these black holes. Formation of the
swallow-tail shape in $F-T$ diagram (continuous line) indicates the
existence of a first-order small-large black hole phase transition for $%
P<P_{c}$. 
%%%%%%%%%%%%%%%%%%%%%%%%%%%%%%%%%%%%%%%%%%%%%%%%%%%%%%%%%%%%%%%%%%%%%%%%%%%%%%%%%%%%%%%
%%%%%%%%%%%%%%%%%%%%%%%%%%%%%%%%%%%%%%%%%%%%%%%%%%%%%%%%%%%%%%%%%%%%%%%%%%%%%%%%%%%%%%%%%%

%%%%%%%%%%%%%%%%%%%%%%%%%%%%%%%%%%%%%%%%%%%%%%%%%%%%%%%%%%%%%%%%%%%%%%%%%%%%%%%%%%%%%%%%%%%%%%%%%%%%%%
\begin{table}[tbp]
\caption{Critical values for $B=0.2$, $\protect\beta =0.04$, $A=0.02$, $%
Q=0.2 $ and $\protect\mu =0.15$.}
\label{tab1}
\begin{center}
\begin{tabular}{||c|c|c|c|c||}
\hline\hline
$J $ & $\upsilon_{c}$ & $T_{c}$ & $P_{c}$ & $\frac{P_{c}\upsilon_{c}}{T_{c}}$
\\ \hline\hline
$0.010$ & $0.9780$ & $0.2052$ & $0.0728$ & $0.3469$ \\ \hline
$0.015$ & $1.0529$ & $0.1948$ & $0.0650$ & $0.3517$ \\ \hline
$0.020$ & $1.1307$ & $0.1847$ & $0.0580$ & $0.3555$ \\ \hline
$0.025$ & $1.2073$ & $0.1754$ & $0.0520$ & $0.3584$ \\ \hline\hline
\end{tabular}%
\end{center}
\end{table}
\begin{table}[tbp]
\caption{Critical values for $B=0.2$, $\protect\beta =0.04$, $A=0.02$, $%
J=0.02$ and $\protect\mu =0.15$.}
\label{tab2}
\begin{center}
\begin{tabular}{||c|c|c|c|c||}
\hline\hline
$Q$ & $\upsilon_{c}$ & $T_{c}$ & $P_{c}$ & $\frac{P_{c}\upsilon_{c}}{T_{c}}$
\\ \hline\hline
$0.15$ & $1.0203$ & $0.2120$ & $0.0753$ & $0.3627$ \\ \hline
$0.20$ & $1.1307$ & $0.1847$ & $0.0580$ & $0.3555$ \\ \hline
$0.25$ & $1.2740$ & $0.1595$ & $0.0438$ & $0.3497$ \\ \hline
$0.30$ & $1.4441$ & $0.1382$ & $0.0331$ & $0.3458$ \\ \hline\hline
\end{tabular}%
\end{center}
\end{table}
\begin{table}[tbp]
\caption{Critical values for $B=0.2$, $\protect\beta =0.04$, $A=0.02$, $%
J=0.02$ and $Q=0.2$.}
\label{tab3}
\begin{center}
\begin{tabular}{||c|c|c|c|c||}
\hline\hline
$\mu $ & $\upsilon_{c}$ & $T_{c}$ & $P_{c}$ & $\frac{P_{c}\upsilon_{c}}{T_{c}%
}$ \\ \hline\hline
$0.14$ & $1.1496$ & $0.1791$ & $0.0544$ & $0.3492$ \\ \hline
$0.15$ & $1.1307$ & $0.1847$ & $0.0580$ & $0.3555$ \\ \hline
$0.16$ & $1.1168$ & $0.1895$ & $0.0613$ & $0.3614$ \\ \hline
$0.17$ & $1.1070$ & $0.1936$ & $0.0642$ & $0.3672$ \\ \hline\hline
\end{tabular}%
\end{center}
\end{table}
%%%%%%%%%%%%%%%%%%%%%%%%%%%%%% 
%%%%%%%%%%%%%%%%%%%%%%%%%%%%%%%%%%%%%%%%%%%%%%%%%%%%%%%%%%

To find the critical point in $P-\upsilon$ diagram, we use the concept of
inflection point which is presented by these two equations 
\begin{equation}
\frac{\partial P}{\partial \upsilon }\bigg|_{\upsilon =\upsilon
_{c},T=T_{c}}=0,~\ \ \ \&~\ \ ~\frac{\partial ^{2}P}{\partial \upsilon ^{2}}%
\bigg|_{\upsilon =\upsilon _{c},T=T_{c}}=0.  \label{Eq27}
\end{equation}

The equation (\ref{Eq26}) is much complicated to determine critical
quantities analytically. But for slowly rotating accelerating black holes
located in the weak electric field, one can obtain critical quantities as
follows 
\begin{eqnarray}
\upsilon _{c} &=&\sqrt{\frac{12\mu ^{\frac{2}{3}}\left( Q^{2}\left[ 3+2\beta
+4A^{2}Q^{2}B^{\frac{4}{3}}\mu ^{-\frac{4}{3}}\right] +\mathcal{G}\right) }{%
B^{\frac{2}{3}}(3+\beta ^{2}-4A^{2}Q^{2}+8BA^{2}Q^{2})}},  \notag \\
&&  \notag \\
T_{c} &=&\frac{B^{\frac{2}{3}}\mu ^{\frac{1}{3}}\upsilon _{c}^{4}\left[
3+\beta ^{2}-4A^{2}Q^{2}(1-2B)\right] -8\mu Q^{2}\upsilon _{c}^{2}\left(
3+2\beta ^{2}+4A^{2}Q^{2}\right) -432J^{2}B^{\frac{10}{3}}\mu ^{-\frac{7}{3}}%
}{B\pi \upsilon _{c}^{5}\left( 3-\beta ^{2}+A^{2}Q^{2}\right) },  \notag \\
&&  \notag \\
P_{c} &=&\frac{A^{2}B^{\frac{4}{3}}\upsilon _{c}^{6}+4B^{\frac{2}{3}}\mu ^{%
\frac{2}{3}}\upsilon _{c}^{4}\left[ 3+\beta ^{2}-4A^{2}Q^{2}(1-2B)\right]
-48\mu ^{\frac{4}{3}}Q^{2}\upsilon _{c}^{2}\left( 3+2\beta
^{2}+4A^{2}Q^{2}\right) -2880J^{2}\mu ^{-2}B^{\frac{10}{3}}}{24\pi B^{\frac{4%
}{3}}\upsilon _{c}^{6}},  \label{Eq28}
\end{eqnarray}%
where $\mathcal{G}=\sqrt{15J^{2}B^{4}\mu ^{-4}\left[ 3+\beta
^{2}-4A^{2}Q^{2}(1-2B)\right] +Q^{4}\left[ 3+2\beta +4A^{2}Q^{2}B^{\frac{4}{3%
}}\mu ^{-\frac{4}{3}}\right] ^{2}}$.

Now we would like to explore the impact of black hole parameters on critical
quantities in Tables. \ref{tab1}-\ref{tab3}. From Table. \ref{tab1}, the
effect of angular momentum is to decrease (increase) critical temperature
and pressure (critical volume). The obtained results in Table. \ref{tab2},
show that the effect of electric charge on critical quantities is similar to
that of angular momentum. Therefore the similar discussions can be used in
this situation as well. Table. \ref{tab3}, provides interesting information
related to the effect of string tension on the critical values. As one can
see, the critical temperature and pressure are increasing functions of this
parameter, whereas the critical volume is a decreasing function of it. Also,
by looking at Tables. \ref{tab1}-\ref{tab3}, one can find that the universal
critical ratio $\left( \frac{P_{c}\upsilon _{c}}{T_{c}}\right) $ is an
increasing function of angular momentum (Table. \ref{tab1}) and string
tension (Table. \ref{tab3}). While the electric charge has a decreasing
effect on it (Table. \ref{tab2}).

\begin{figure}[tbh]
\centering
\subfloat[$J=0.02$ and $ Q=0.2$]{
        \includegraphics[width=0.32\textwidth]{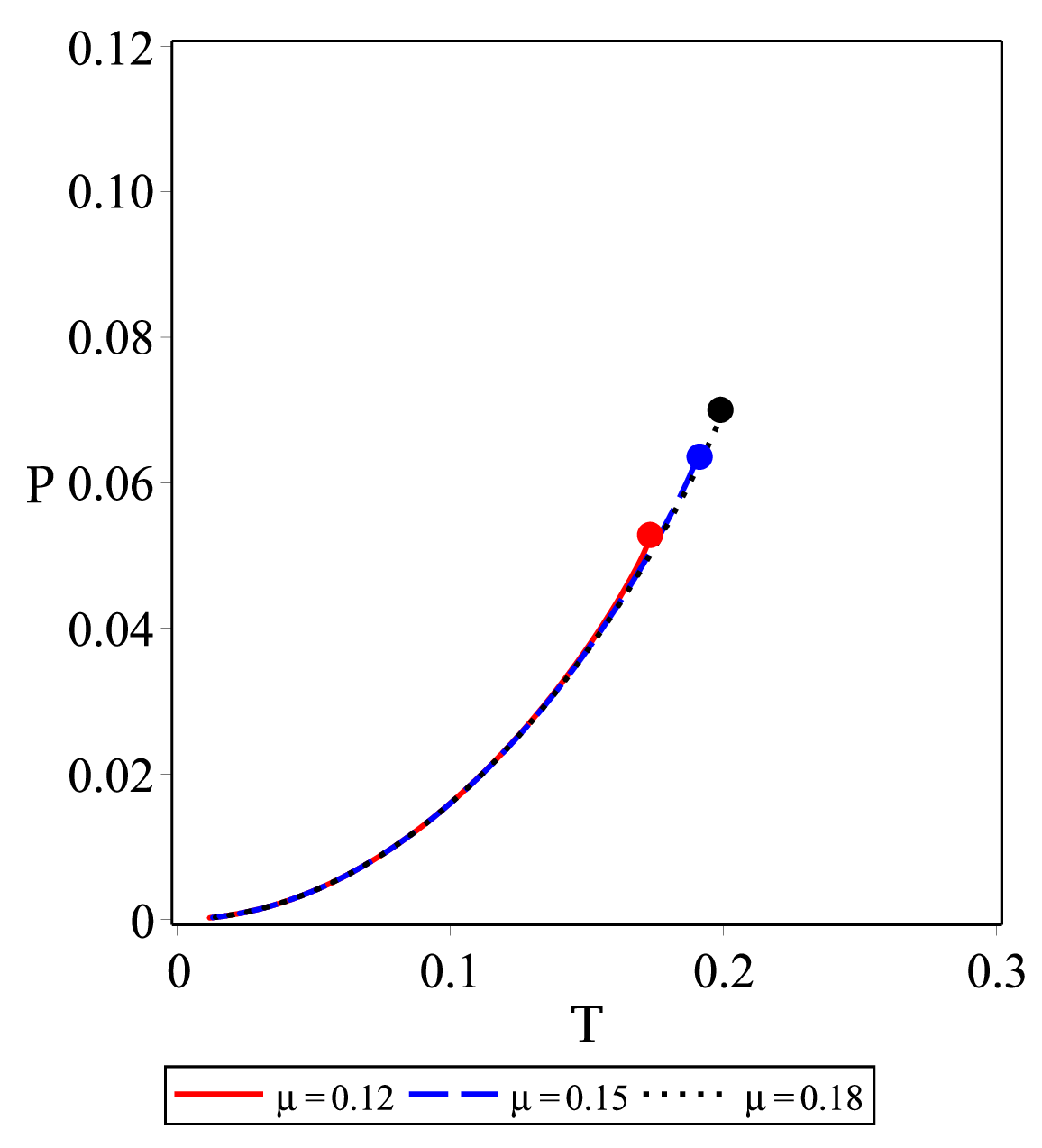}} 
\subfloat[$J=0.02$ and $ Q=0$ ]{
        \includegraphics[width=0.32\textwidth]{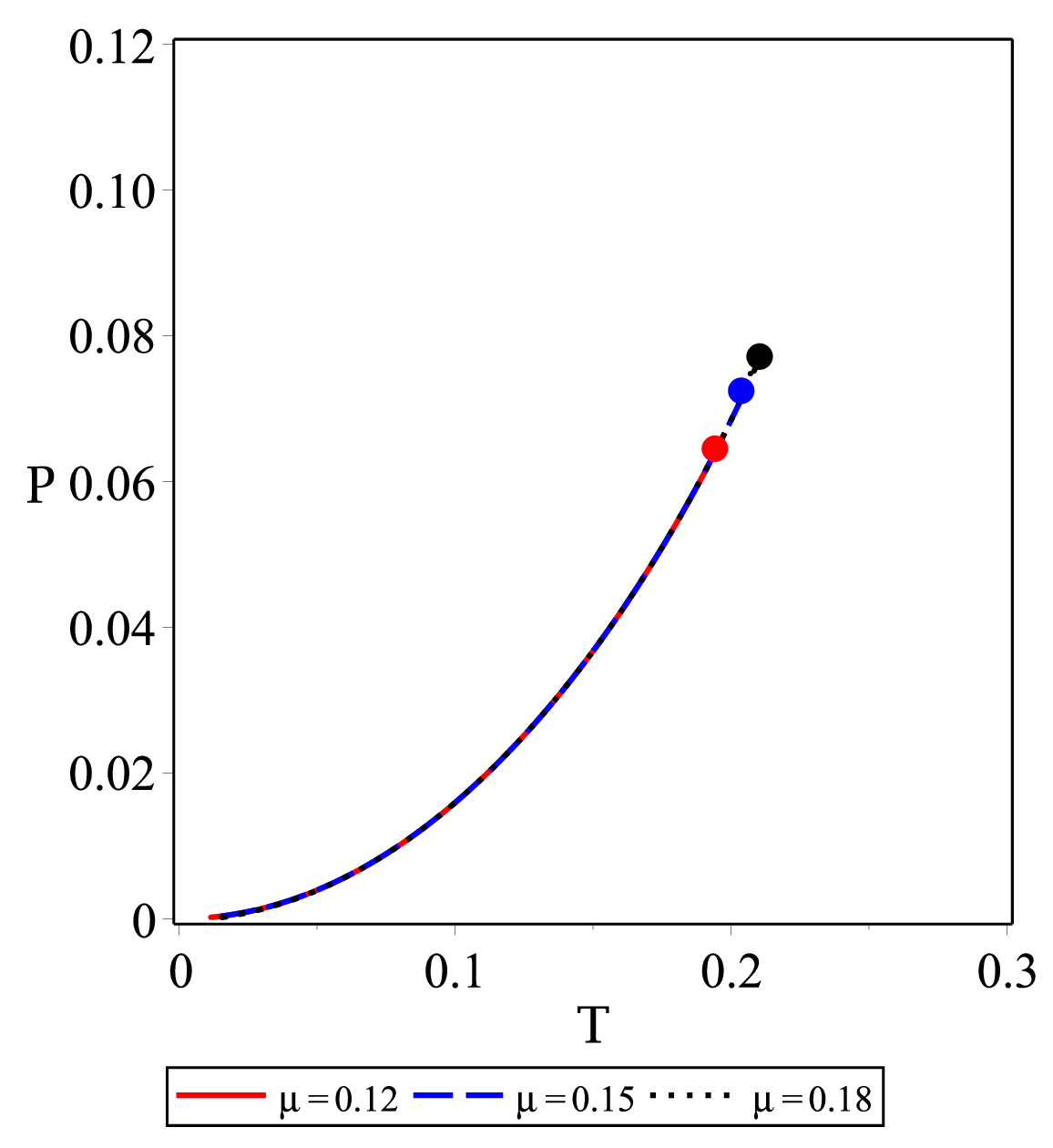}} 
\subfloat[$J=0$ and $ Q=0.2$ ]{
        \includegraphics[width=0.32\textwidth]{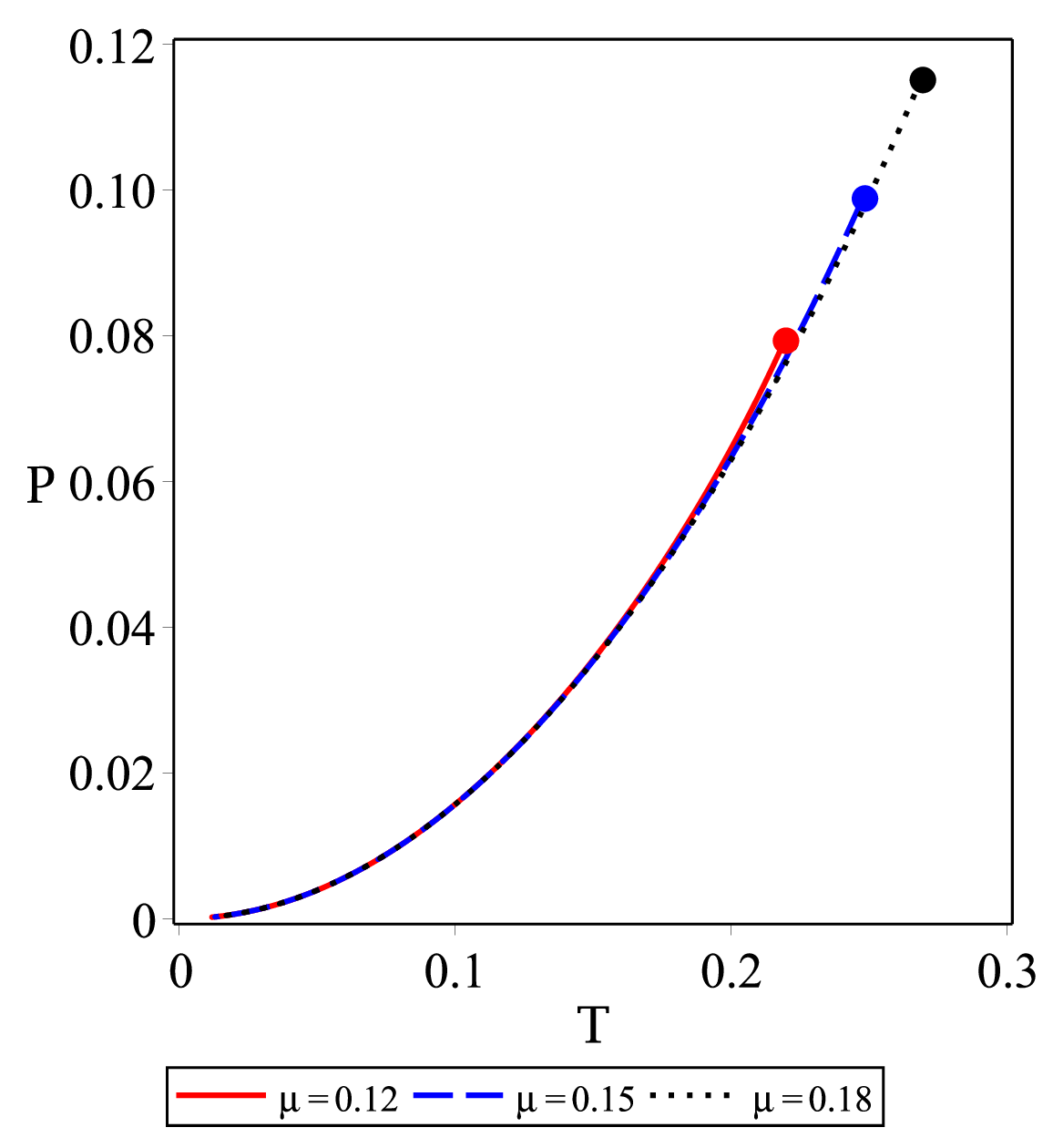}} \newline
\caption{$P$ - $T$ diagram for $B=0.2$, $\protect\beta =0.04$, $A=0.02$ and
different values of the string tension. Small circles in the endpoint of
each line represent the critical points.}
\label{FigPTn}
\end{figure}

To compare the critical temperature and pressure of the charged rotating
accelerating black hole to those of its non-rotating and uncharged
counterparts, we have plotted Fig. \ref{FigPTn}. As one can see, the charged rotating accelerating black hole (Fig. \ref{FigPTn}(a)) has smaller the critical
temperature and pressure than uncharged (Fig. \ref{FigPTn}(b))
and non-rotating (Fig. \ref{FigPTn}(c)) accelerating black holes. In
addition, by comparing Fig. \ref{FigPTn}(c) and Fig. \ref{FigPTn}(b) to each
other, one can find that a charged accelerating black hole has
bigger the critical points than a rotating accelerating black hole.

\begin{figure}[tbh]
\centering
\subfloat[ $ Q=0.2 $ and $\mu =0.15$]{
        \includegraphics[width=0.32\textwidth]{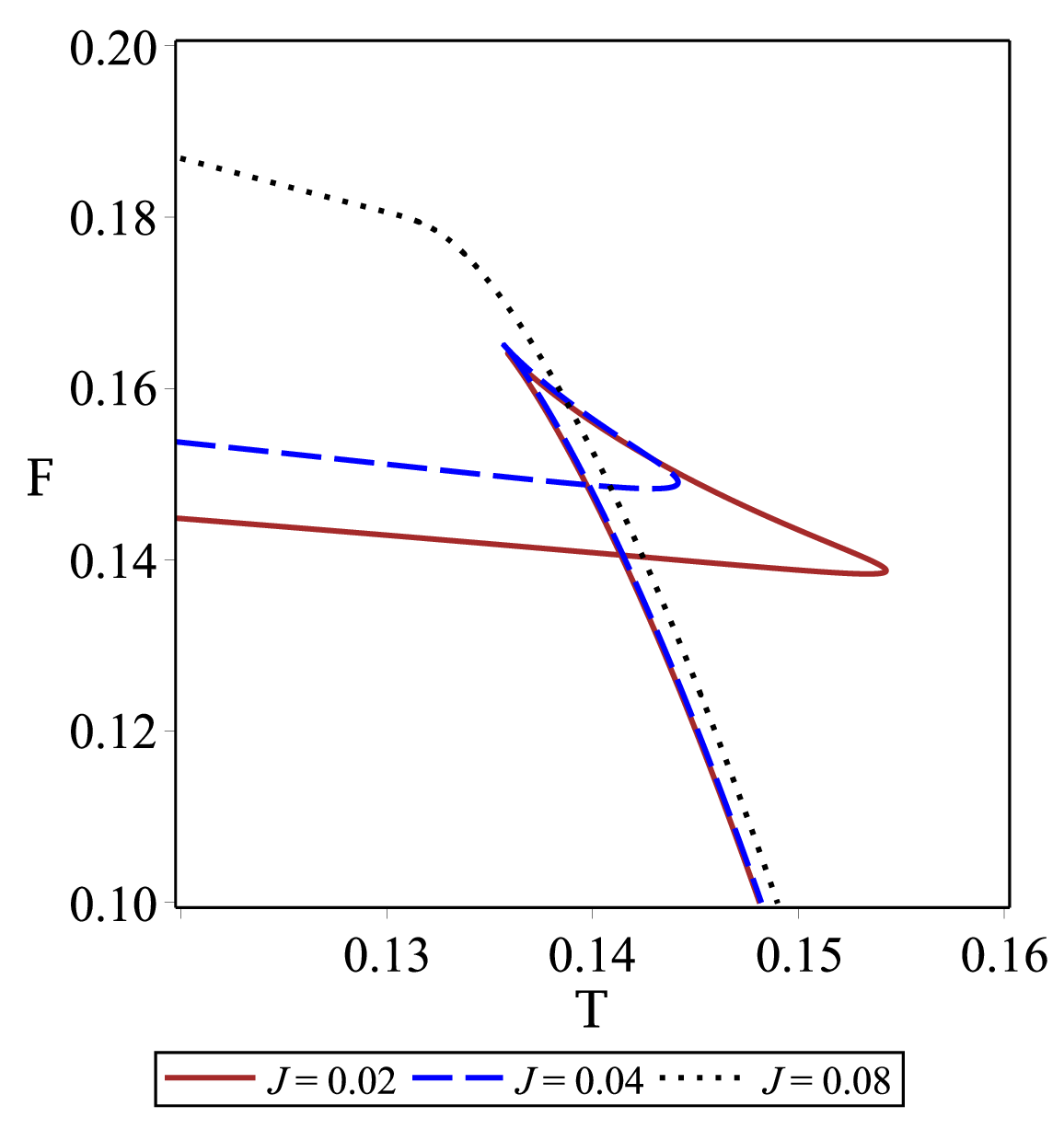}} 
\subfloat[$J=0.02$ and $\mu =0.15$]{
        \includegraphics[width=0.32\textwidth]{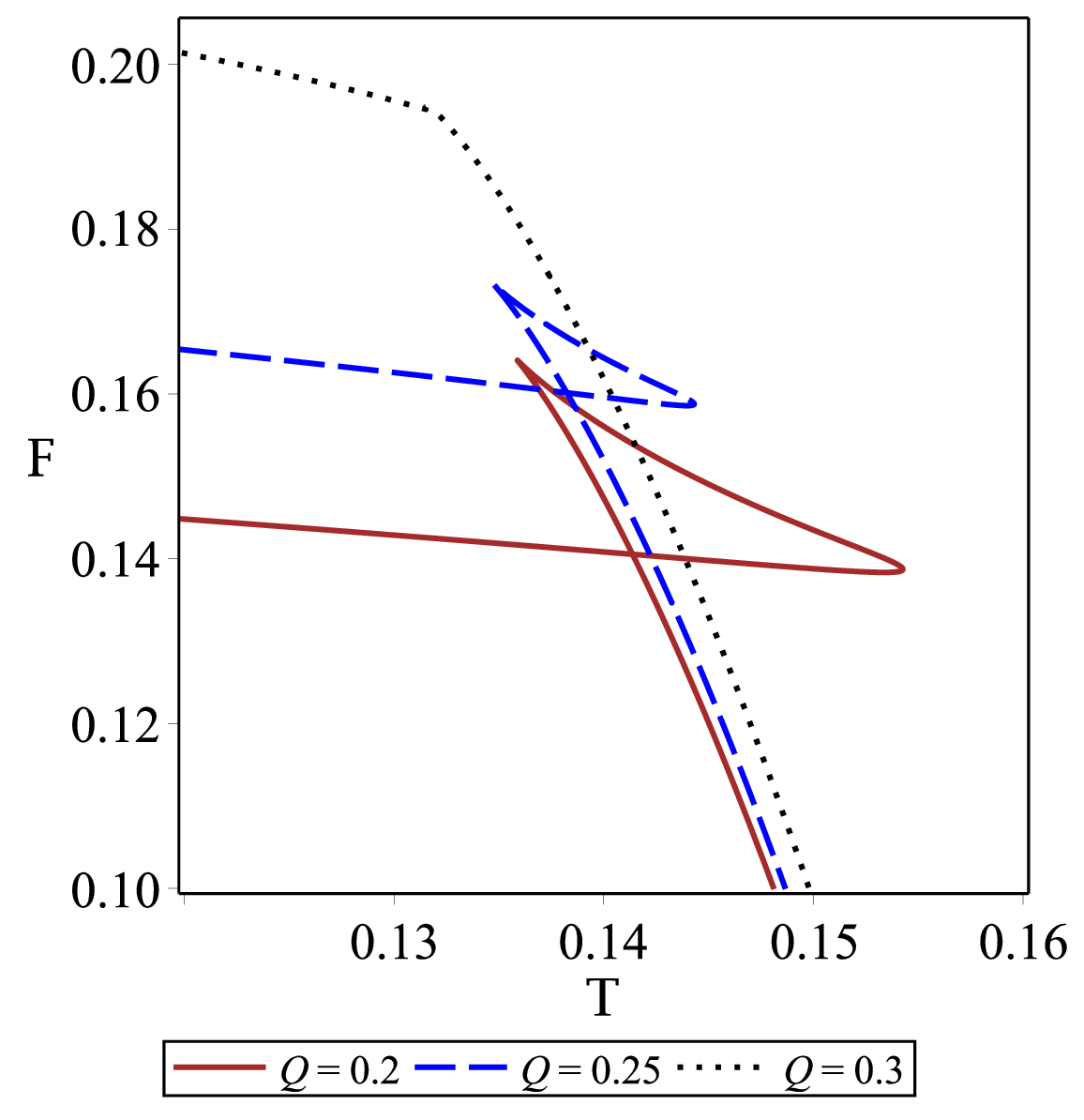}} 
\subfloat[$J=0.02$ and $ Q=0.2 $]{
        \includegraphics[width=0.32\textwidth]{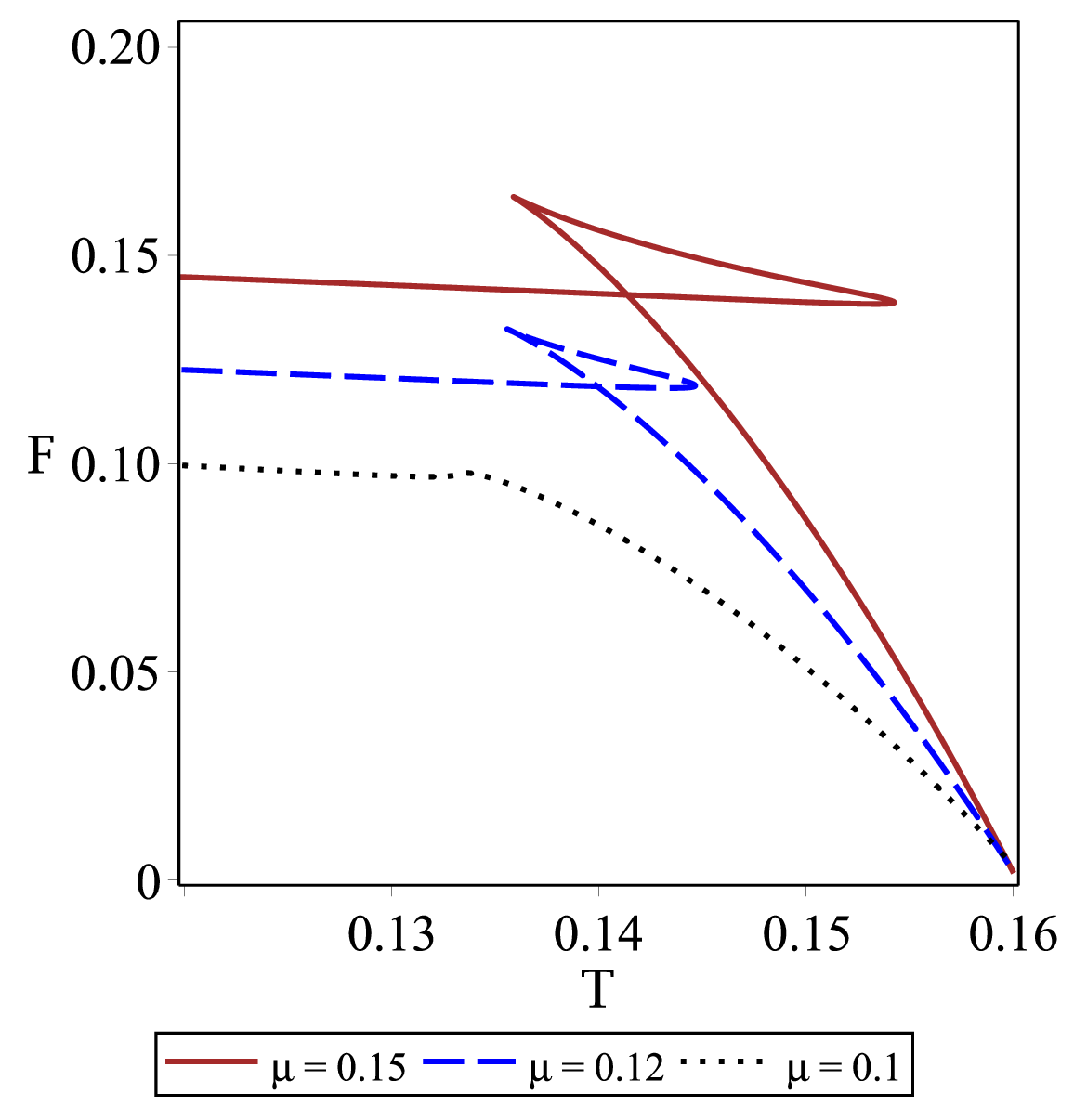}} \newline
\caption{$F$ - $T$ diagrams for $B=0.2$, $\protect\beta =0.04$, $A=0.02$, $%
P=0.03 $ and different values of the angular momentum (left panel),
different values of the electric charge (middle panel) and different values
of the string tension (right panel).}
\label{Fig7a}
\end{figure}

\begin{figure}[tbh]
\centering
\subfloat[ $J=0.02$, $ Q=0.2 $ and $\mu =0.1$]{
        \includegraphics[width=0.32\textwidth]{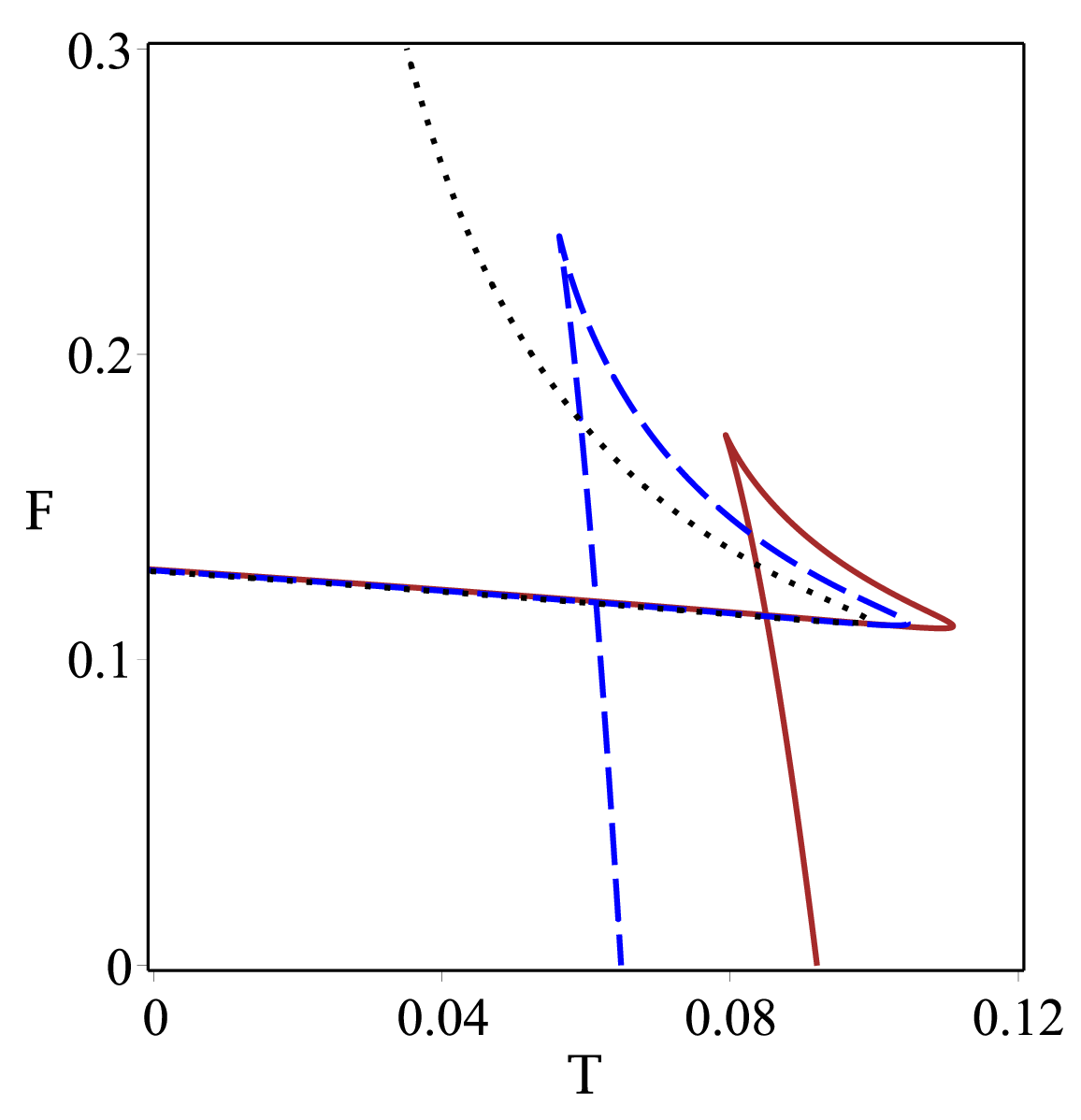}} 
\subfloat[$J=0.02$, $ Q=0.2 $ and $\mu =0.05$]{
        \includegraphics[width=0.32\textwidth]{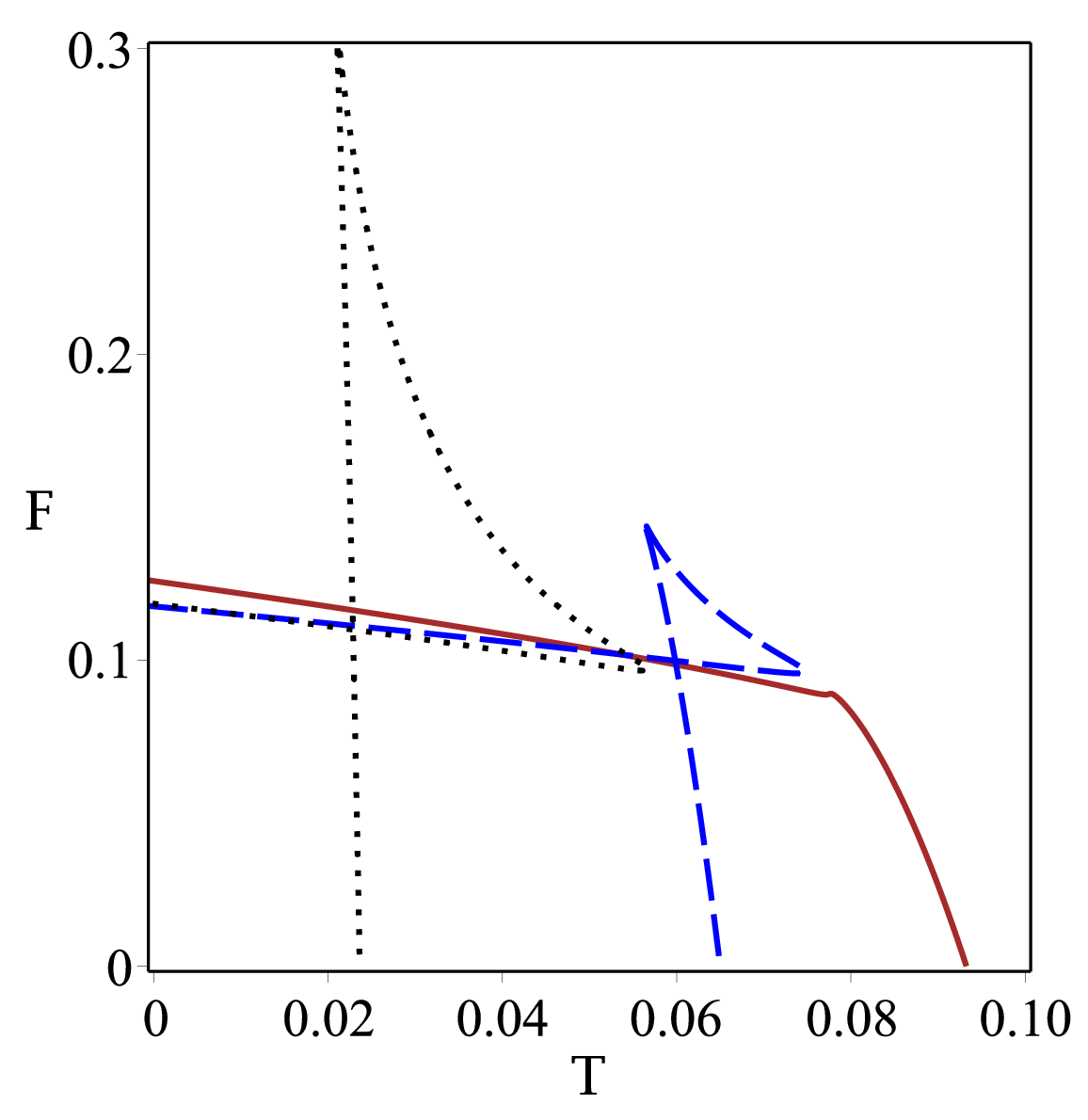}} 
\subfloat[$\mu =0.1$]{
        \includegraphics[width=0.32\textwidth]{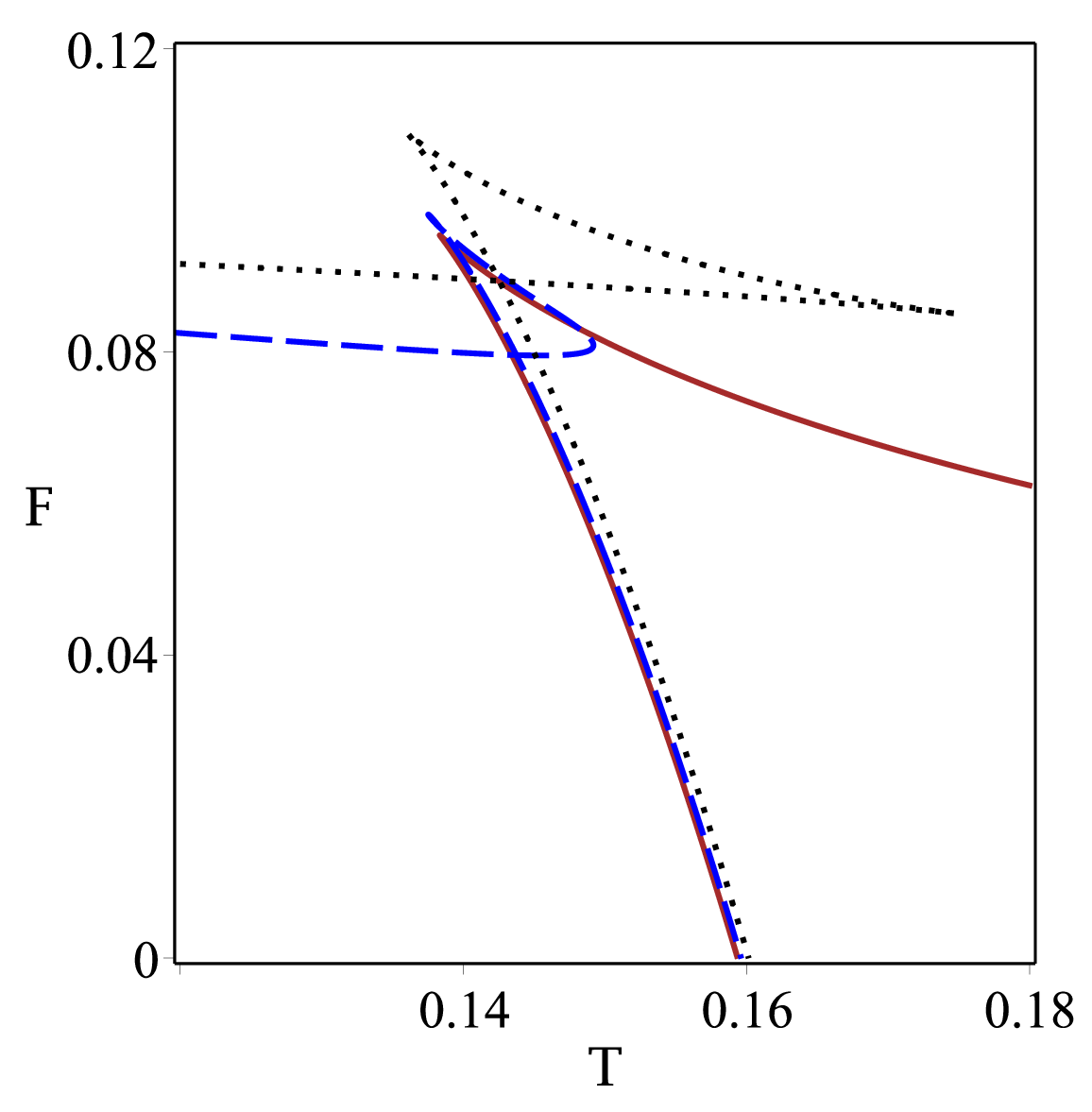}} \newline
\caption{$F$ - $T$ diagrams for $B=0.2$, $\protect\beta =0.04$ and $A=0.02$. 
\textbf{Left and middle}: $P=0.01 $ (continues line), $P=0.005 $ (dashed
line) and $P=0.0005 $ (dotted line). \textbf{Right}: $J=Q=0 $ (continues
line), $J=0.02 $, $Q=0 $ (dashed line) and $J=0 $, $Q=0.2 $ (dotted line).}
\label{Fig7b}
\end{figure}

As it was mentioned, for pressures and temperatures smaller than the
critical pressure and temperature, one can observe a first-order small/large
black hole phase transition. By looking at Figs. \ref{Fig7a}(a) and \ref%
{Fig7a}(b), one can find that for large values of angular momentum and
electric charge, the free energy is a decreasing function of the temperature and
swallow-tail shape does not appear. In this case, just a single stable phase
exists for black holes. Decreasing of $J $ and $Q $, the swallow-tail
shape appears which shows that the black hole undergoes a first-order phase
transition. From Fig. \ref{Fig7a}(c), one can see that the effect of string
tension is the opposite of that of the angular momentum and electric charge,
meaning that for small values of the string tension no first-order phase
transition can be observed. But a remarkable point is that by decreasing
of pressure one can observe the first-order phase transition for small
string tension as well (see Figs. \ref{Fig7b}(a) and \ref{Fig7b}(b)). In addition to the
first-order phase transition a zeroth-order phase transition is also
observed for the accelerating black holes \cite{Accel6,Accel7}. But according to our analysis, the
zeroth-order phase transition is observed only in absence of the angular
momentum and electric charge (see continues curve in Fig. \ref{Fig7b}(c)).
It is worth pointing out that our study is a little different from \cite%
{Accel6,Accel7} as the authors defined some dimensionless quantities such as 
$\tilde{m}=mA$, $\tilde{e}=eA$, $\tilde{A}=A\ell$ and $\tilde{r}=\frac{r}{%
\ell}$, while we have only considered $\mathcal{B}=mA$ and $\beta=A\ell$ as
fixed parameters in our work (for more details see appendix B). According to
results obtained in \cite{Accel6,Accel7}, string tension (electric charge
and angular momentum) has (have) an increasing (a decreasing) effect on the
bicritical and critical point which is similar to our work, where we
observed that the critical pressure and temperature are increasing
(decreasing) functions of the string tension (electric charge and angular
momentum). 
%%%%%%%%%%%%%%%%%%%%%%%%%%%%%%%%
%%%%%%%%%%%%%%%%%%%%%%%%%%%%%%%%%%%

%%%%%%%%%%%%%%%%%%%%%%%%%%%%%%%%%%%%%%%%%%%%%%%%%%%%%%%%%%%%%%%%%%%%%%%%%%%%%%%%%%%%%
\begin{figure*}[tbh]
\centering
\includegraphics[width=0.35\linewidth]{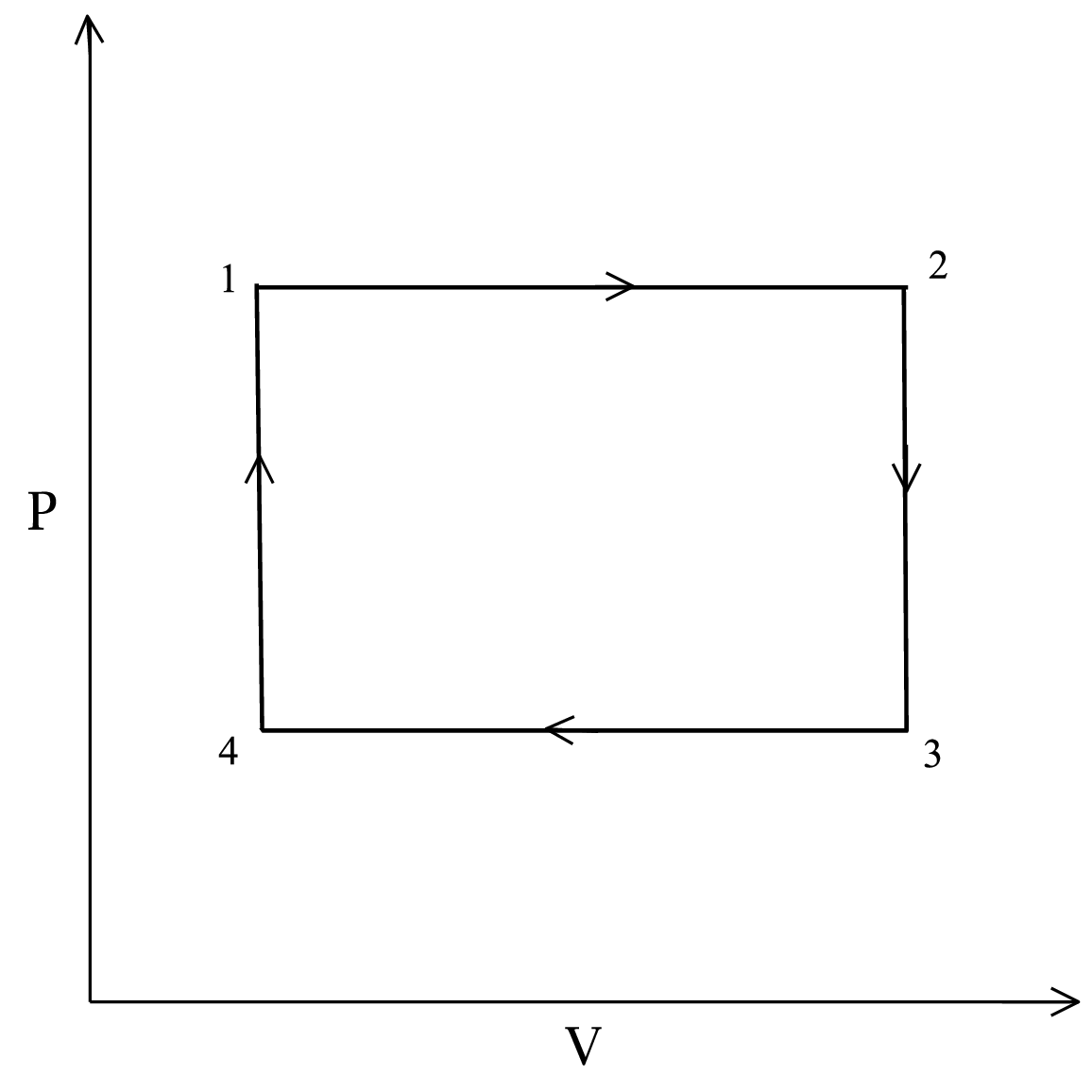}
\caption{$P$ - $V$ diagram.}
\label{Fig11}
\end{figure*}
%%%%%%%%%%%%%%%%%%%%%%%%%%%%%%%%%%%%%%%%%%%%%%%%%%%%%%%%%%%%

\section{The corresponding holographic heat engine}

In this section, we would like to investigate another interesting quantity
called the heat engine efficiency. We should point out that we conduct our
studies in a classical framework. Considering black holes as thermodynamic
systems in the extended phase space, it is natural to assume them as heat
engines \cite%
{Heat2,Heat4,Heat5,Heat6,Heat7,Heat12,Heat13,Heat14,Heat15,Heat16,Heat17,Heat18}%
. In fact, with such a phase space, it is possible to extract the mechanical
work via the term $VdP$ in the first law. A heat engine is a physical system
that takes some heat ($Q_{H}$) from a warm reservoir, converts some of this
thermal energy to useful work ($W$), and transfers the remaining heat energy 
$(Q_{C}=Q_{H}-W)$ to the cold reservoir. The efficiency of the heat engine
is obtained as 
\begin{equation}
\eta =\frac{W}{Q_{H}}=1-\frac{Q_{C}}{Q_{H}}.  \label{Eq29}
\end{equation}

The cycle of the heat engine is defined as a closed path in the $P-V$
diagram. So, the equation of state of the black hole has a significant
impact on the efficiency. The heat capacity is one of the important
thermodynamic quantities in calculating efficiency which is obtained by the
standard thermodynamic relations as follows 
\begin{equation*}
C_{P}=T\left( \frac{\partial S}{\partial T}\right) _{P},~\ \ \&~~\
C_{V}=T\left( \frac{\partial S}{\partial T}\right) _{V}.
\end{equation*}

%%%%%%%%%%%%%%%%%%%%%%%%%%%%%%%%%%%%%%%%%%
\begin{figure}[!htb]
\centering
\subfloat[]{
        \includegraphics[width=0.32\textwidth]{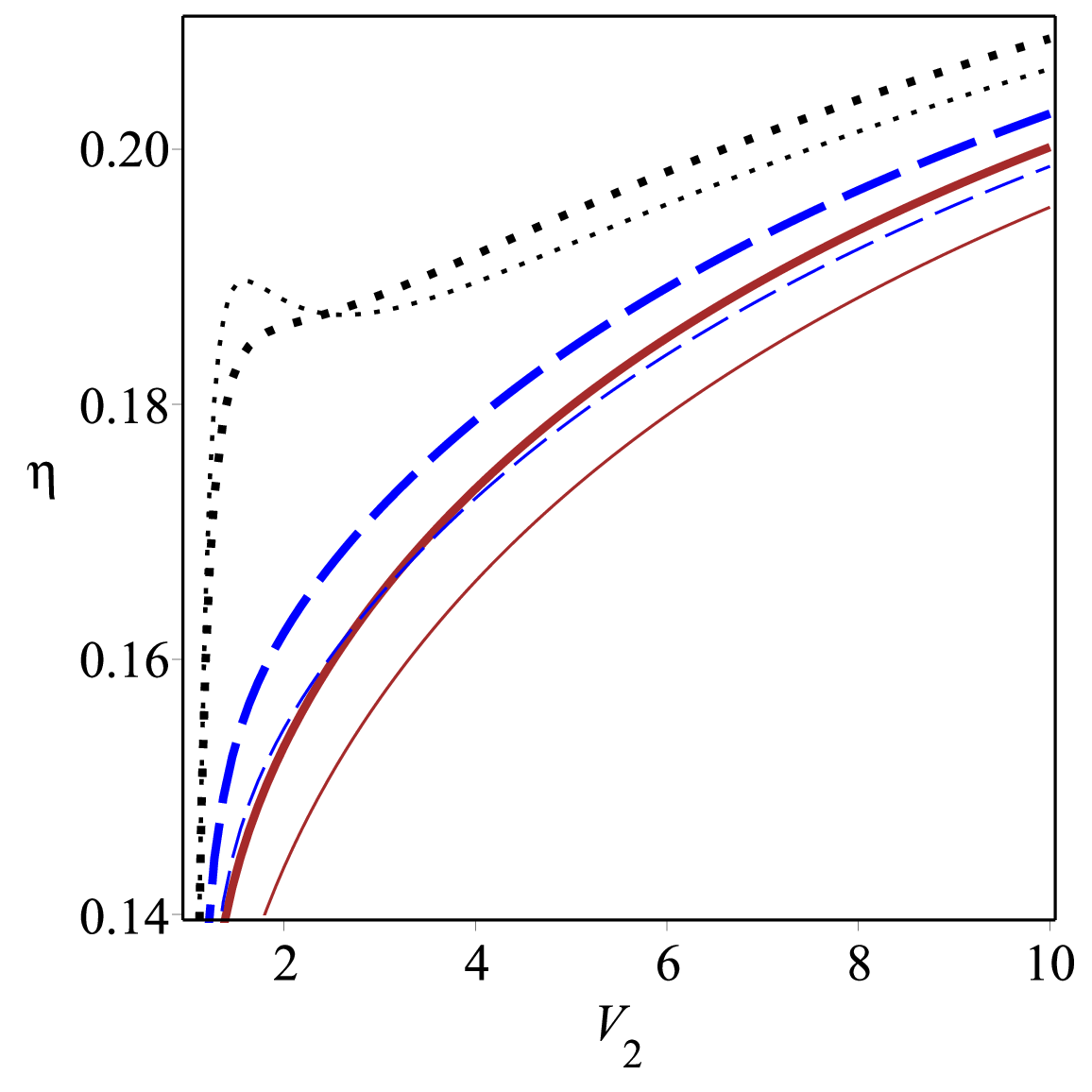}} 
\subfloat[]{
        \includegraphics[width=0.33\textwidth]{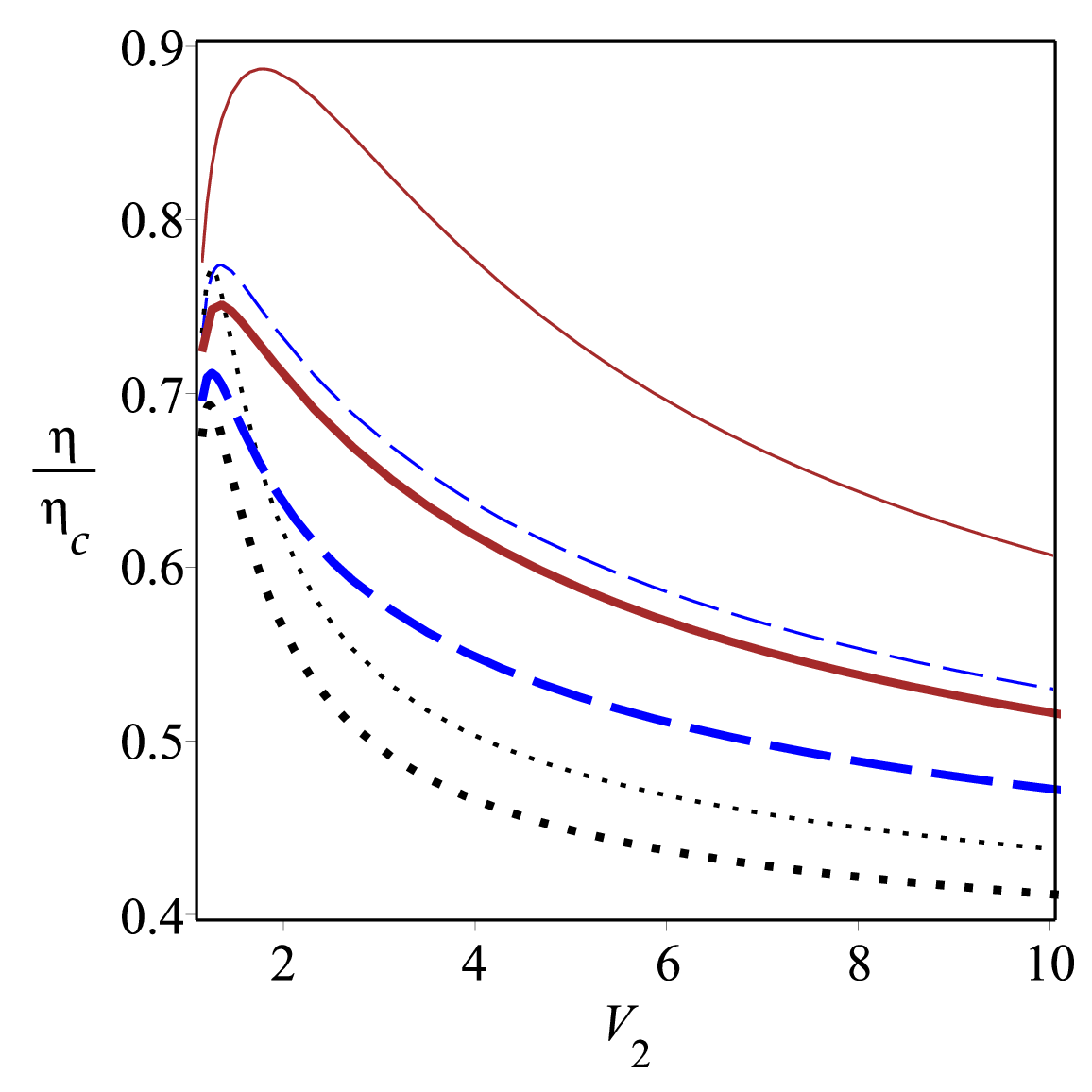}} \newline
\subfloat[]{
        \includegraphics[width=0.32\textwidth]{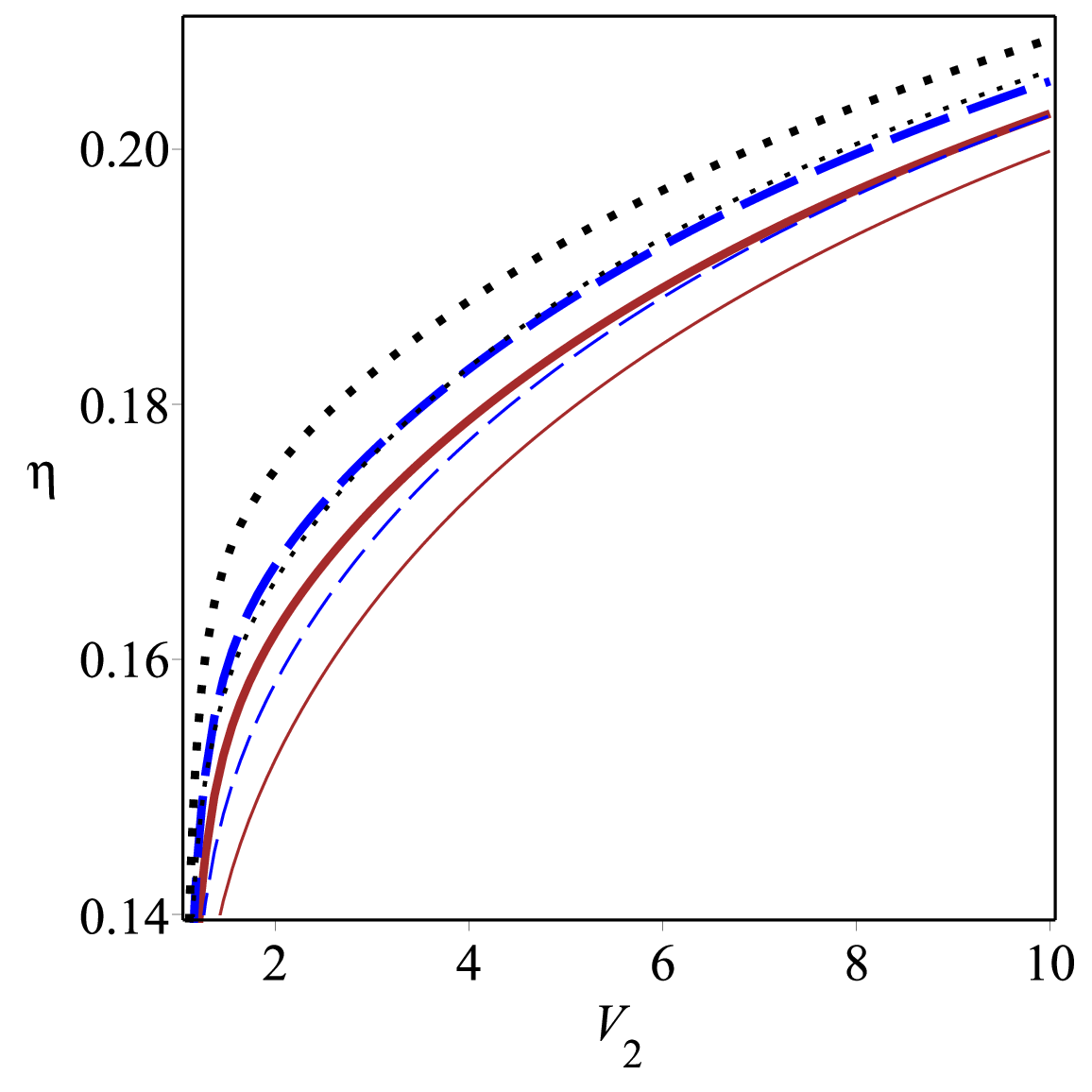}} 
\subfloat[]{
        \includegraphics[width=0.33\textwidth]{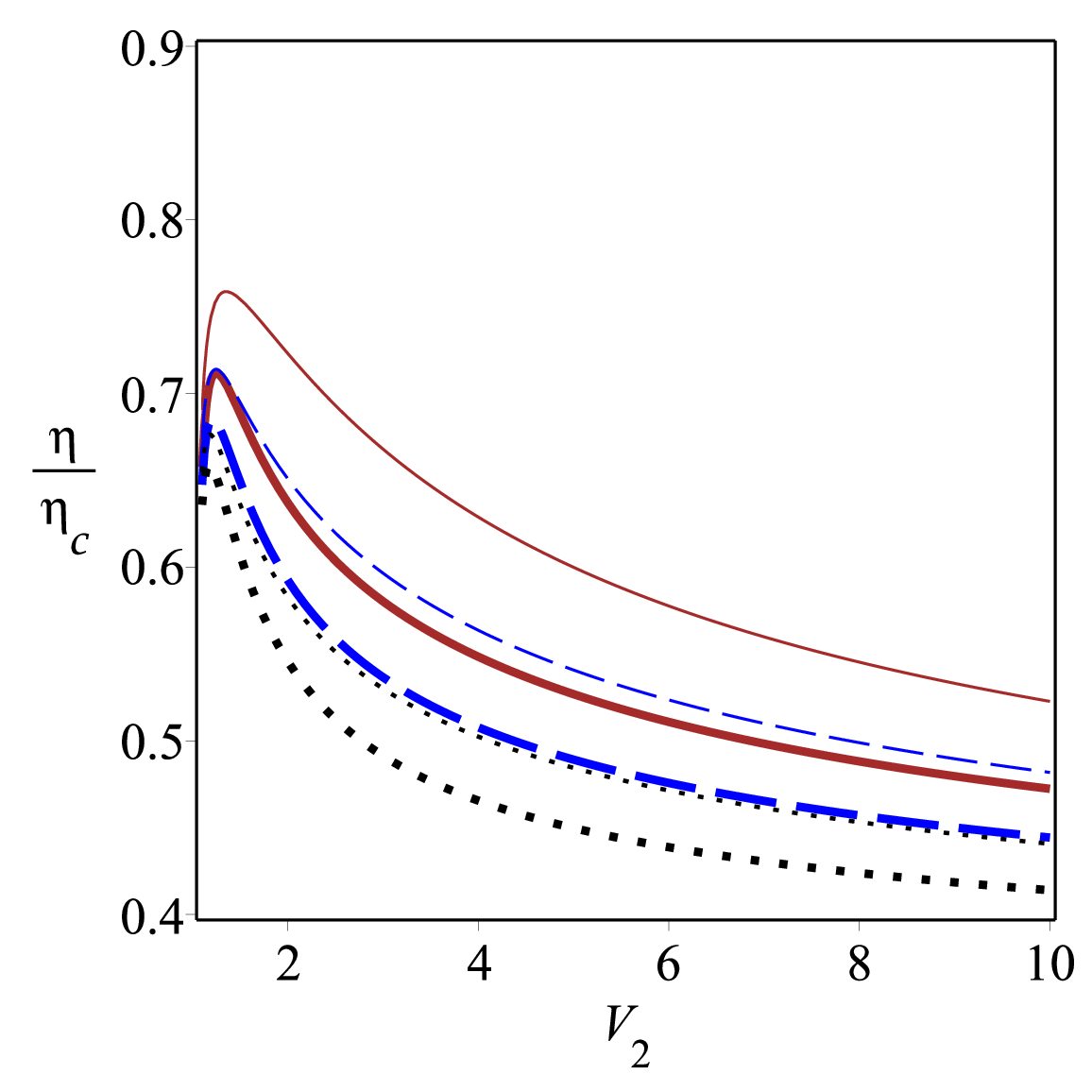}} \newline
\caption{Variation of $\protect\eta $ and $\frac{\protect\eta}{\protect\eta %
_{C}}$ versus $V_{2}$ for $B=0.2 $, $\protect\beta=0.04 $, $A=0.02 $, $%
\protect\mu =0.15$, $P_{1}=0.07$, $P_{4}=0.05$ and $V_{1}=1$. \textbf{Up
panels}: bold lines for $Q=0.2$ and thin lines for $Q=0$; $J=0.01$
(continues line), $J=0.03$ (dashed line) and $J=0.05$ (dotted line). \textbf{%
Down panels}: bold lines for $J=0.03$ and thin lines for $J=0$; $Q=0.2$
(continues line), $Q=0.25$ (dashed line) and $Q=0.3$ (dotted line).}
\label{Fig12}
\end{figure}

%%%%%%%%%%%%%%%%%%%%%%%%%%%%%%%%%%%%%%%%%%%%%%%%%%%%%%%%%%%%
For static black holes, thermodynamic volume is proportional to entropy. So,
the heat capacity at constant volume ($C_{V}=T\frac{\partial S}{\partial T} 
{\large \mid _{V}}$) is zero and calculating efficiency is straightforward
in this case \cite{CVJohnson,CVJohnson2,Heat2}. But for rotating black
holes, such a condition will not be satisfied due to the existence of
rotation effects. Consequently, usual methods cannot be applied to find the
efficiency of such black holes.

One of the most important problems regarding the AdS black holes with $%
C_{V}\neq 0$ comes from the fact that one does not know what the limits of
integration are in determining the heat. Using of the rectangular
cycles, one is able to obtain these limits by studying the limiting behavior
of these cycles. Furthermore, a rectangular cycle is the most natural cycle
to consider for all AdS black holes as it can be generalized to an algorithm
that allows for more complicated cycles to be numerically or even exactly
computed. In Ref. \cite{Chakraborty1a}, Chakraborty and Johnson suggested a
circular/elliptical cycle to investigate the efficiency of the black hole
heat engine. Since all thermodynamic quantities will be changed from point
to point on the circle, this cycle should be equally difficult for black
holes with $C_{V}\neq 0$, unless special conditions are considered to
simplify problems somewhat.

For rotating cases, the approach introduced in Ref. \cite{RAHennigar} is a
valid method for calculating efficiency. Here, we employ a rectangle cycle
to study the holographic heat engine for the black hole solution (see Fig. %
\ref{Fig11}). For cases with $C_{V}\neq 0$, $Q_{H}$ and $Q_{C}$ are
expressed in the following forms, 
\begin{eqnarray}
Q_{C} &=&M(V_{2},P_{1})-M(V_{4},P_{4})-\Delta PV_{2},  \notag \\
&&  \notag \\
Q_{H} &=&M(V_{2},P_{1})-M(V_{4},P_{4})-\Delta PV_{1},  \label{Eq30}
\end{eqnarray}%
where $\Delta P=P_{1}-P_{4}$. Employing Eqs. (\ref{Eq29}) and (\ref{Eq30}),
the efficiency is obtained as follows, 
\begin{equation}
\eta =\frac{\Delta P\Delta V}{\Delta M_{T}+\Delta U_{L}}.  \label{Eq31}
\end{equation}

The Carnot efficiency which is the maximum allowed efficiency by
thermodynamic laws is obtained as, 
\begin{equation}
\eta _{C}=1-\frac{T_{C}}{T_{H}}=1-\frac{T_{4}(P_{4},V_{1})}{%
T_{2}(P_{1},V_{2})}.  \label{Eq33}
\end{equation}

%%%%%%%%%%%%%%%%%%%%%%%%%%%%%%%%%%%%%%%%%%
\begin{figure}[tbh]
\centering
\subfloat[$J=0.03$ and $ Q=0.2 $]{
        \includegraphics[width=0.32\textwidth]{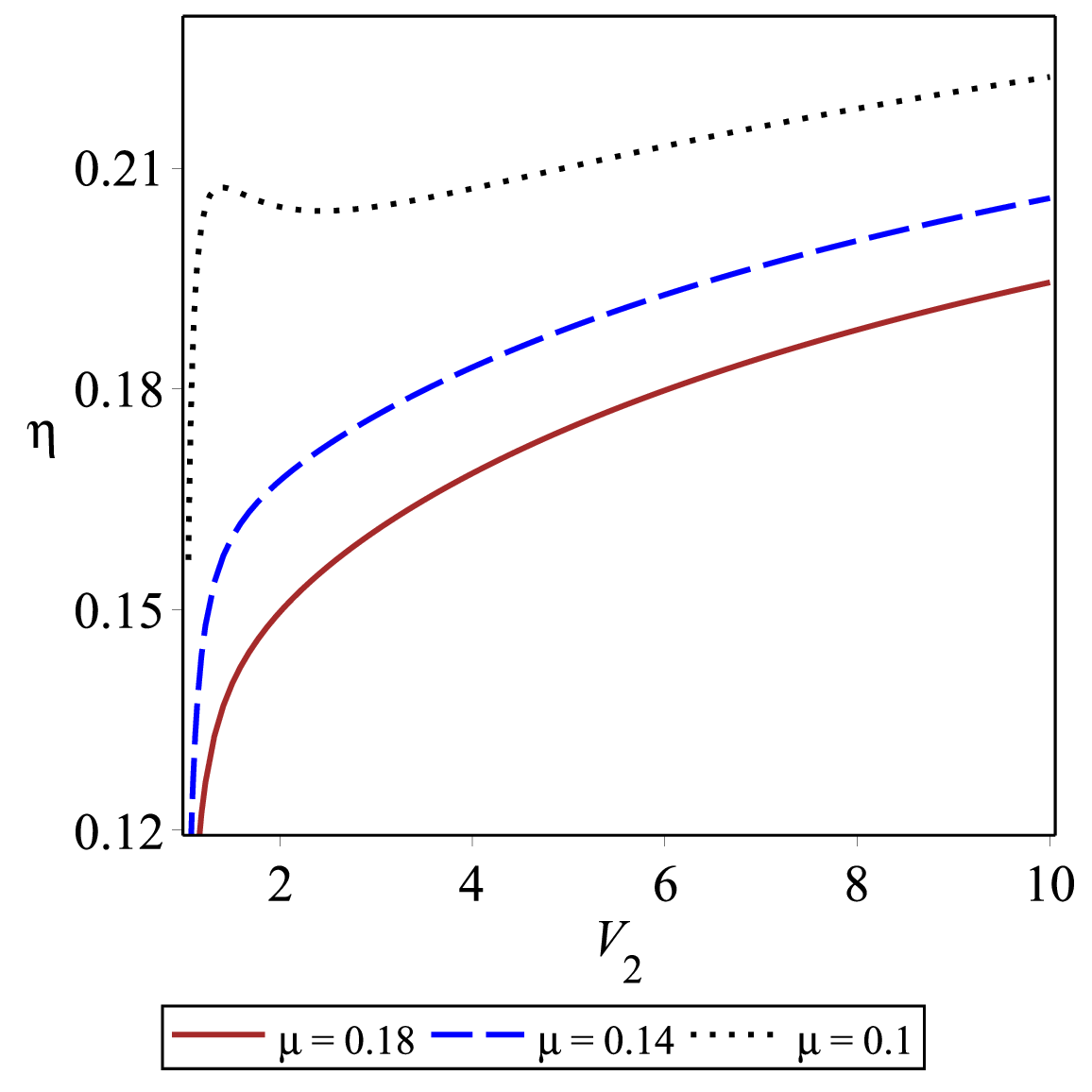}} 
\subfloat[$J=0.03$ and $ Q=0 $]{
        \includegraphics[width=0.32\textwidth]{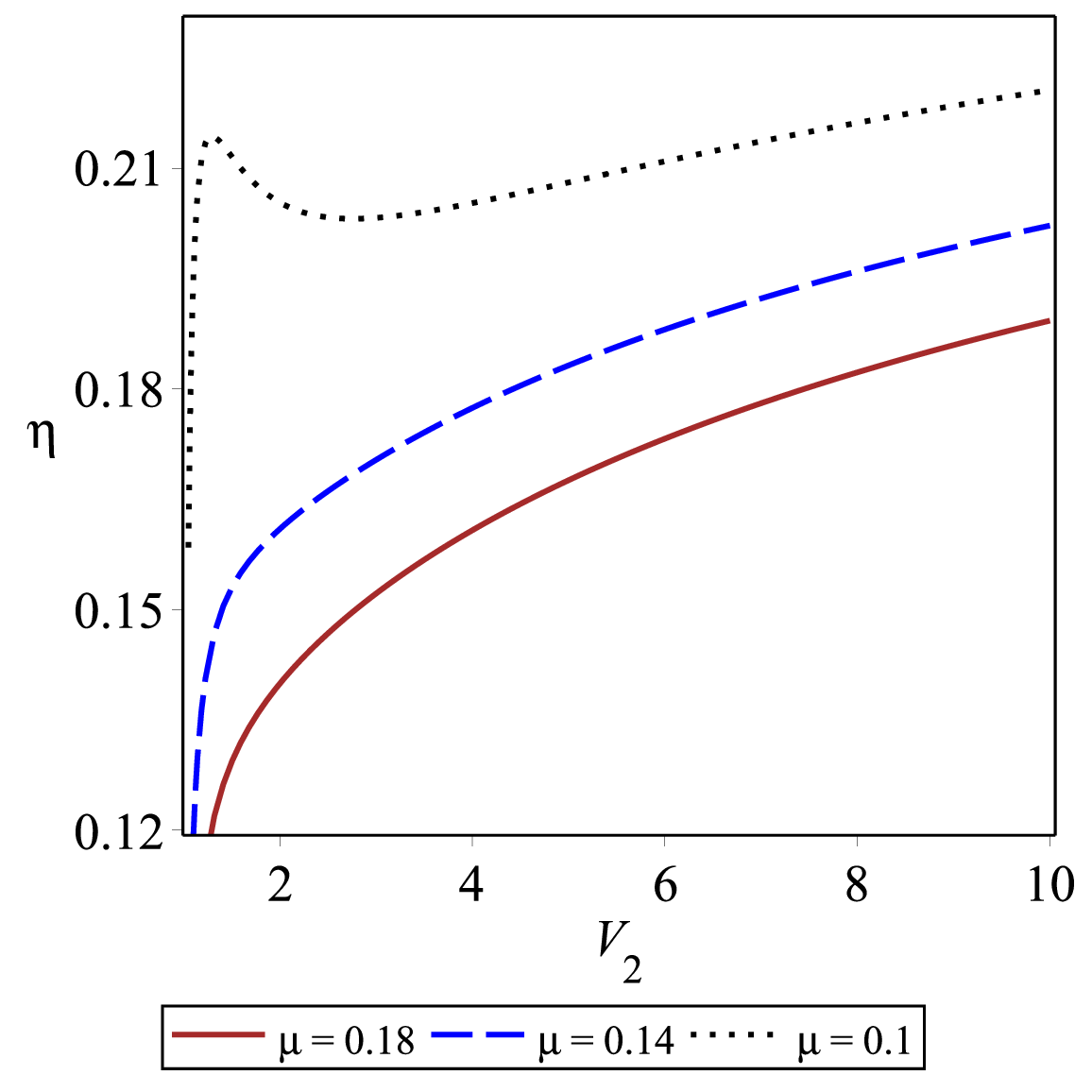}} 
\subfloat[$J=0$ and $ Q=0.2 $]{
        \includegraphics[width=0.32\textwidth]{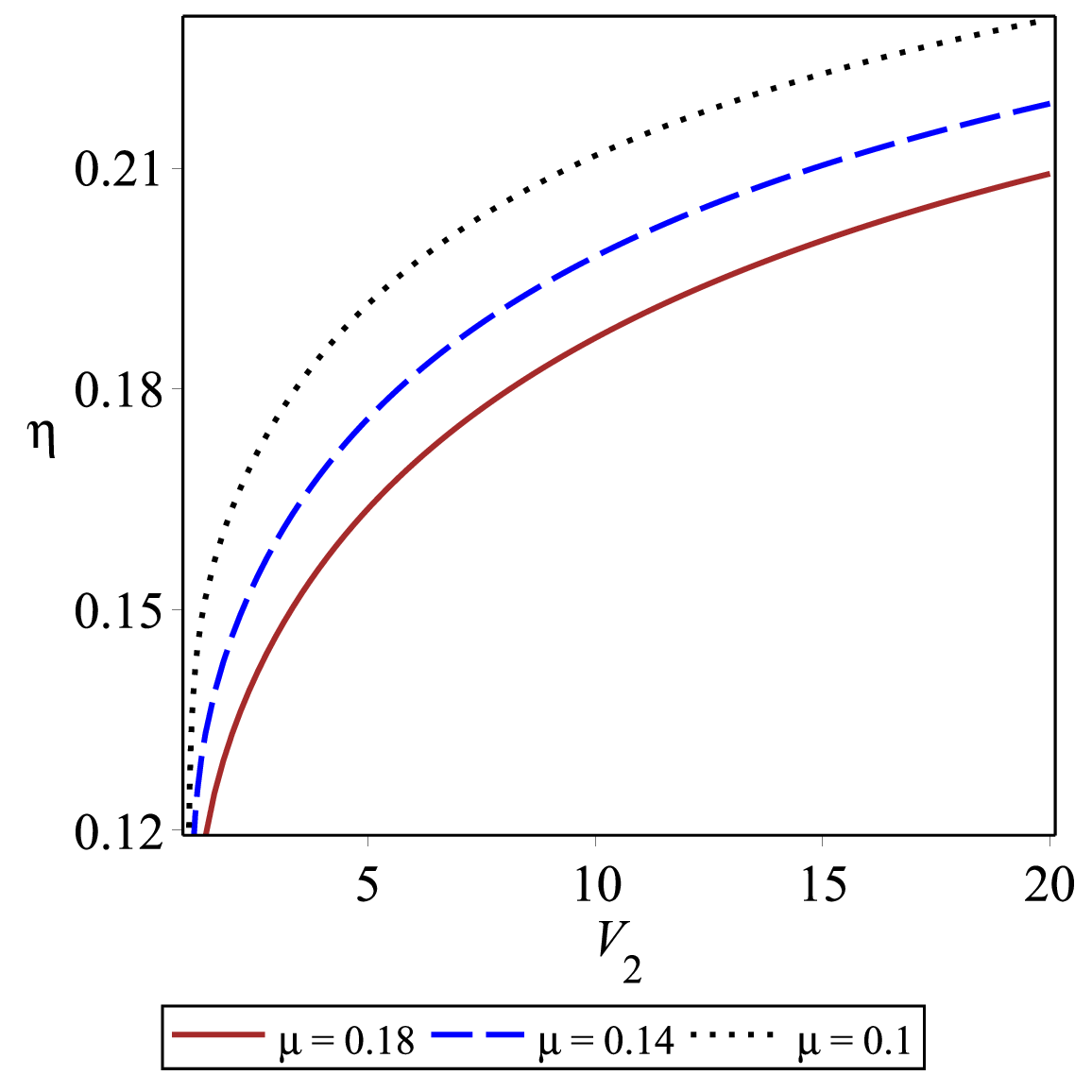}} \newline
\subfloat[$J=0.03$ and $ Q=0.2 $]{
        \includegraphics[width=0.32\textwidth]{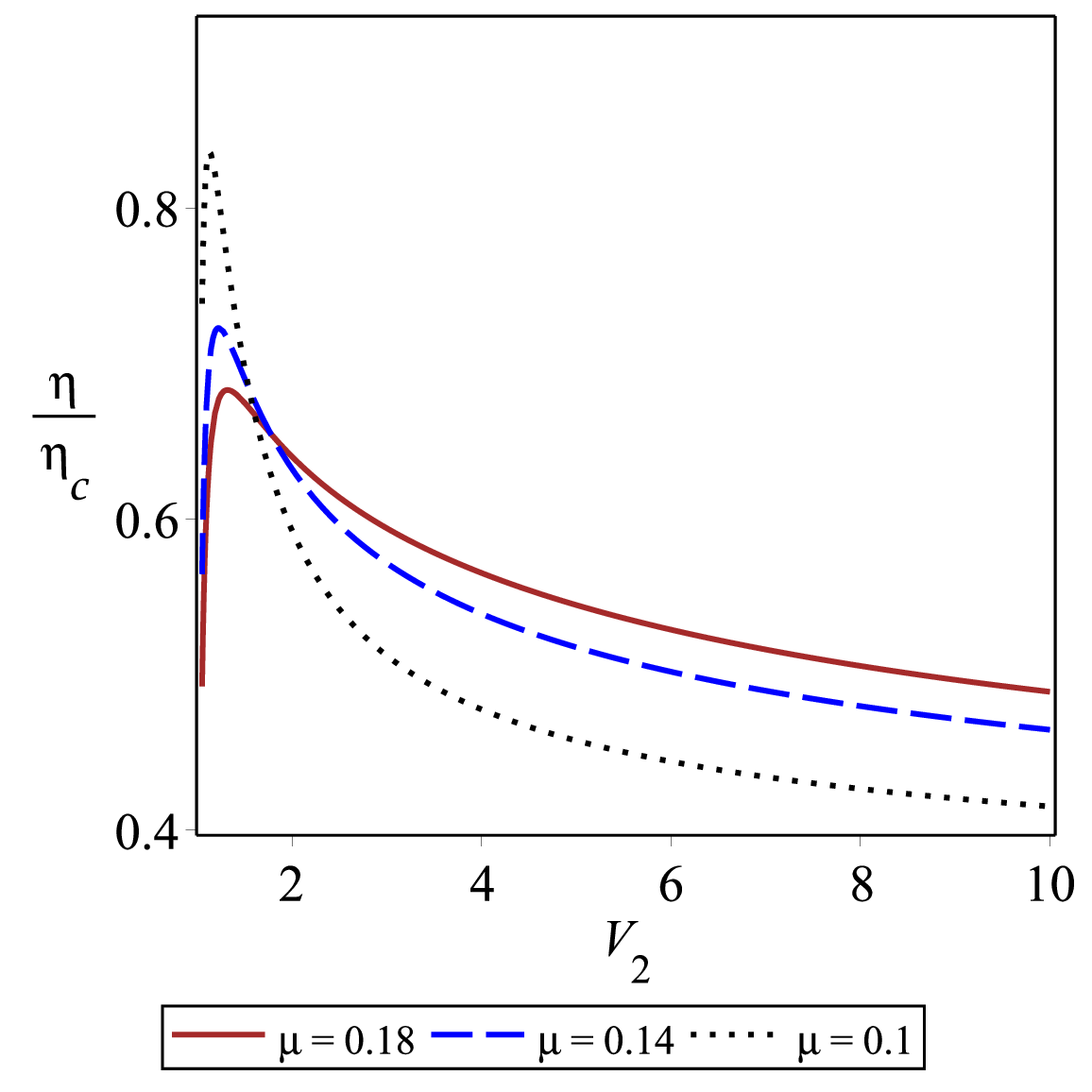}} 
\subfloat[$J=0.03$ and $ Q=0 $]{
        \includegraphics[width=0.32\textwidth]{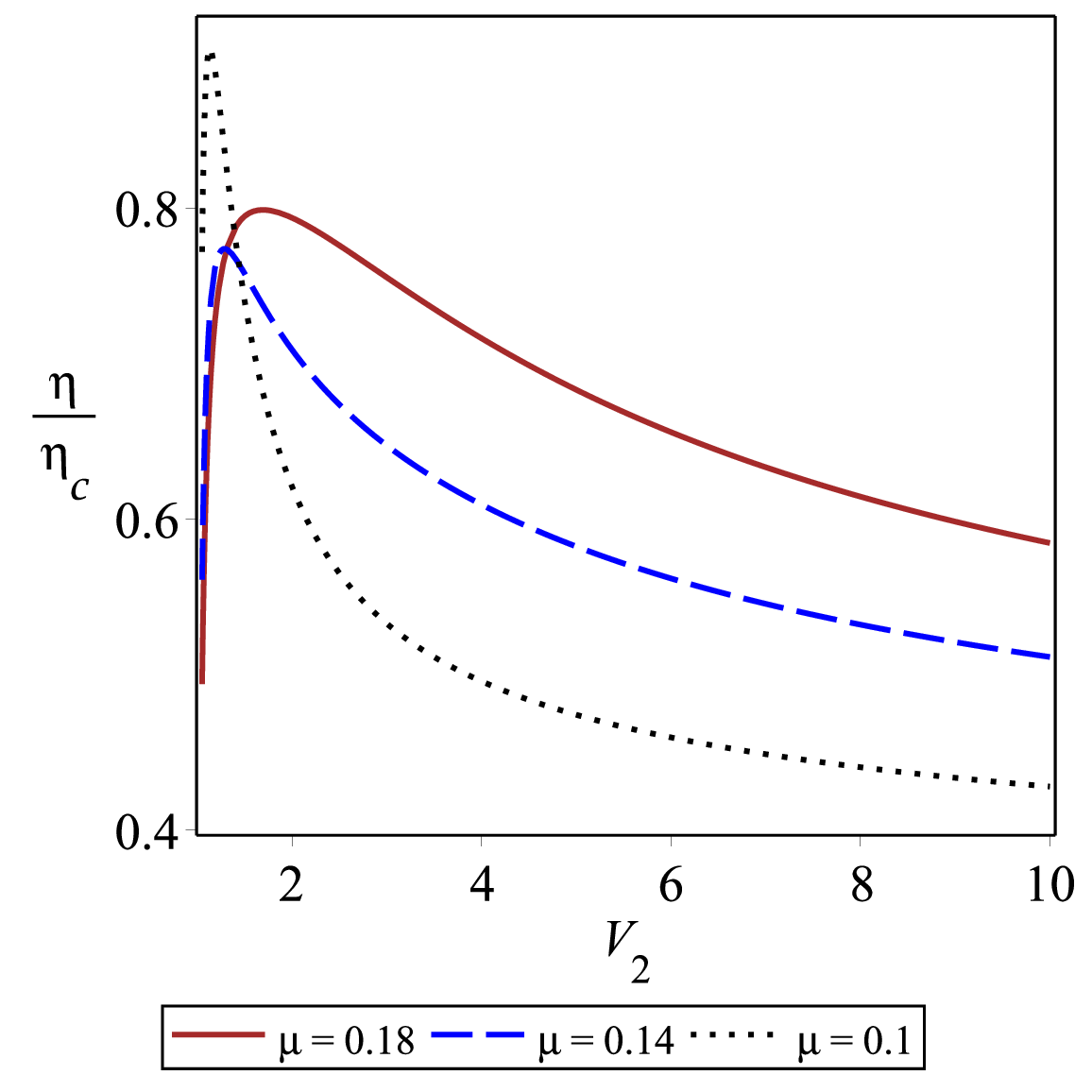}} 
\subfloat[$J=0$ and $ Q=0.2 $]{
\includegraphics[width=0.32\textwidth]{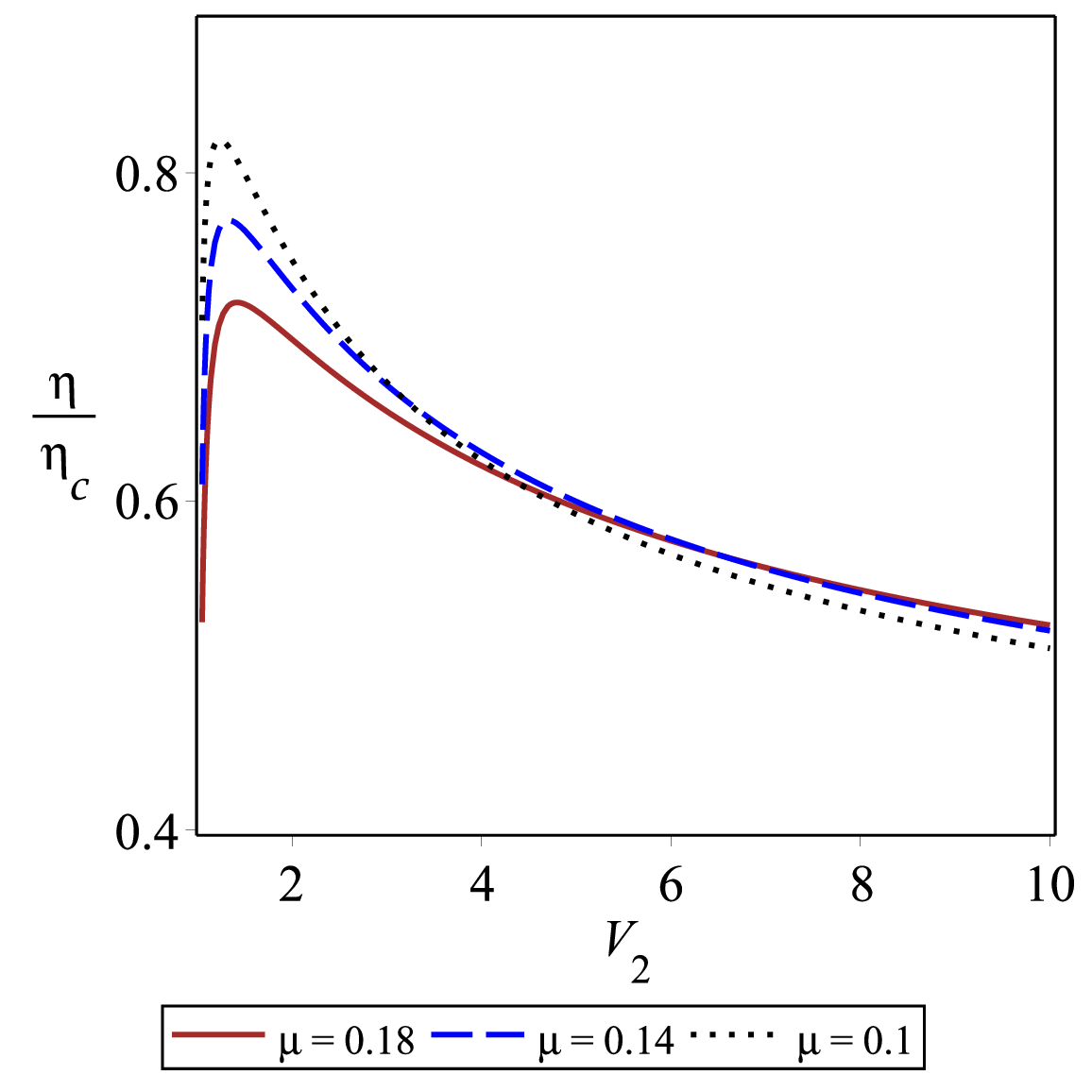}} \newline
\caption{Variation of $\protect\eta $ and $\frac{\protect\eta }{\protect\eta %
_{C}}$ versus $V_{2}$ for $B=0.2$, $\protect\beta =0.04$, $A=0.02$, $%
P_{1}=0.07$, $P_{4}=0.05$, $V_{1}=1$ and different values of string tension. 
\textbf{Left}: in the presence of rotation and electric charge. \textbf{%
Middle}: in the absence of electric charge. \textbf{Right}: in the absence
of rotation parameter.}
\label{Fig13}
\end{figure}
%%%%%%%%%%%%%%%%%%%%%%%%%%%%%%%%%%%%%%%%%%%%%%%%%%%

The efficiency of the engine ($\eta $) and Carnot efficiency ($\eta _{C}$)
are given in appendix C (see Eq. (\ref{Eq214}) and Eq. (\ref{Eq215})). Now
we analyze the behavior of the heat engine efficiency $\eta $ and the ratio $%
\frac{\eta }{\eta _{C}}$ under variation of black hole parameters. In Fig. %
\ref{Fig12}, we investigate the effects of angular momentum and electric
charge on $\eta $ and the ratio $\frac{\eta }{\eta _{C}}$ with fixed string
tension and pressure $P_{1}$, $P_{4}$. As we see, $\eta $ ($\frac{\eta }{%
\eta _{C}}$) is an increasing (a decreasing) function of these two
parameters. Comparing up and down panels, one can find that variation of $J$
has a stronger effect on $\eta $ and $\frac{\eta }{\eta _{C}}$ than the
electric charge. For small values of $J$, the efficiency monotonically
increases as the volume $V_{2}$ grows (see continues and dashed lines of
Fig. \ref{Fig12}(a)). While for large values of $J$, the efficiency curve
has a local minimum value which shows that there exists a finite value of
the volume $V_{2}$ at which the black hole heat engine works at the lowest
efficiency (see the dotted line of Fig. \ref{Fig12}(a)). Taking a close look
at the left panels of Fig. \ref{Fig12}, one can find that the charged rotating
accelerating black holes have a bigger efficiency than their non-rotating
and uncharged counterparts (compare bold and thin lines). Just in the region
of volume $V_{2}$ near $V_{1}$, their efficiency becomes smaller than the rapidly rotating accelerating black hole (compare bold-dotted and
thin-dotted lines in Fig. \ref{Fig12}(a)). In the right panels of Fig. \ref%
{Fig12}, we compare the heat engine efficiency with the Carnot efficiency.
As we see, $\frac{\eta }{\eta _{C}}<1$ always holds for all values of the
angular momentum and electric charge. For a small volume difference $\Delta
V $, the heat engine efficiency is close to the Carnot efficiency, whereas
in the limit of that the volume $V_{2}$ goes to infinity, the efficiency
becomes very smaller than Carnot efficiency. 
%%%%%%%%%%%%%%%%%%%%%%%%%%%%%%%%%%%%%%%%%%%%%%%%%%%
\begin{figure}[!htb]
\centering
\subfloat[]{
        \includegraphics[width=0.40\textwidth]{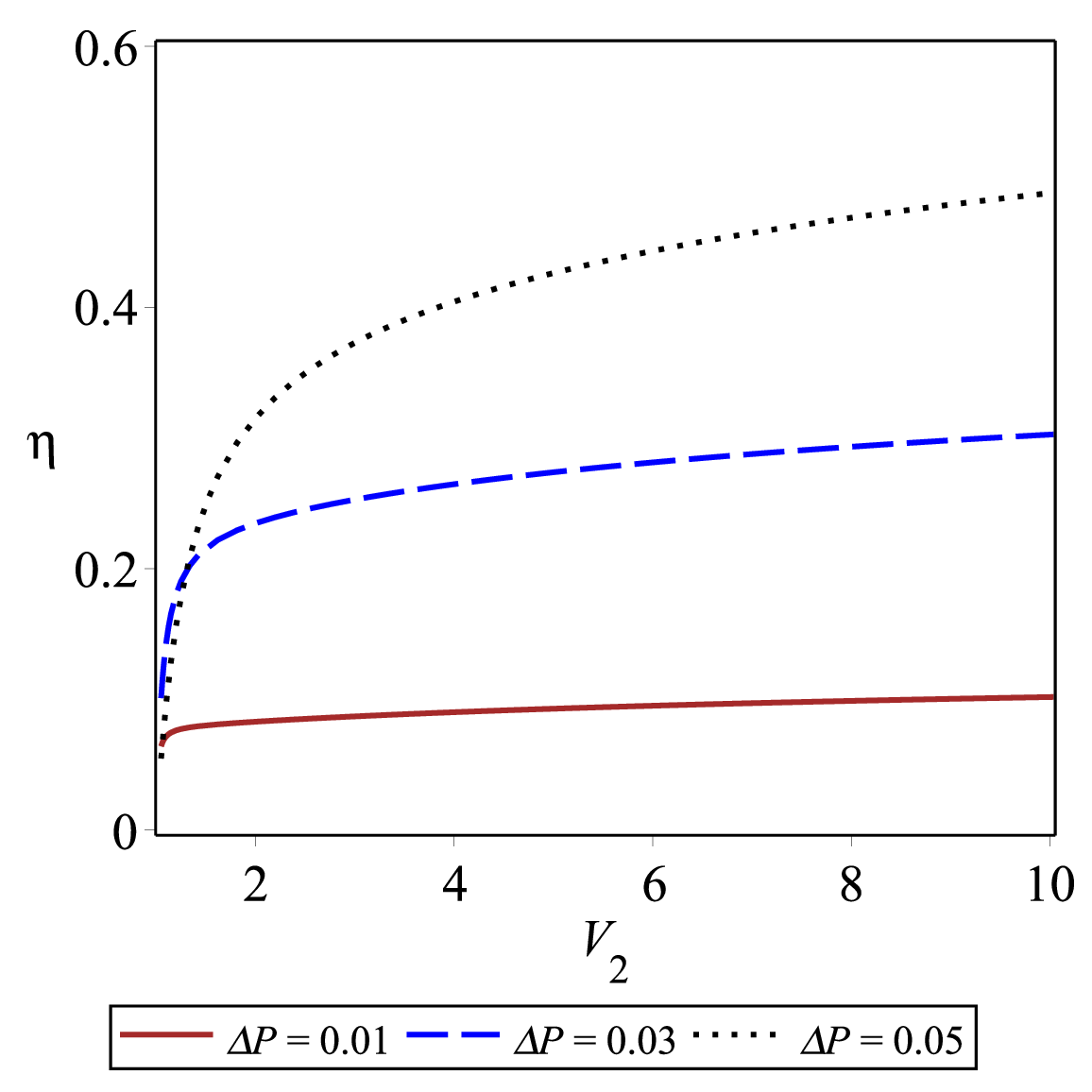}} 
\subfloat[]{
        \includegraphics[width=0.40\textwidth]{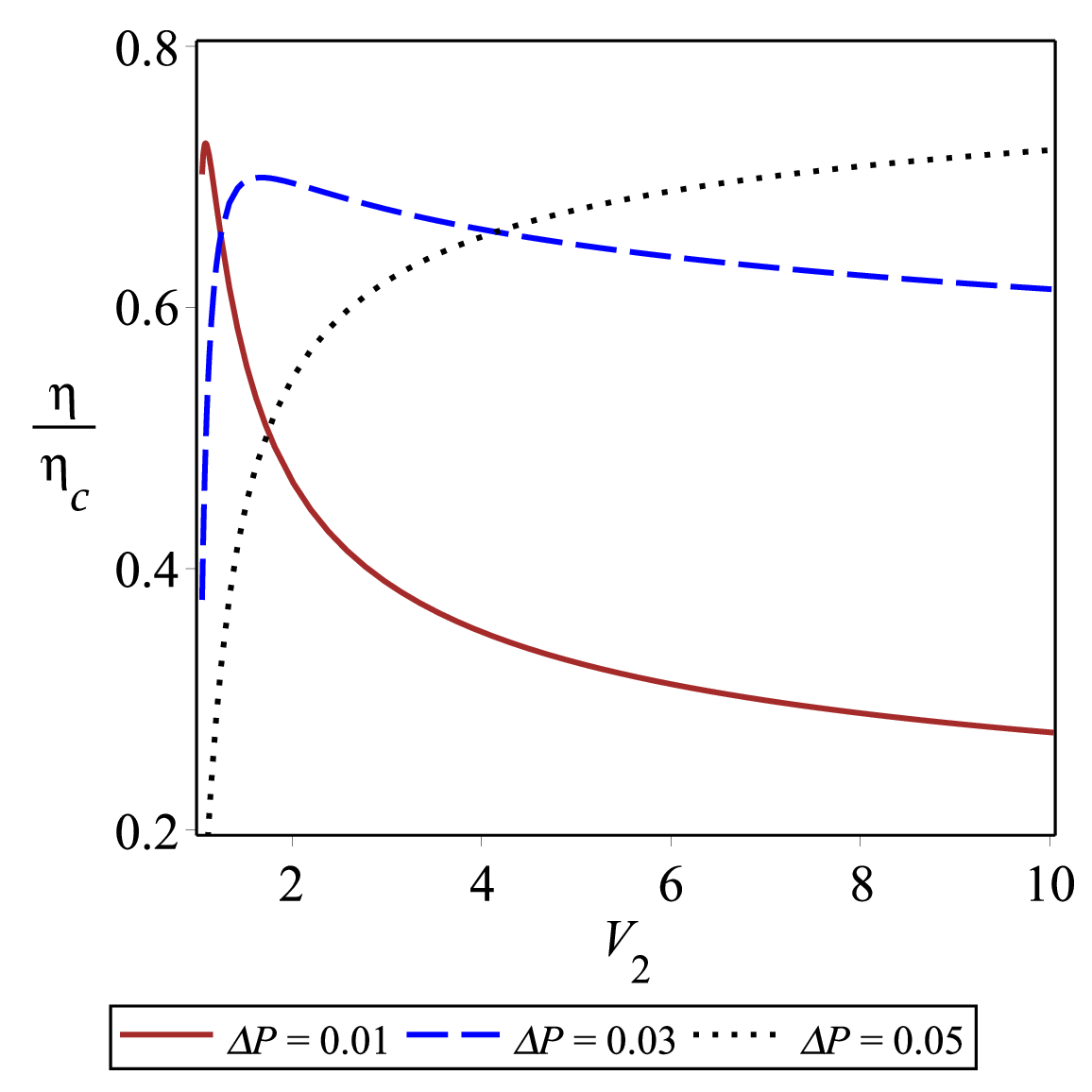}} \newline
\caption{Variation of $\protect\eta $ and $\frac{\protect\eta}{\protect\eta %
_{C}}$ versus $V_{2}$ for $B=0.2 $, $\protect\beta=0.04 $, $A=0.02 $, $%
J=0.03 $, $Q=0.2 $, $\protect\mu=0.15 $, $P_{1}=0.07$ and $V_{1}=1$.}
\label{Fig14}
\end{figure}
%%%%%%%%%%%%%%%%%%%%%%%%%%%%%%%%%%%%%%%%%%%%%%%%%%%

To show the effect of string tension on $\eta $ and $\frac{\eta }{\eta _{C}}$%
, we have plotted Fig. \ref{Fig13}. From up panels of this figure, we see
that increasing the string tension makes the decreasing of heat engine
efficiency. Since the string tension is directly related to the acceleration
parameter ($\mu =\frac{mA}{K}$), one can say that increasing the
acceleration parameter leads to the decreasing of efficiency. Comparing Fig. %
\ref{Fig13}(c) with Figs. \ref{Fig13}(a) and \ref{Fig13}(b), we find that
the effect of string tension becomes significant in presence of the angular
momentum. In absence of $J$ (charged accelerating black hole), the
efficiency monotonously increases with the growth of volume $V_{2}$ for all
values of the string tension (see Fig. \ref{Fig13}(c)). In other words, the
increase of volume difference between small and large black holes will make
the heat engine efficiency increase. In presence of $J$, the efficiency is
an increasing function of the volume $V_{2}$ for large values of $\mu $ (see
continues and dashed lines of Figs. \ref{Fig13}(a) and \ref{Fig13}(b)),
whereas, for small $\mu $, the efficiency curve acquires a minimum in a
finite value of the volume $V_{2}$ (see the dotted line of Figs. \ref{Fig13}%
(a) and \ref{Fig13}(b)). This means that for tiny acceleration, the heat
engine of the black hole works at the lowest efficiency for a certain value
of the volume $V_{2}$. It is worth pointing out that according to our
analysis the efficiency is smaller than one ($\eta <1 $) for all values of
charged rotating accelerating black hole parameters. To more precisely
compare the efficiency of charged rotating accelerating black holes to that
of uncharged or non-rotating accelerating ones, we have plotted the graphs
up to $V_{2}=10$ here. Down panels of Fig. \ref{Fig13}, illustrate the effect
of string tension on the ratio $\frac{\eta }{\eta _{C}}$. As we see, for
small (large) volume difference, increasing the acceleration leads to a
lower (higher) ratio $\frac{\eta }{\eta _{C}}$. We also observe that in the
region of volume $V_{2} $ near $V_{1}$, the ratio $\frac{\eta }{\eta _{C}}$
will be greater than one for small enough string tension which is
inconsistent with the second law.

In Fig. \ref{Fig14}, we have displayed the variation of $\eta $ and $\frac{%
\eta }{\eta _{C}}$ as a function of the volume $V_{2}$ for different values
of the pressure difference ($\Delta P$) with fixed angular momentum,
electric charge and string tension. According to Fig. \ref{Fig14}(a), the
efficiency is an increasing function of $\Delta P$. For all values of $%
\Delta P$, the efficiency monotonically increases as the volume $V_{2}$
grows and then tends to the saturation value. Fig. \ref{Fig14}(b), shows the
effect of pressure on the ratio $\frac{\eta }{\eta _{C}}$. We see that for
small volume difference, $\frac{\eta }{\eta _{C}}$ decreases by
increasing of $\Delta P$, while the opposite is true for large volume
difference. So, in the region of volume $V_{2}$ near $V_{1}$, the heat
engine efficiency is close to the Carnot efficiency for small pressure
difference (see continues line of Fig. \ref{Fig14}(b)). Whereas, in the
limit of that the volume $V_{2}$ goes to infinity, the efficiency approaches
the Carnot efficiency for large $\Delta P$ (see the dotted line of Fig. \ref%
{Fig14}(b)).

\section{Conclusion}

In this paper, we investigated the thermodynamic behavior of charged
rotating accelerating AdS black holes. First, we explored the impact of
angular momentum, electric charge and string tension on thermodynamic
quantities. We found that the temperature is a decreasing (an increasing)
function of the angular momentum and electric charge (string tension). The
total mass is an increasing function of these three parameters. Regarding
the entropy, we observed that increasing the angular momentum and string
tension lead to increasing of entropy, whereas the electric charge has an
opposite effect. Studying the effects of these parameters on thermal
stability/instability of the system, we noticed that the regions of
stability decrease as the angular momentum and electric charge (string
tension) decrease (increases).

Then, we considered the cosmological constant as a thermodynamic
pressure and investigated the possibility of van der Waals-like phase
transition for these black holes. We extracted the critical quantities and found
that the critical volume is an increasing (a decreasing) function of the
angular momentum and electric charge (string tension). Whereas, the opposite
behavior is observed for the critical temperature and pressure.

Finally, by considering the charged accelerating black holes as  working substances, we studied
the holographic heat engine by using a rectangle heat cycle in the $P-V$
plot. Investigating the black hole heat engine efficiency and comparing its
results with the Carnot efficiency led to the following interesting results:

I) The efficiency (the ratio $\frac{\eta}{\eta _{C}}$) is an increasing (a
decreasing) function of the angular momentum and electric charge. The
condition $\frac{\eta}{\eta _{C}} < 1$ will be satisfied for all values of
these two parameters.

II) Charged rotating accelerating black holes have a bigger efficiency than
their non-rotating and uncharged counterparts, except in the region of
volume $V_{2} $ near $V_{1} $, the efficiency of rapidly rotating
accelerating black holes becomes bigger than them.

III) Increasing the acceleration makes decreasing of the heat engine
efficiency. For small (large) volume difference, the ratio $\frac{\eta }{%
\eta _{C}}$ is a decreasing (an increasing) function of the string tension.
The efficiency will be bigger than Carnot efficiency for sufficiently small
string tension which is forbidden by the thermodynamic second law. This
result may suggest that the string tension must be constrained to preserve
the thermodynamics laws.

IV) The effect of string tension on the efficiency of rotating black holes
is more noticeable than that of charged black holes.

V) The efficiency increases by increasing pressure difference. For all
values of pressure, the efficiency is always smaller than Carnot efficiency
which is consistent to the second law. In the region of volume $V_{2}$ near $%
V_{1}$, the heat engine efficiency is close to the Carnot efficiency for
small pressure difference. When the volume $V_{2}$ goes to infinity, the
efficiency approaches the Carnot efficiency for large pressure difference.

Note added: concurrently with our work, W. Ahmed et al., have studied the heat
engine efficiency of such black holes through a circular cycle in Ref. \cite%
{Wasif}. They have investigated the effects of black hole parameters on
efficiency and obtained similar results with our work. It is worthwhile to
mention that we did not only study the impact of black hole parameters and
pressure on the heat engine efficiency but also we compared it to Carnot
efficiency and investigated the conditions to preserve the thermodynamic
second law.\newline

\section*{Acknowledgements}

We are grateful to the anonymous referees for the insightful comments and
suggestions, which have allowed us to improve this paper significantly. BEP
thanks the University of Mazandaran.

\section*{Data Availability}

This manuscript does not have any associated data.

\begin{center}
\textbf{Appendix}

\textbf{A: The extremal black hole}
\end{center}

To study the extremal black holes, we first obtain the value of the electric
charge, angular momentum and string tension in this limit. Since the
temperature relation (\ref{Eq9}) is very complicated to determine $Q_{e}$, $%
J_{e}$ and $\mu_{e}$ (index (e) denotes to extremal) analytically, we
perform our calculations once in the absence of the angular momentum and
again in the absence of the electric charge. For a charged accelerating
case, $Q_{e}$ and $\mu_{e}$ are obtained as

\begin{equation}
Q_{e}=\sqrt{\frac{(3r_{e}^{2}+\ell^{2}-2\beta^{2}r_{e}^{2}-A^{2}r_{e}^{4}),%
\mu^{2}r_{e}^{2}}{\ell^{2}-2\beta^{2}r_{e}^{2}}},
\end{equation}

\begin{equation}
\mu_{e}=\sqrt{\frac{(\ell^{2}-2\beta^{2}r_{e}^{2})B^{2}Q^{2}}{%
(3r_{e}^{2}+\ell^{2}-2\beta^{2}r_{e}^{2}-A^{2}r_{e}^{4})r_{e}^{2}}}.
\end{equation}

\begin{figure}[tbh]
\centering
\subfloat[charged accelerating BH]{
        \includegraphics[width=0.28\textwidth]{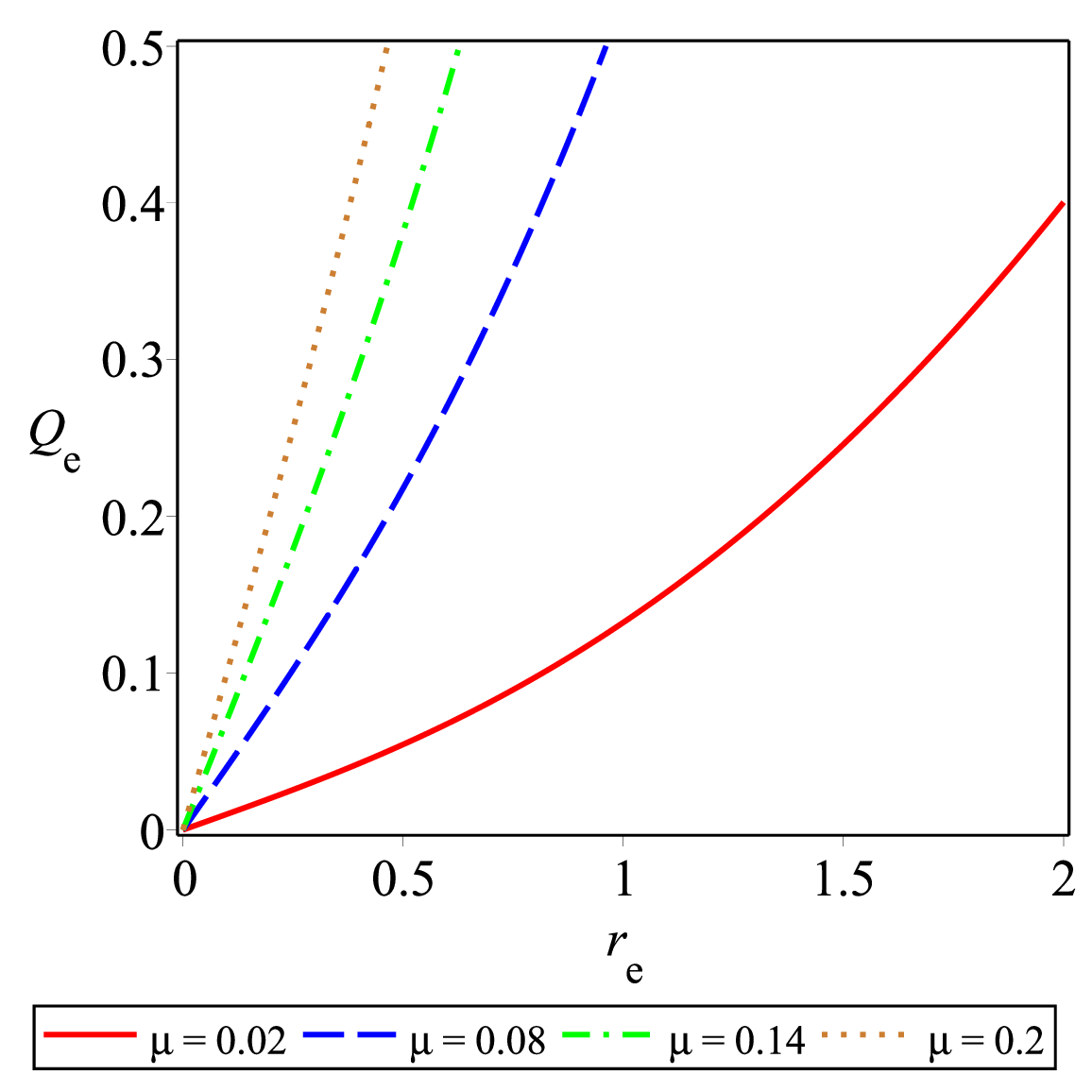}} 
\subfloat[charged accelerating BH]{
        \includegraphics[width=0.28\textwidth]{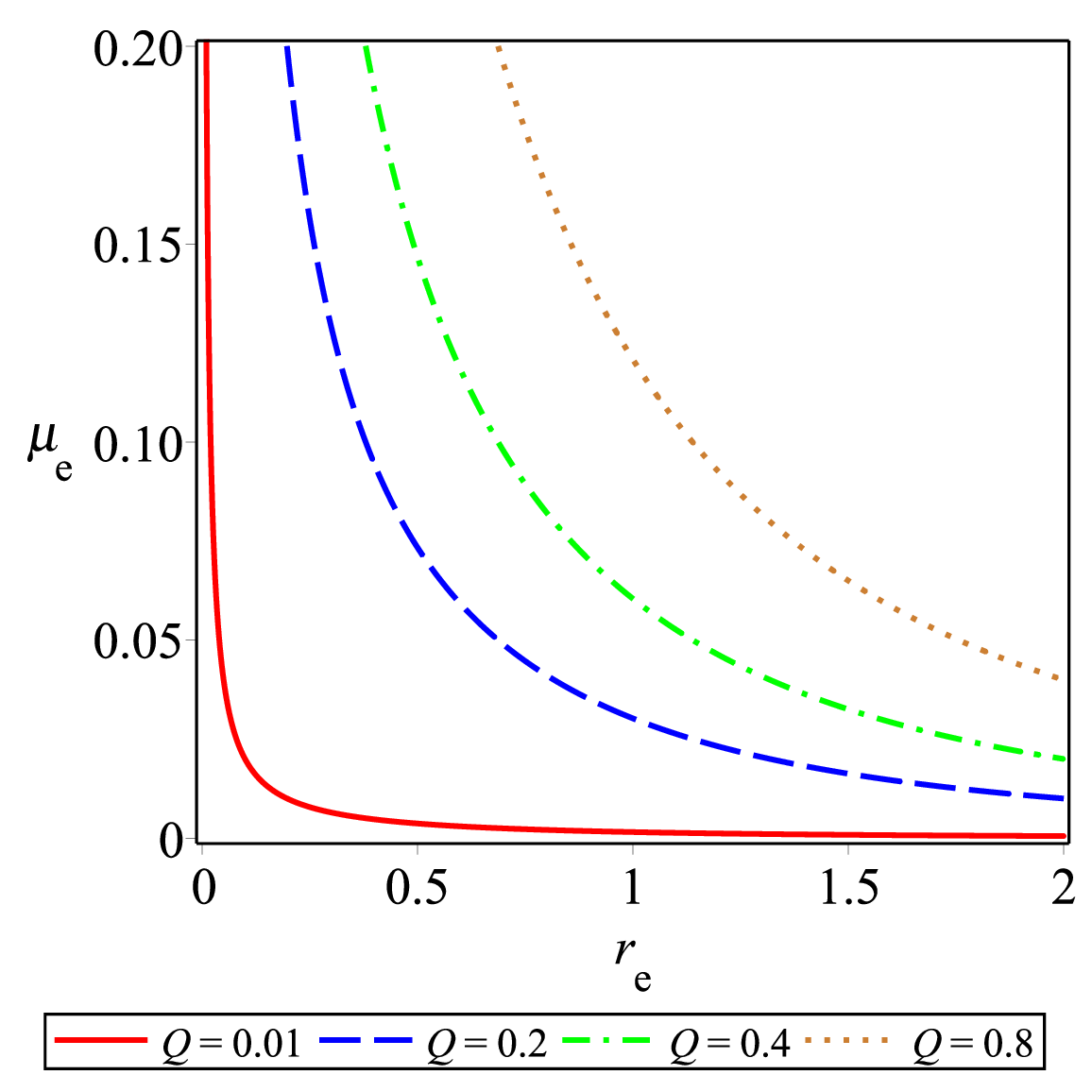}}\newline
\subfloat[rotating and accelerating BH]{
        \includegraphics[width=0.28\textwidth]{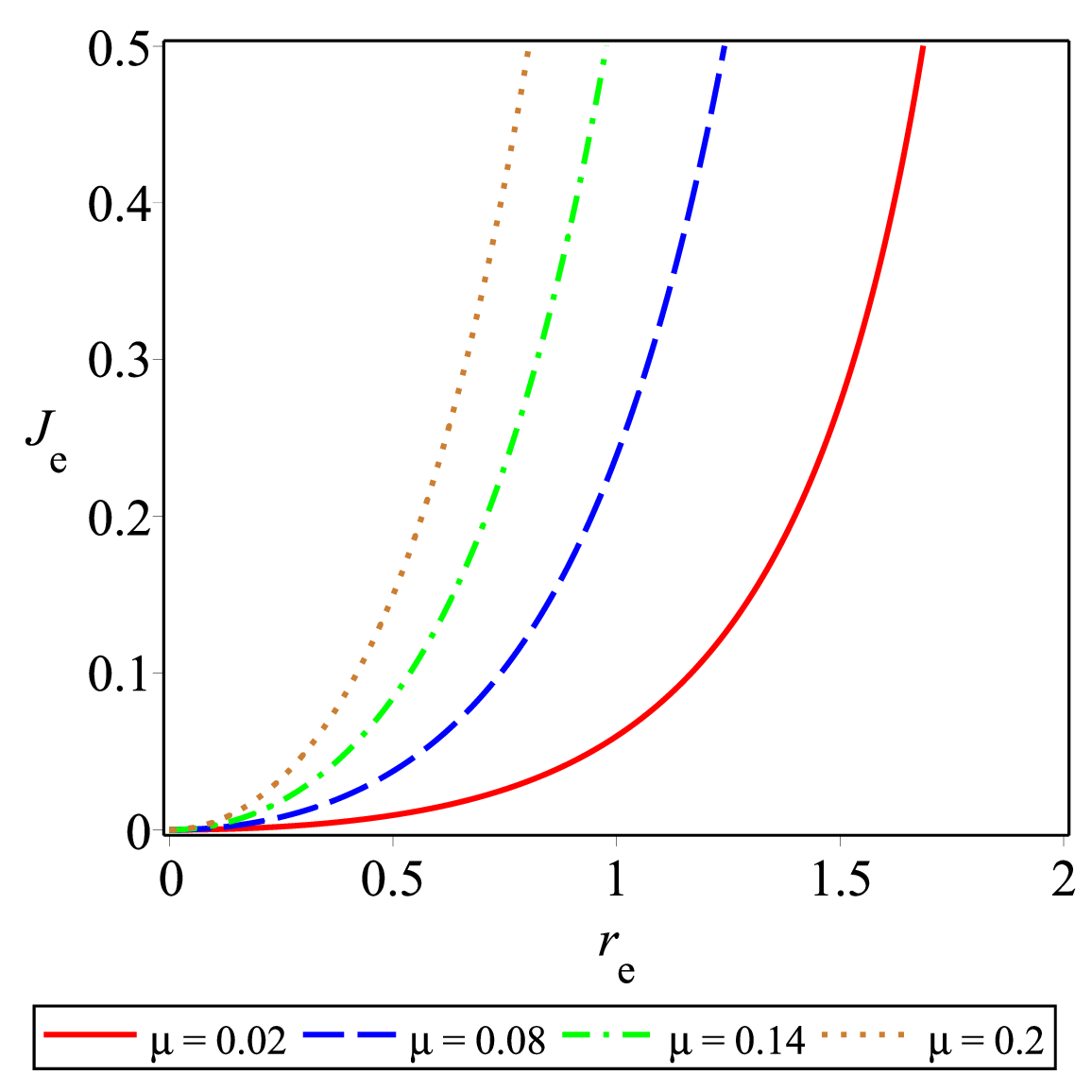}} 
\subfloat[rotating and accelerating BH]{
        \includegraphics[width=0.28\textwidth]{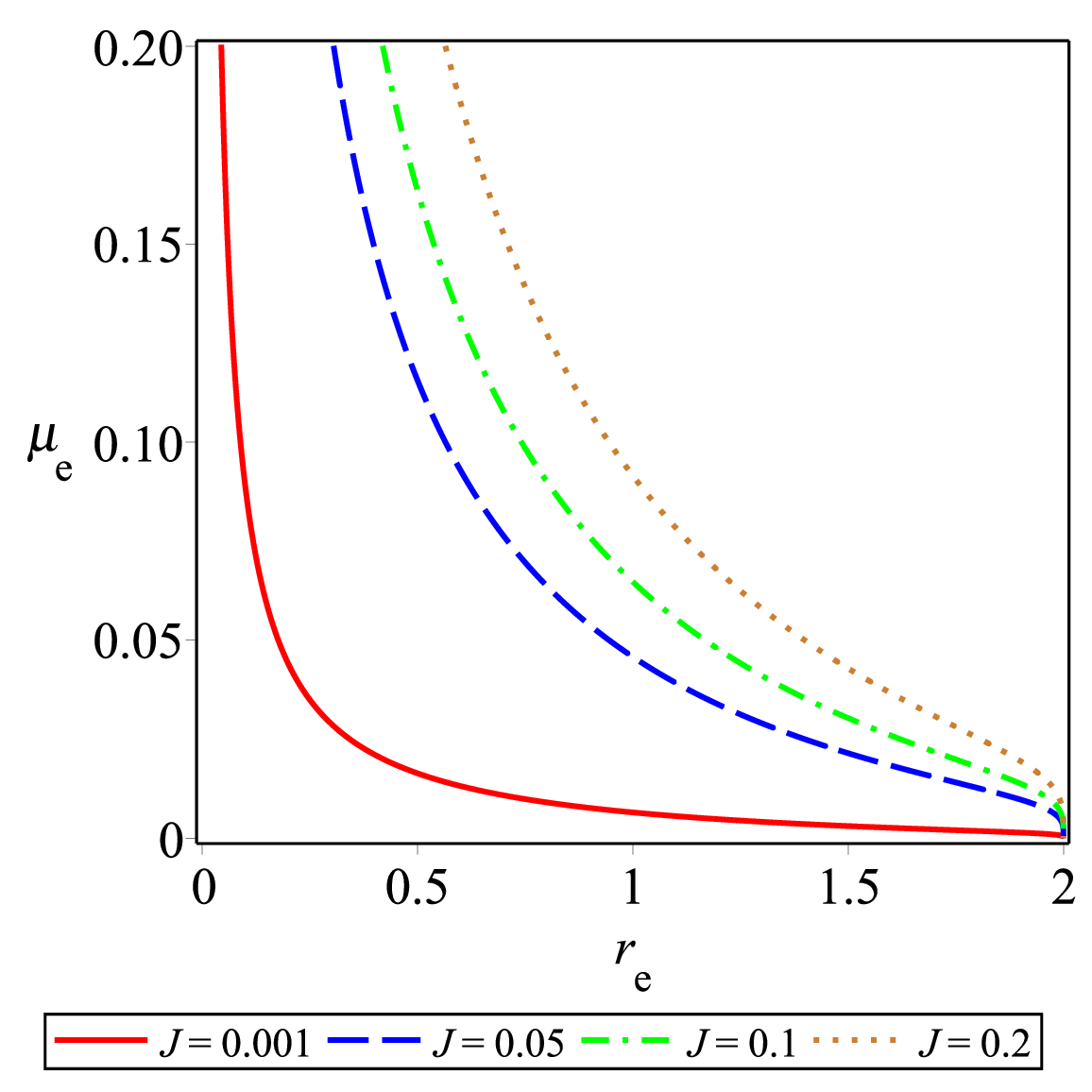}} \newline
\caption{The behavior of $Q_{e} $, $J_{e} $ and $\protect\mu_{e} $ as a
function of the extremal radius $r_{e} $ for $B=0.2 $, $\ell=2 $ and $A=0.02 
$.}
\label{Figex}
\end{figure}

Qualitative behavior $Q_{e}$ and $\mu_{e}$ with respect to the extremal
radius is depicted in Figs. \ref{Figex}(a) and \ref{Figex}(b). As we see, $%
Q_{e}$ ($\mu_{e}$) is an increasing (a decreasing) function of $r_{e}$ and
its qualitative behavior never changes by varying the string tension (the
electric charge).

Regarding the rotating and accelerating case, $J_{e}$ and $\mu_{e}$ are
given by %=============================
%============================ 
\begin{equation}
J_{e}=\frac{\mu^{2}r_{e}(r_{e}^{2}+\ell^{2})}{2B^{2}\ell^{2}}\sqrt{\frac{%
3r_{e}^{4}+\ell^{2}r_{e}^{2}+2A^{2}r_{e}^{6}-A^{2}\ell^{2}r_{e}^{4}}{%
\ell^{2}-r_{e}^{2}}},
\end{equation}

\begin{equation}
\mu_{e}=\sqrt{\frac{2JB^{2}\sqrt{\ell^{2}-r_{e}^{2}}}{r_{e}(r_{e}^{2}+%
\ell^{2})\sqrt{3r_{e}^{4}+\ell^{2}r_{e}^{2}+2A^{2}r_{e}^{6}-A^{2}%
\ell^{2}r_{e}^{4}}}}.
\end{equation}

From Figs. \ref{Figex}(c) and \ref{Figex}(d), one can find that qualitative
behavior of $J_{e}$ and $\mu_{e}$ is similar to $Q_{e}$ and $\mu_{e}$ for
the charged accelerating case. So, similar discussion can be used in this
case. Fig. \ref{Figex}, shows that for a fixed $r_{e}$ both $Q_{e}$ and $%
J_{e} $ increase as the string tension increases. This reveals the fact that
when a black hole is pulled by stronger string tension, it should be located
in a more powerful electric field or rotate quickly to have a physical
solution.

\begin{figure}[tbh]
\centering
\subfloat[Reissner-Nordström BH]{
        \includegraphics[width=0.28\textwidth]{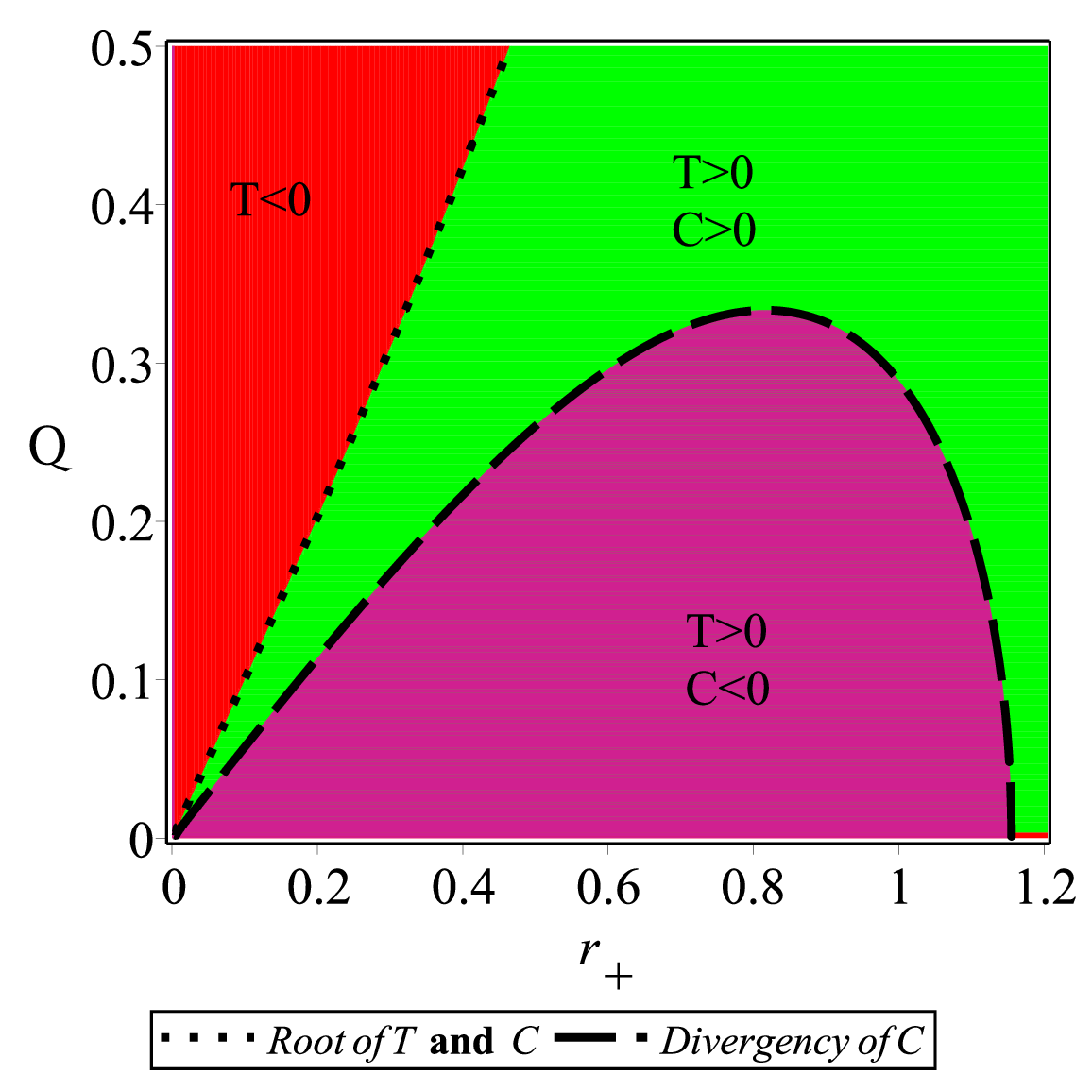}} 
\subfloat[Kerr BH]{
        \includegraphics[width=0.28\textwidth]{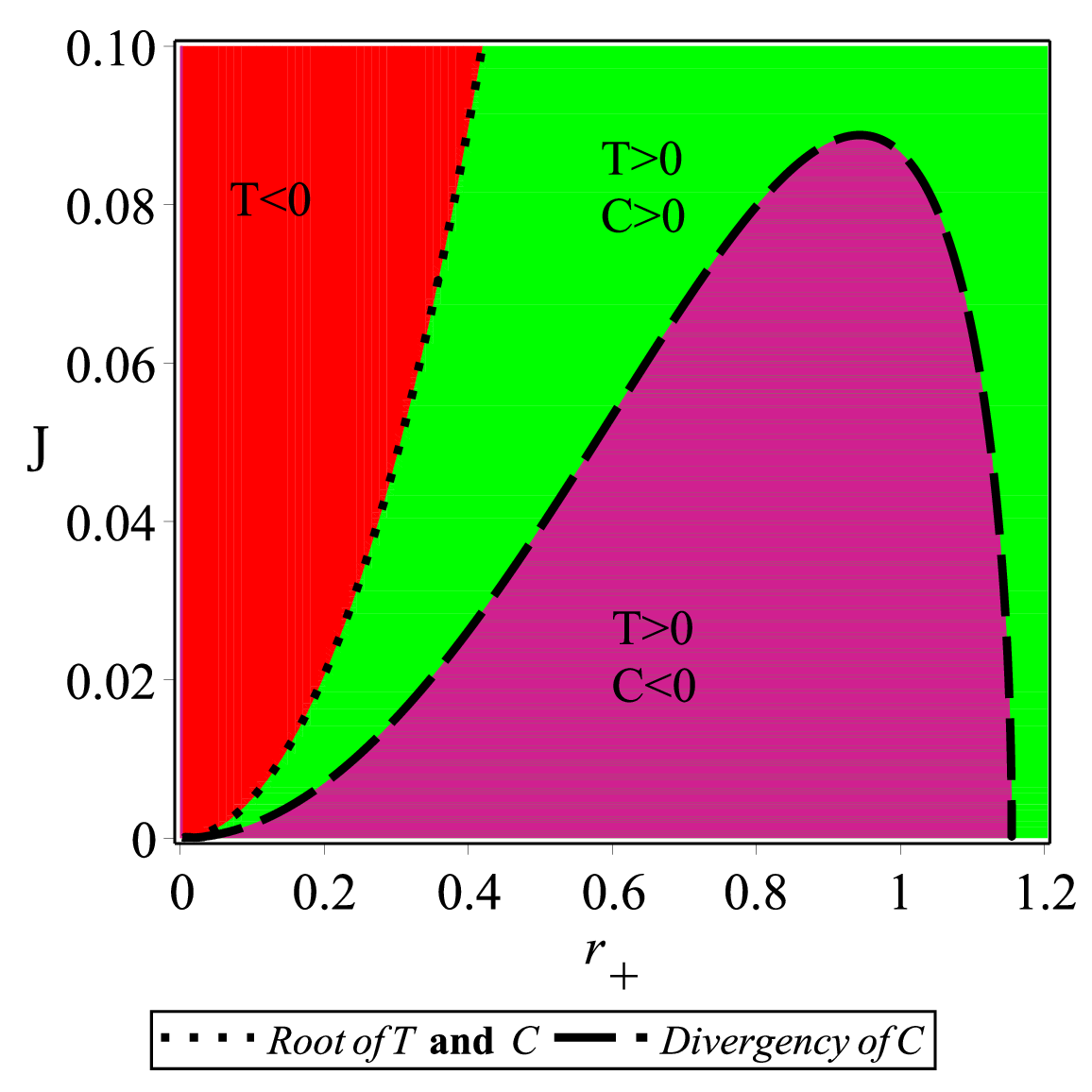}}\newline
\subfloat[ Kerr-Newman BH]{
        \includegraphics[width=0.28\textwidth]{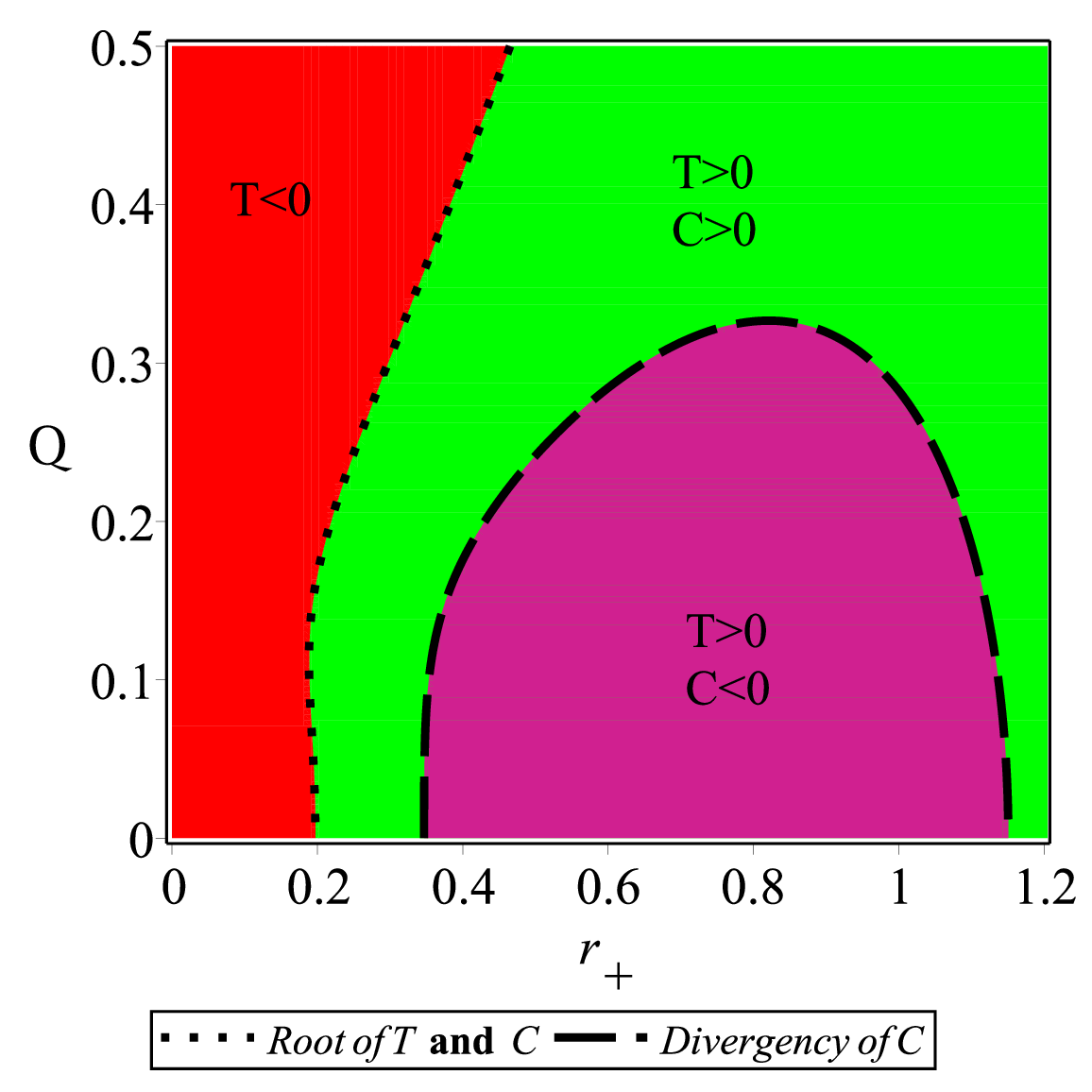}} 
\subfloat[Kerr-Newman BH]{
        \includegraphics[width=0.28\textwidth]{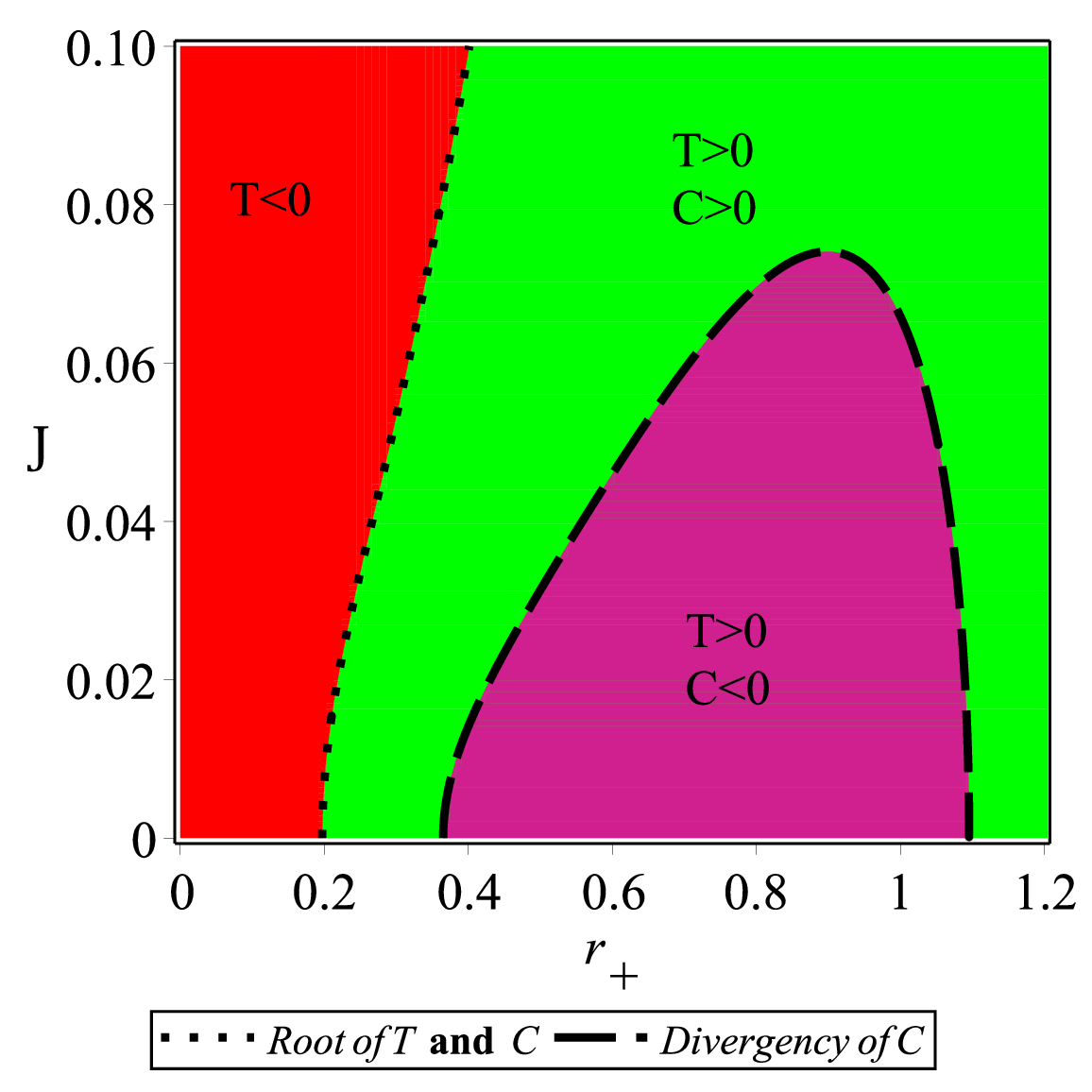}} \newline
\caption{Thermally stable and/or unstable regions of the black holes.}
\label{FigSt}
\end{figure}

\begin{figure}[tbh]
\centering
\subfloat[charged accelerating BH]{
        \includegraphics[width=0.28\textwidth]{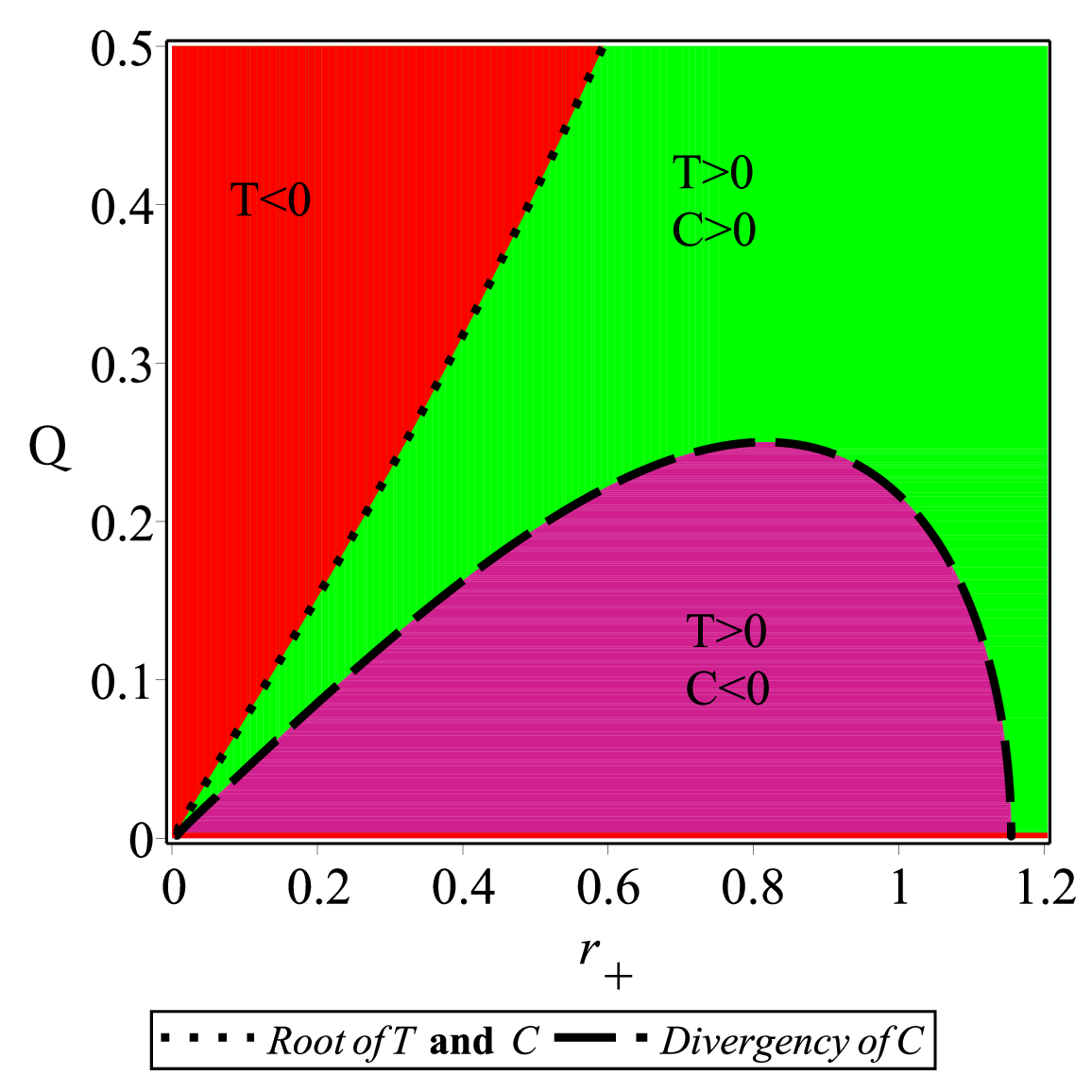}} 
\subfloat[charged accelerating BH]{
        \includegraphics[width=0.28\textwidth]{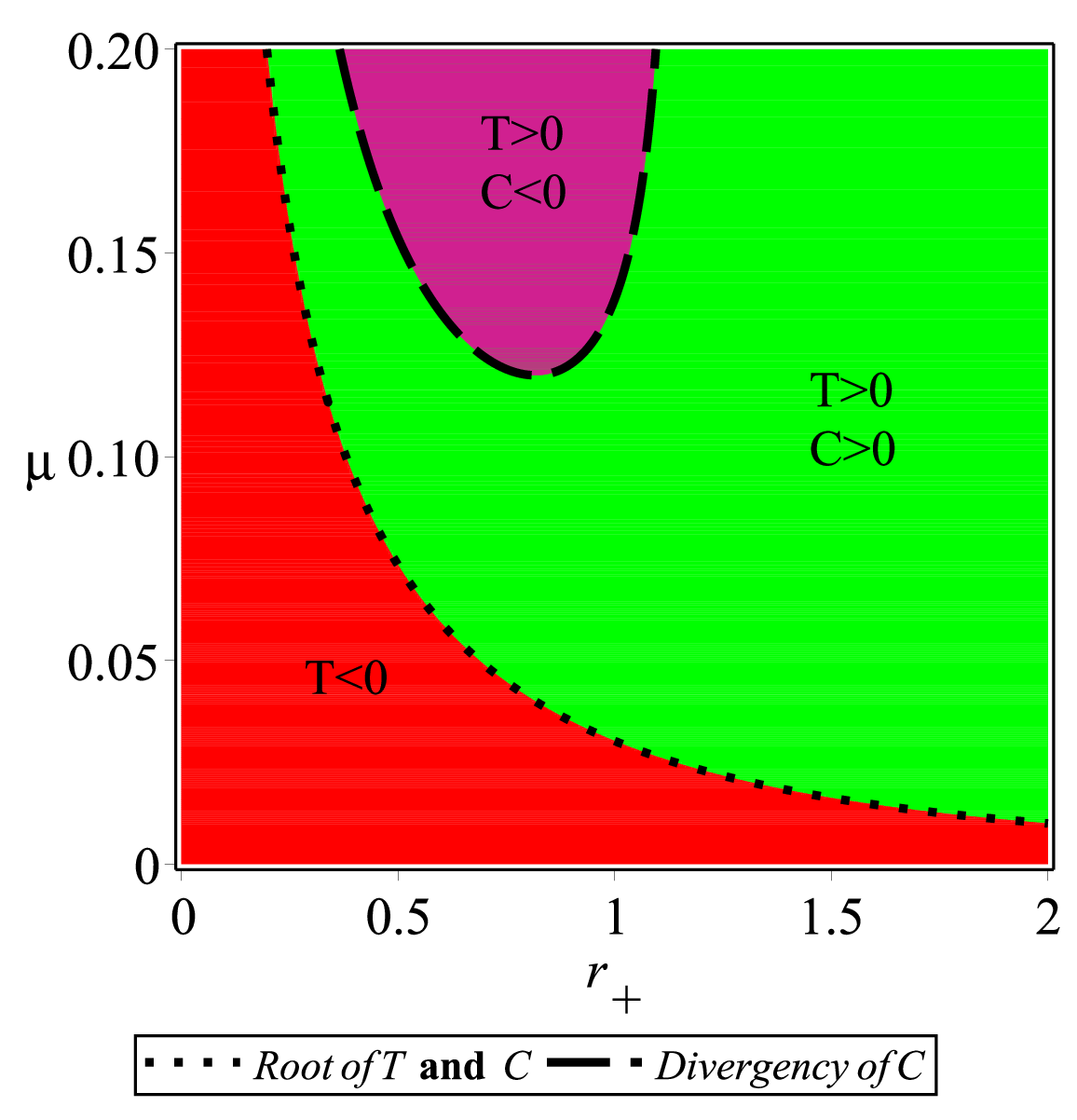}}\newline
\subfloat[ rotating and accelerating BH]{
        \includegraphics[width=0.28\textwidth]{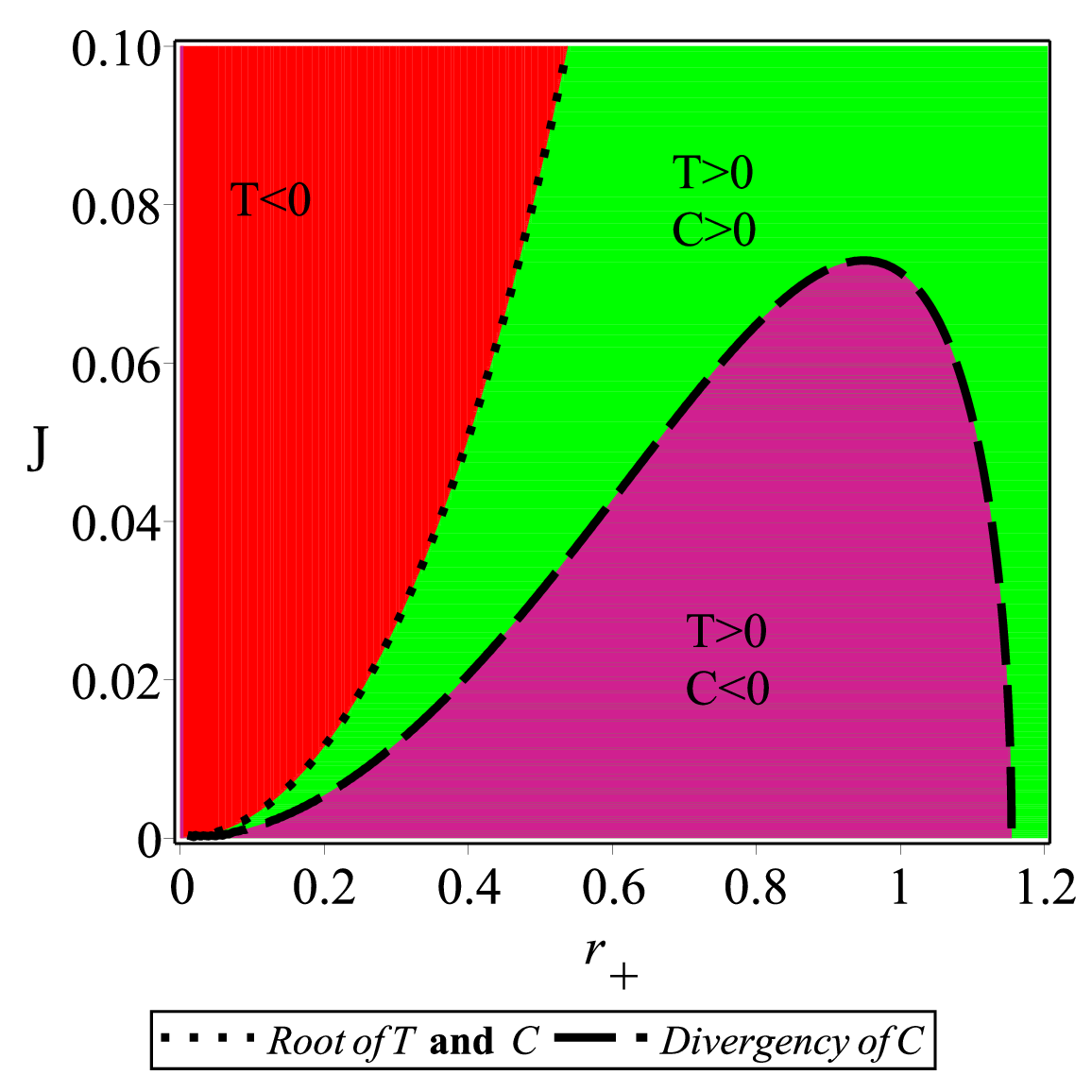}} 
\subfloat[rotating and accelerating BH]{
        \includegraphics[width=0.28\textwidth]{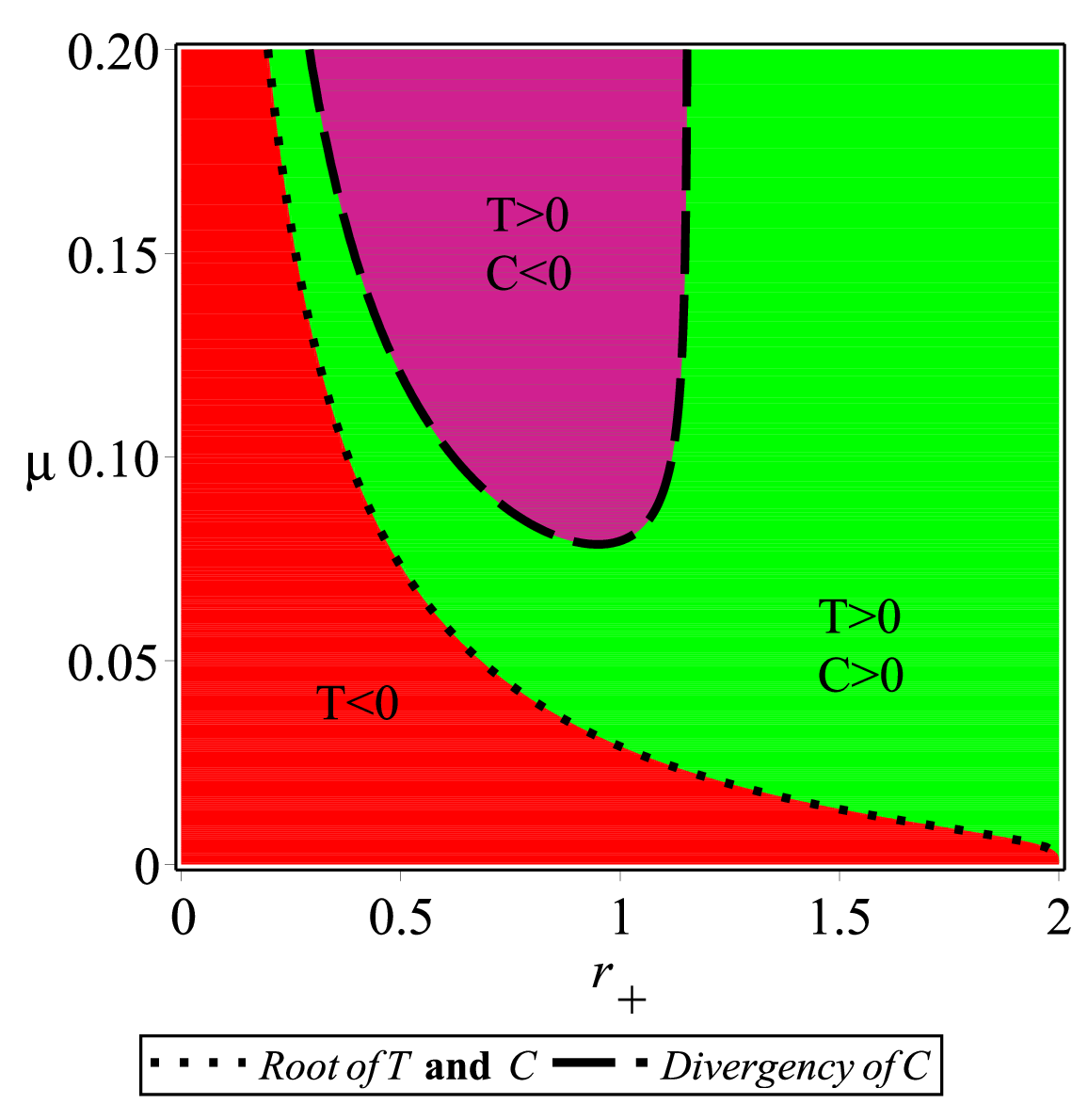}} \newline
\caption{Thermally stable and/or unstable regions of the black holes for $%
B=0.2 $, $\ell=2 $ and $A=0.02 $.}
\label{FigSt1}
\end{figure}

Now, we demonstrate that the qualitative behavior of these two parameters is
independent of the existence of acceleration. To do so, we depict thermally
stable/unstable regions in Fig. \ref{FigSt} (in the absence of acceleration)
and Fig. \ref{FigSt1} (in the presence of acceleration). Comparing Figs. \ref%
{FigSt}(a) and \ref{FigSt1}(a), one can notice that qualitative behavior of $%
Q_{e}$ is the same in the absence or presence of the acceleration. The only
difference is that the unphysical region increases with the existence of the
acceleration. Similar explanation can be used for $J_{e}$ (compare Figs. \ref%
{FigSt}(b) and Fig. \ref{FigSt1}(c), together). Comparing Figs. \ref{FigSt}(a) and Fig. %
\ref{FigSt}(c), we see that for a black hole located in a weak electric
field, the unphysical region increases by adding the rotation parameter to
the black hole and vice versa. We can use such discussion for rotating black
holes as well. A black hole with small rotation exits in its physical state
if an electric charge adds to it (compare Figs. \ref{FigSt}(b) and Fig. \ref%
{FigSt}(d), together).

Also, down panels of Fig. \ref{FigSt}, illustrate that the unstable region
decreases by adding the electric charge and rotation parameter to the black
hole. Such explanation is true regarding the string tension as well (compare
Figs. \ref{FigSt1} and Fig. \ref{Fig6}, together). This shows that by adding each of
these parameters to the black hole the system approaches more stability.%
\newline

\begin{center}
\textbf{B: Comparing the obtained results with \cite{Accel6,Accel7}}
\end{center}

First we compare our analysis to what was investigated in \cite{Accel6}. For
a more accurate comparison, we consider the angular momentum to be zero.
Regarding the fixed parameters, as we have already mentioned, our study is a
little different from \cite{Accel6} as the authors defined some
dimensionless quantities such as $\tilde{m}=mA$, $\tilde{e}=eA$, $\tilde{A}%
=A\ell$ and $\tilde{r}=\frac{r}{\ell}$, while we have only considered $%
\mathcal{B}=mA$ and $\beta=A\ell$ as fixed parameters. To study the effect
of the admissible region of parameter space, we first determine the
admissible parameter space in the context of our studies. Studying the
admissible space of the parameters is done by exploring the following three
conditions:

I) Existence of a black hole in the bulk. We can find it by investigating
the condition for an extremal black hole 
\begin{equation}
f(r_{e})=0=f^{\prime}(r_{e}),
\end{equation}
which results into the following relations 
\begin{eqnarray}
\mathcal{B}&=& \frac{r_{e}\mu\left( \mu (1-A^{2}r_{e}^{2})+\sqrt{%
\mu^{2}-A^{2}Q^{2}-2A^{2}\mu^{2}r_{e}^{2}}\right) }{AQ^{2}(1-A^{2}r_{e}^{2})}%
, \\
\beta &=&\frac{Ar_{e}}{\sqrt{2}}\frac{\sqrt{4A^{2}Q^{2}+A^{2}%
\mu^{2}r_{e}^{2}-3\mu^{2}+\sqrt{9\mu^{4}-8A^{2}\mu^{2}Q^{2}-6A^{2}r_{e}^{2}%
\mu^{4}}}}{(1-A^{2}r_{e}^{2})\sqrt{\mu^{2}-A^{2}Q^{2}}}.
\end{eqnarray}

\begin{figure*}[tbh]
\centering
\subfloat[$ \mu=0.15 $, $A=0.02$ and $ Q=0.2
$]{\includegraphics[width=0.28\linewidth]{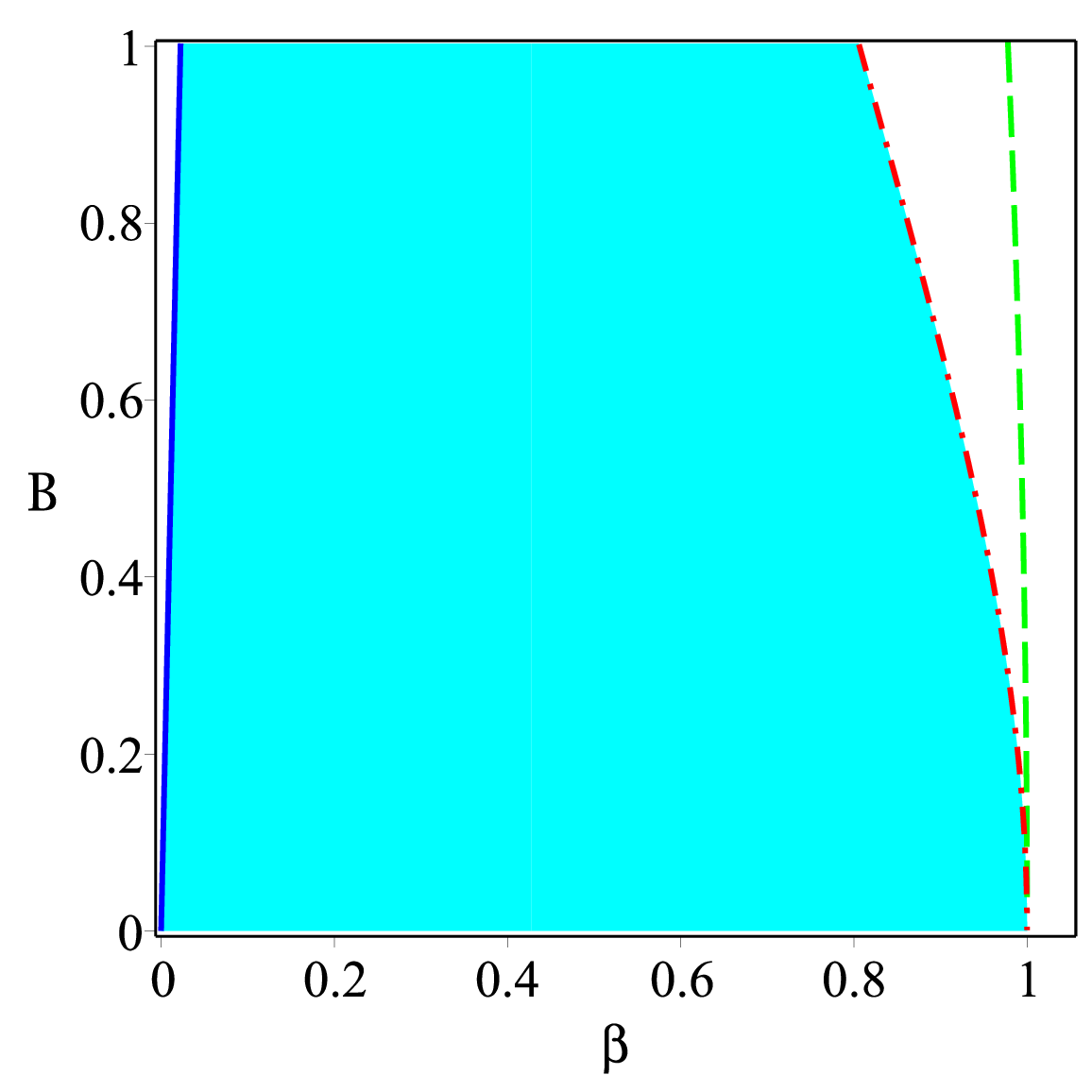}} 
\subfloat[$ \mu=0.15 $,
$A=0.1$ and $ Q=1 $]{\includegraphics[width=0.28\linewidth]{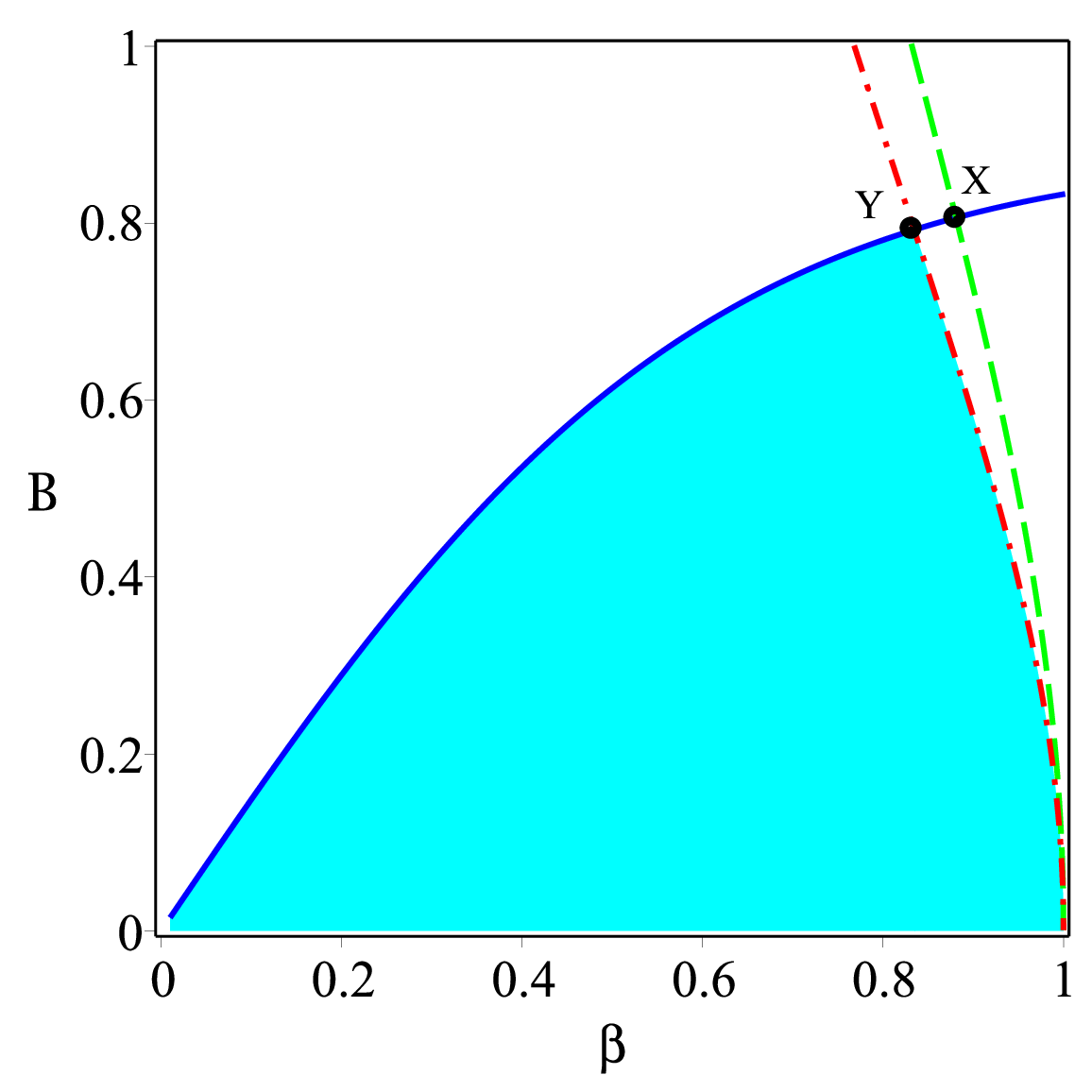}} 
\subfloat[$ \mu=0.1 $, $A=0.1$ and $ Q=1
$]{\includegraphics[width=0.28\linewidth]{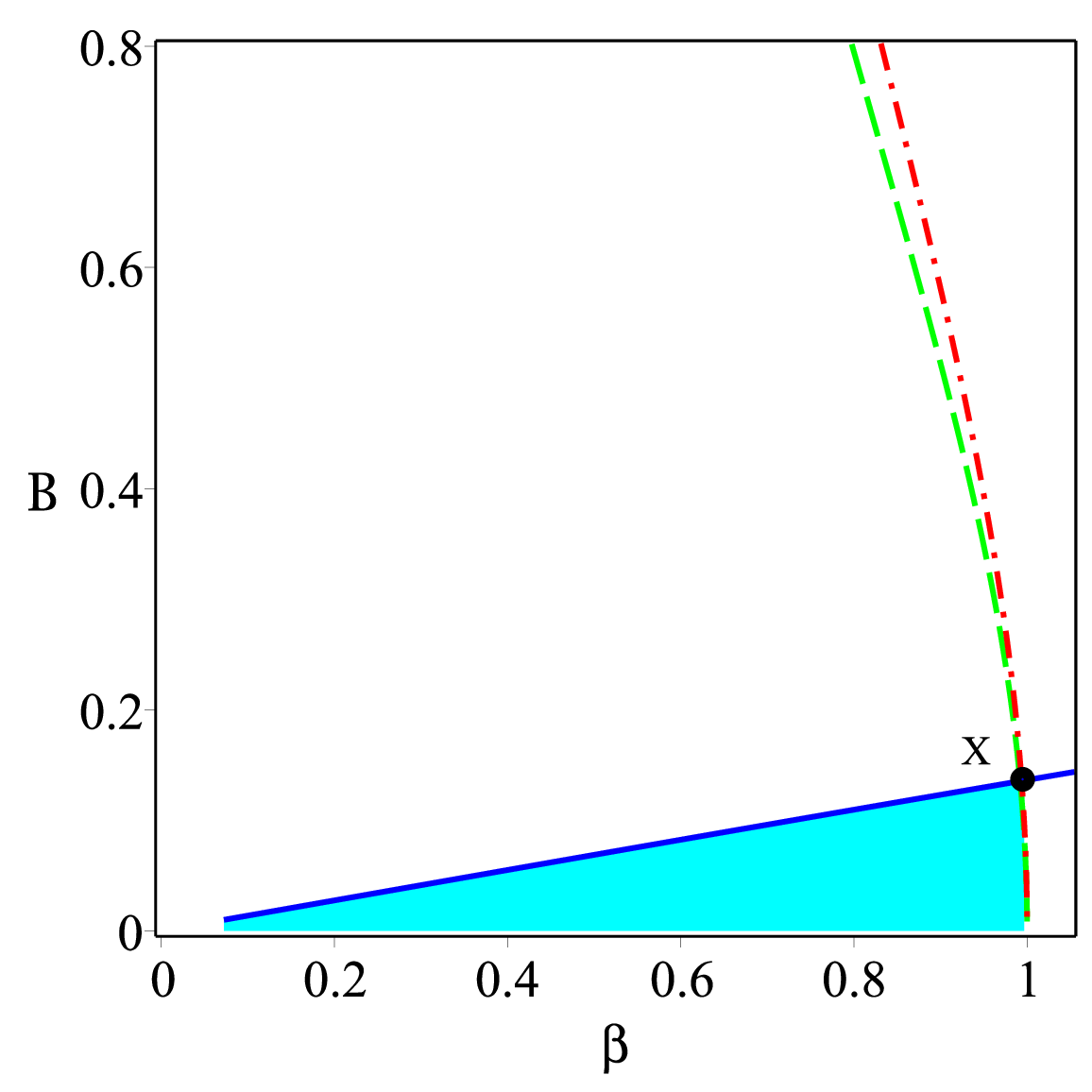}}\newline
\subfloat[$ \mu=0.01 $, $A=0.02$ and $ Q=0.2
$]{\includegraphics[width=0.28\linewidth]{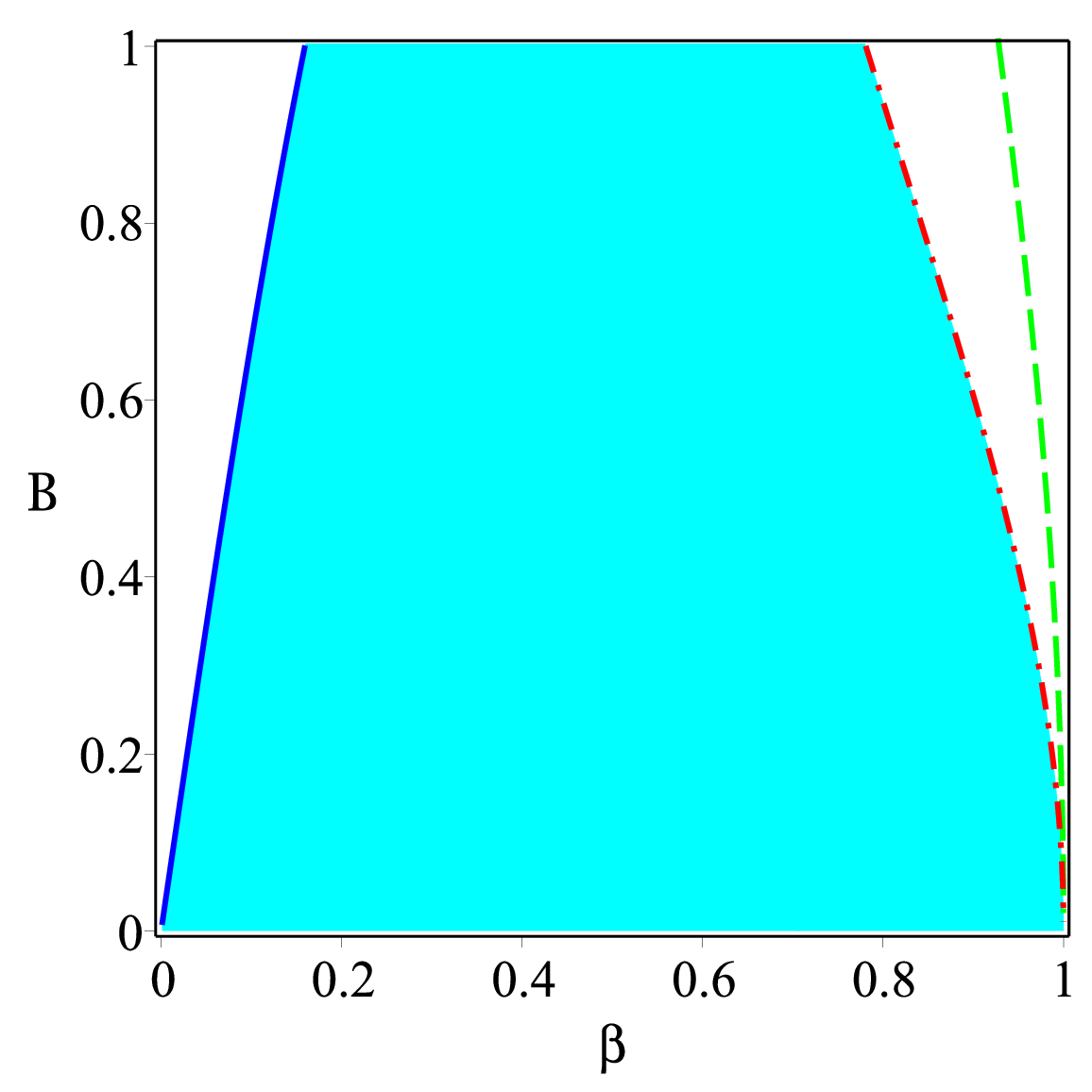}} 
\subfloat[$
\mu=0.005 $, $A=0.02$ and $ Q=0.2
$]{\includegraphics[width=0.28\linewidth]{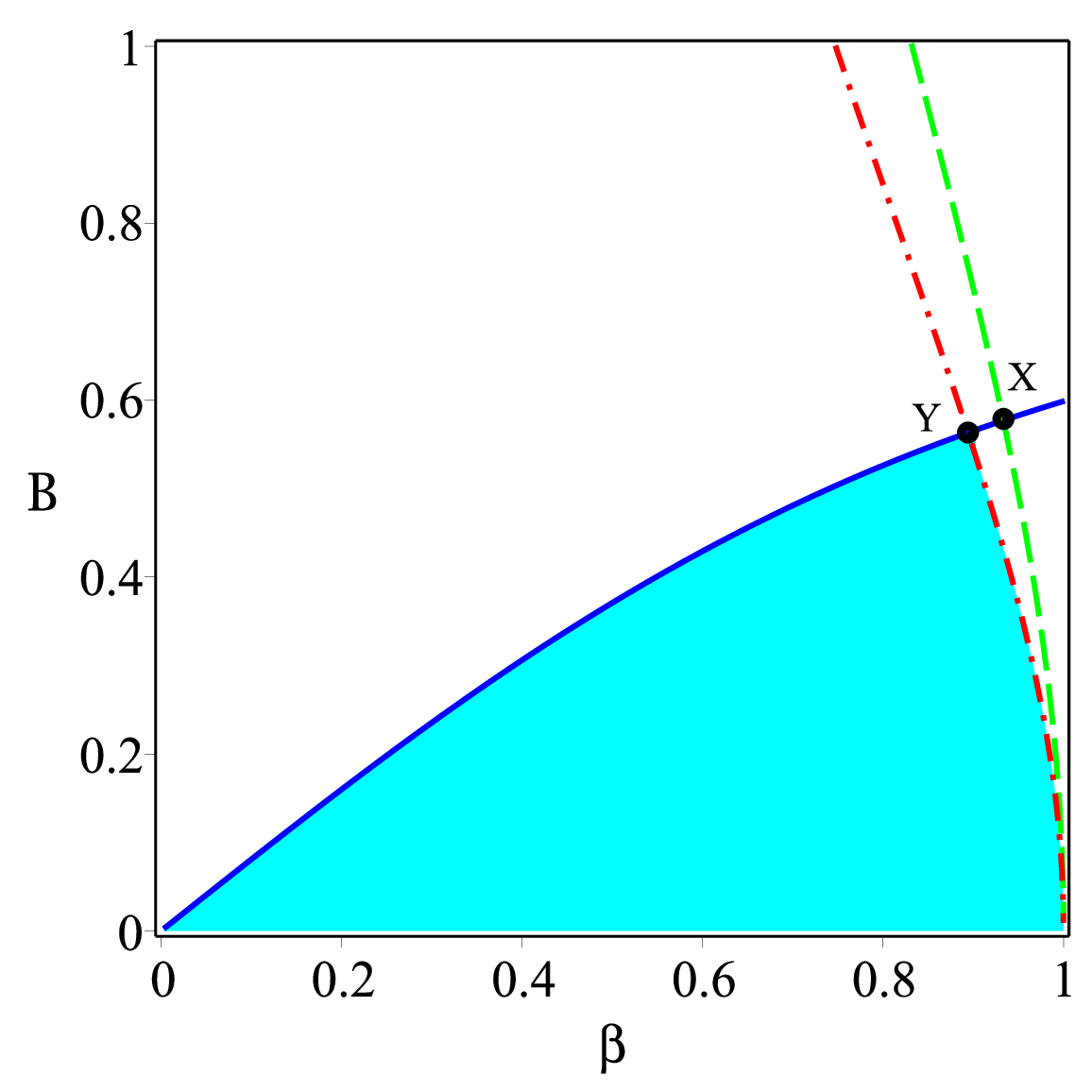}} 
\subfloat[$
\mu=0.004 $, $A=0.02$ and $ Q=0.2
$]{\includegraphics[width=0.28\linewidth]{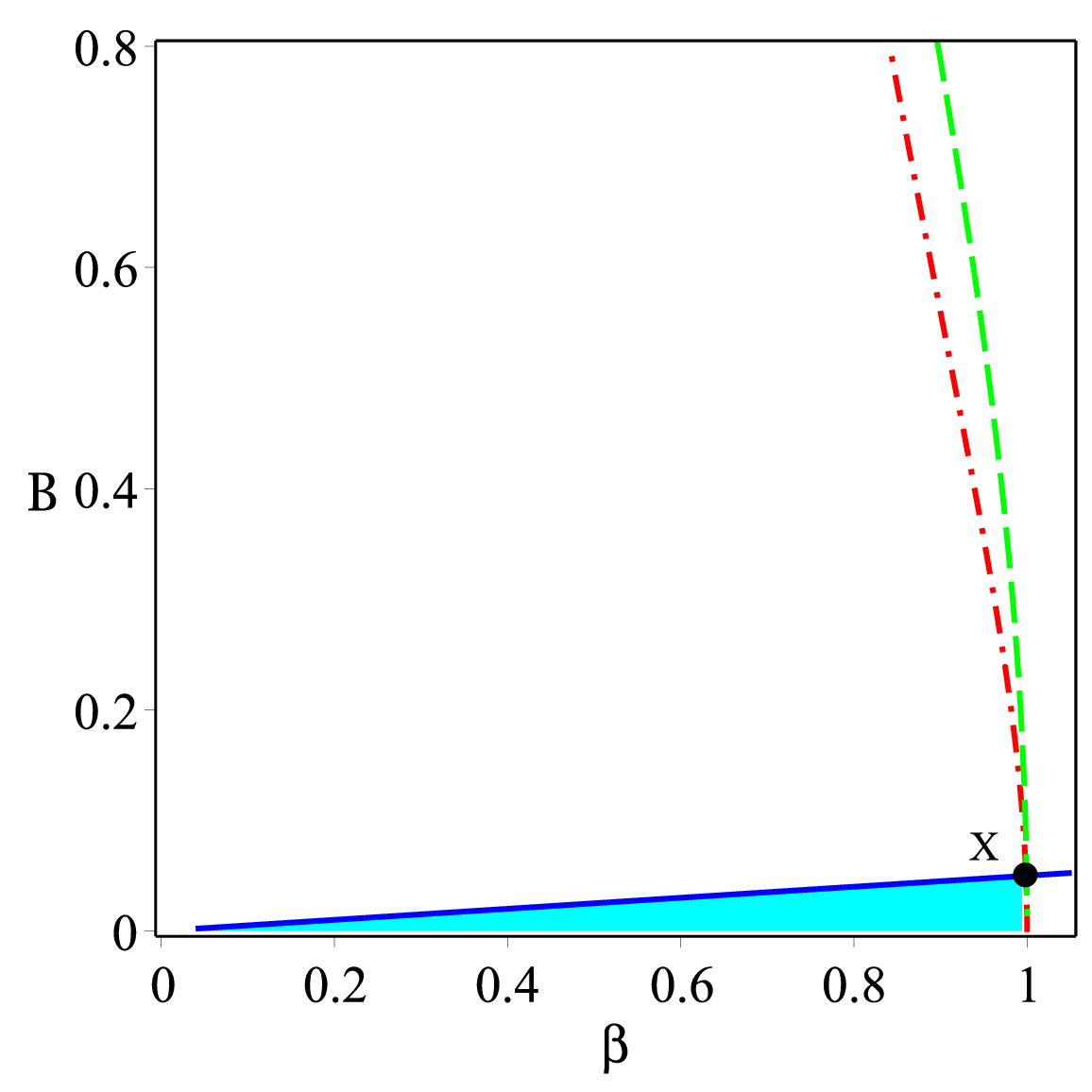}}
\caption{The admissible parameter space (denoted by shaded areas) is
depicted in the $B - \protect\beta $ plane. The blue curve is the boundary
for the existence of black holes in the bulk. The green dashed curve
corresponds to $\protect\alpha=0 $, and the red dash-dotted curve shows the
no-acceleration horizon condition.}
\label{Figmuq}
\end{figure*}

The resultant curve is displayed in Fig. \ref{Figmuq}, by the blue line
which shows a bound for the existence of black hole, denoting the
extremal limit. Below this line, a black hole (with two horizons) is
present, whereas no black hole exists above it.

II) Having well-defined thermodynamics. Such a condition is guaranteed by a
positive normalization factor. To study this condition, we plot the $%
\alpha=0 $ curve which leads to the following relation 
\begin{equation}
\mathcal{B}=\frac{\mu \sqrt{1-\beta^{2}}}{AQ\beta},
\end{equation}
the boundary is displayed in Fig. \ref{Figmuq}, by the green dashed curve.

%%%%%%%%%%%
%%%%%%%%%%%%%%%%%%%%%%%%%%%%%%%%%
III) Removing acceleration horizon. The sufficient condition for the slow
acceleration regime is that $f(r)$ does not develop any roots on the
boundary. This condition can be satisfied when 
\begin{equation}
f(x=-y)=0=f^{\prime}(x=-y),
\end{equation}
where $x=\frac{1}{Ar} $ and $y=\cos \theta $. This yields the following
relations
\begin{eqnarray}
\mathcal{B} &=&\frac{\mu^{2} \beta (y^{2}-1)+\mu \sqrt{\left((1-y^{2})
(\mu^{2}-A^{2}Q^{2})\beta^{2} +A^{2}Q^{2}\right)(1-y^{2}) }}{A^{2}Q^{2}\beta
y(1-y^{2})}, \\
\beta &=&\frac{1}{2}\frac{\mu^{2}(1-3y^{2})-4A^{2}Q^{2}(1-2y^{2})+ 2\mu\sqrt{%
\mu^{2}(1-3y^{2})^{2}+4A^{2}Q^{2}y^{2}(1-2y^{2})}}{(1-y^{2})\sqrt{%
\mu^{2}-A^{2}Q^{2}}}.
\end{eqnarray}

The corresponding parameter space can be plotted parametrically, for $y \in
[ -1, 1]$. The resultant curve which determines the boundaries of the
admissible regions is illustrated by the red dot-dash curve in Fig. \ref{Figmuq}.

In Ref. \cite{Accel6}, was shown that the snapping of the swallowtail is a result
of the existence of a critical slice through the point $X$ in the parameter
space. Black holes whose parameters are in the neighborhood of this point
are called mini-entropic. To study the mini-entropic black holes, we need to
examine the intersection of three curves. As we see from Fig. \ref{Figmuq}%
(a), for small (large) values of the electric charge and acceleration
parameter (string tension), the three curves do not intersect in the
admissible region. For the large electric charge, acceleration parameter and
string tension, one can observe two salient intersection points $X$ and $Y$
which prevents the formation of mini-entropic black holes (see Fig. \ref%
{Figmuq}(b)). These two points coincide with each other for values of $\mu$
in an intermediate range (see Fig. \ref{Figmuq}(c)). It is worth mentioning
that $B$ and $\beta $ will be negative for $\mu <0.1$ in such a condition.
As one can see from Figs. \ref{Figmuq}(d) and \ref{Figmuq}(e), the mini-entropic
black holes cannot be formed for very small values of string tension, unless
for $\mu <0.004$ (see Fig. \ref{Figmuq}(f)). It should be noted that real
values of $B$ and $\beta $ cannot be observed in this case. From Figs. \ref%
{Figmuq}(c) and \ref{Figmuq}(f), one can see that point $X$ lies outside of
the admissible region in parameter space but it is possible to come
arbitrarily close to it. %%%%%%%%%%%
%%%%%%%%%%
\begin{figure*}[tbh]
\centering
\includegraphics[width=0.28\linewidth]{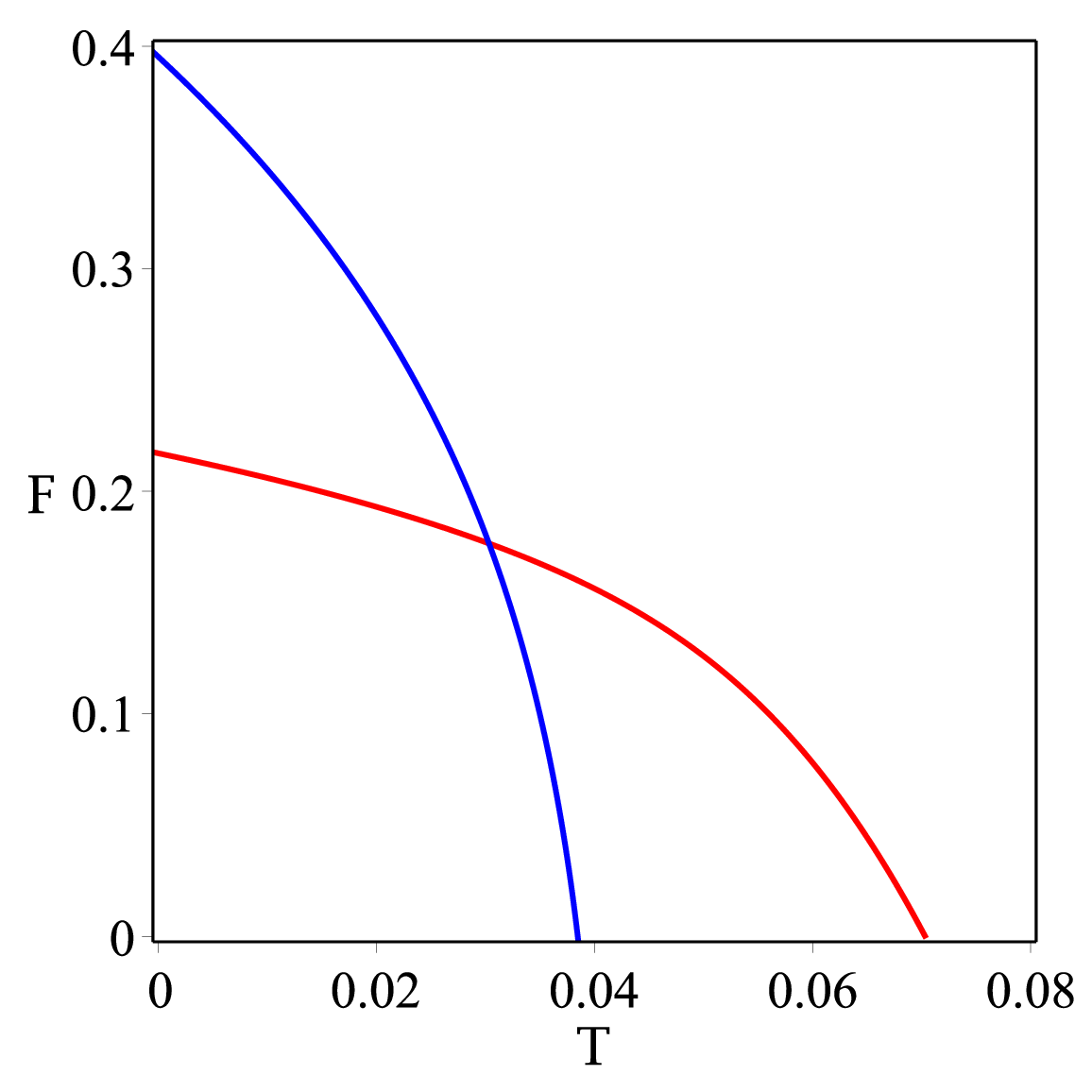}
\caption{The $F-T $ curves for $\mathcal{B}=0.05$, $\protect\beta =0.95$, $%
A=0.02$, $Q=0.2$, $\protect\mu=0.004 $ and $P=0.005 $ (red curve) and $%
B=0.15 $, $\protect\beta =0.95$, $A=0.1$, $Q=1$, $\protect\mu=0.1 $ and $%
P=0.001 $ (blue curve).}
\label{FigFTX}
\end{figure*}
%%%%%%%%%%%%%%%%%%%%%%%%%%%%%%%%%
For the mini-entropic black holes, their area is finite while their volume
diverges in this region (as it was already mentioned, the point $X$ is
located on $\alpha=0 $ curve and thermodynamic volume would diverge in such
a situation). Computing the isoperimetric ratio $\mathcal{R} $ 
\begin{eqnarray}
\mathcal{R}&=&\left( \frac{3V}{\mathcal{V}_{2}}\right) ^{\frac{1}{3}}\left( 
\frac{\mathcal{V}_{2}}{\mathcal{A}}\right) ^{\frac{1}{2}} \\
&=&\left( \frac{1+\varsigma}{\alpha}\right) ^{\frac{1}{3}}\left(
1-A^{2}r_{+}^{2}\right) ^{\frac{1}{2}} \geq 1,  \notag
\end{eqnarray}
where $\mathcal{V}_{2}=\frac{4\pi}{K} $ denotes the volume of the unit
2-sphere and 
\begin{equation*}
\varsigma=2A^{2}r_{+}^{2}+\frac{3A^{2}}{16\pi P}+ \frac{9A^{2}}{128\pi
^{2}P^{2}r_{+}^{2}}+\frac{9A^{2}\mathcal{B}^{2}Q^{2}}{128\pi ^{2}P^{2}\mu
^{2}r_{+}^{4}}.
\end{equation*}
%%%%%%%%%%%%%%%%%%%%%%%%%%%%%%%%

As we see the isoperimetric ratio diverges as point $X$ is approached. It is
obvious that the admissible region of parameter space and the obtained
results in our work is different from \cite{Accel6}. In our investigation,
the mini-entropic black holes only appear under special conditions. To
investigate the phase transition of mini-entropic black holes, we have
depicted Fig. \ref{FigFTX}. We find that these black holes are
thermodynamically stable and no phase transition is observed for them which
is unlike \cite{Accel6} where a zeroth-order phase transition was observed
around point $X$.

\begin{figure*}[tbh]
\centering
\includegraphics[width=0.28\linewidth]{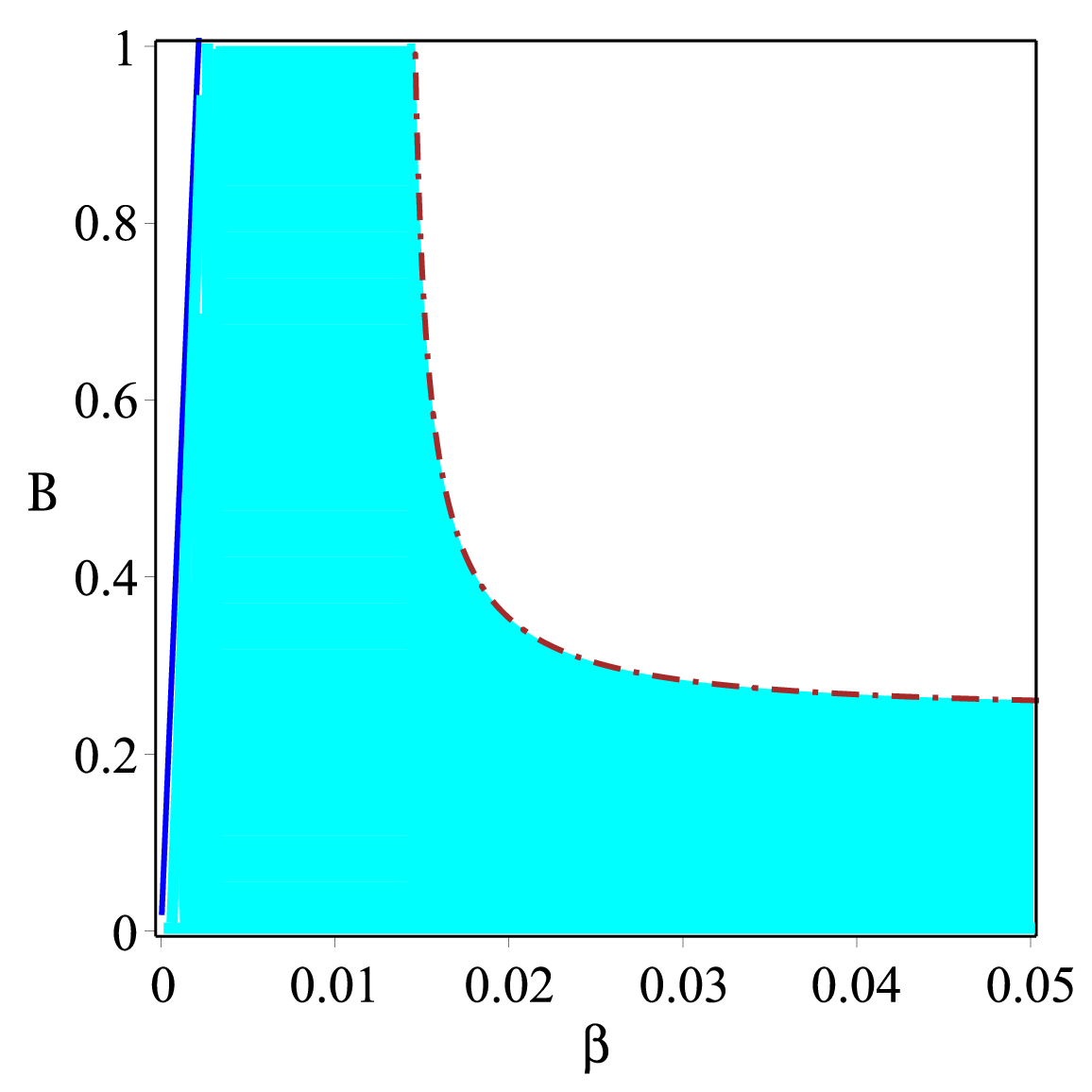}
\caption{The admissible parameter space (denoted by shaded areas) is
depicted in the $B - \protect\beta $ plane. The blue curve is the boundary
for the existence of black holes in the bulk, and the red dash-dotted curve
shows the no-acceleration horizon condition. The curve related to $\protect%
\alpha=0 $ is far from this region.}
\label{FigMX9}
\end{figure*}

Now, we compare our investigation with what was obtained in \cite{Accel7}.
Here, we consider the electric charge to be zero. In Ref. \cite{Accel7} similar
to Ref. \cite{Accel6}, some dimensionless quantities were defined such as $\tilde{%
m}=mA$, $\tilde{a}=aA$, $\tilde{A}=A\ell$ and $\tilde{r}=\frac{r}{\ell}$.
Drawing admissible parameter space in this case, we noticed that none of the
three mentioned curves intersect each other (see Fig. \ref{FigMX9}). For more
clarity, we drew the shape up to $\beta=0.05$. The curve related to $%
\alpha=0 $ is far from this region. To avoid repetition, we skip to writing
the related equations. We see from Fig. \ref{FigMX9}, a
mini-entropic black hole cannot be observed in our investigation for
rotating and accelerating black holes. As we have pointed out, no
zeroth-order phase transition occurs in our analysis, so one cannot study
the "fine splitting" phenomena for these black holes unlike \cite{Accel7}.

%%%%%%%%%%
%%%%%%%%%%%%%%%%%%%%%%%%%%%%%%%%

\begin{center}
\textbf{C: the heat engine efficiency}
\end{center}

As it was mentioned, the efficiency is obtained by following relation, 
\begin{equation}
\eta =\frac{\Delta P\Delta V}{\Delta M_{T}+\Delta U_{L}},  \label{Eq214}
\end{equation}%
where $\Delta V=V_{2}-V_{1}$, $U=M-PV$ and 
\begin{eqnarray}
\Delta M_{T} &=&M(V_{2},P_{1})-M(V_{1},P_{1}), \\
&&  \notag \\
\Delta U_{L} &=&U(V_{1},P_{1})-U(V_{1},P_{4}).  \label{Eq213}
\end{eqnarray}

The total mass of such black holes in the extended phase space is given by

\begin{equation*}
M=\frac{\alpha \mu }{B\Xi }\left( \frac{1}{2}(1+H)DV^{\frac{1}{3}}+\frac{%
2DJ^{2}B^{4}V^{\frac{1}{3}}}{\mu ^{4}y^{2}}+\frac{(1+4B)Q^{2}}{DV^{\frac{1}{3%
}}}+\frac{4R}{3\left( 1-A^{2}D^{2}V^{\frac{2}{3}}\right) }\right) ,
\end{equation*}%
where 
\begin{eqnarray*}
D &=&\frac{1}{2}\left( \frac{6B}{\pi \mu }\right) ^{\frac{1}{3}}, \\
&& \\
R &=&(1+3H)\pi ^{2}D^{3}PV+\frac{4\pi ^{2}J^{2}D^{3}B^{4}PV}{\mu ^{4}y^{2}},
\\
&& \\
y &=&D^{2}V^{\frac{2}{3}}+Q^{2}+4Q^{2}B+\frac{8\pi PD^{4}V^{\frac{4}{3}}}{3},
\\
&& \\
H &=&\frac{A^{2}Q^{2}}{6}-\frac{\beta ^{2}}{6}-\frac{2A^{2}D^{2}V^{\frac{2}{3%
}}}{3}-\frac{4J^{2}B^{4}}{3\mu ^{4}y^{2}}-\frac{2J^{2}B^{4}}{3D^{2}\mu
^{4}V^{\frac{2}{3}}y}-\frac{3AB}{64\pi ^{2}P^{2}D^{3}V}.
\end{eqnarray*}

Also, the Carnot efficiency is expressed as 
\begin{equation}
\eta _{C}=1-\frac{T_{C}}{T_{H}}=1-\frac{T_{4}(P_{4},V_{1})}{%
T_{2}(P_{1},V_{2})}.  \label{Eq215}
\end{equation}

In the extended phase space, one can obtain temperature as follows 
\begin{equation*}
T=\frac{1}{Z_{1}}\left( \frac{32\pi J^{2}D^{3}B^{4}PV}{3\mu ^{4}y^{2}}-\frac{%
(1+4B)Q^{2}}{DV^{\frac{1}{3}}}-\frac{4DJ^{2}B^{4}V^{\frac{1}{3}}}{\mu
^{4}y^{2}}+A^{2}D^{3}VZ_{2}+Z_{3}\right) ,
\end{equation*}%
where 
\begin{eqnarray*}
Z_{1} &=&4\pi \alpha \left( (1+2H)D^{2}V^{\frac{2}{3}}+\frac{%
4D^{2}J^{2}B^{4}V^{\frac{2}{3}}}{\mu ^{4}\left[ D^{2}V^{\frac{2}{3}%
}+Q^{2}+4Q^{2}B+\frac{8}{3}\pi PD^{4}V^{\frac{4}{3}}\right] ^{2}}\right) , \\
&& \\
Z_{2} &=&\frac{(1+4B)Q^{2}}{D^{2}V^{\frac{2}{3}}}+\frac{16}{3}\pi PD^{2}V^{%
\frac{2}{3}}-1, \\
&& \\
Z_{3} &=&8(1+3H)\pi PVD^{3}+(1+H)DV^{\frac{1}{3}}.
\end{eqnarray*}

\end{document}